\documentclass[11pt,showkeys,superscriptaddress,nofootinbib,amssymb,amsmath,aps,prc,eqsecnum]{revtex4}
\usepackage[dvips]{graphicx}
\usepackage{latexsym,epsfig,bm,times,psfrag}
\usepackage{color}
\usepackage{dcolumn}                 
\usepackage{subfigure}

\begin{document}

\title{The effect of strong electric field on the evolution of charmonium in quark gluon plasma}
\author{Biaogang Wu}\email{wubg@impcas.ac.cn}
\affiliation{Institute of Modern Physics, Chinese Academy of Sciences, Lanzhou 730000, China}
\affiliation{University of Chinese Academy of Sciences, Beijing 100049, China}
\author{Baoyi Chen}\email{baoyi.chen@tju.edu.cn}
\affiliation{Department of Physics, Tianjin University, Tianjin 300350, China}
\author{Xingbo Zhao}\email{xbzhao@impcas.ac.cn}
\affiliation{Institute of Modern Physics, Chinese Academy of Sciences, Lanzhou 730000, China}
\affiliation{University of Chinese Academy of Sciences, Beijing 100049, China}

\date{\today}
\begin{abstract}
In ultra-relativistic heavy-ion collisions, the strong electric field can be produced by the colliding nuclei. The magnitude of the electric field $E$ is on the order of $eE\sim m_{\pi}^2$ at the early stage of the collision. In quark gluon plasma (QGP), such a strong electric field can have a significant impact on the evolution of charmonia. We employ the time-dependent Schr\"odinger  equation to study the evolution of charmonium states in the strong electric field generated by the moving charges. The electric field can result in transitions between charmonium states with different angular momenta. In order to see this effect, we make comparisons between the yields of $ J/\psi$, $\psi'$ and $\chi_c$ with and without the electric field. The results show that in the early stage of the collision the electric field induces significant dissociation of $J/\psi$. In the meantime, $\chi_c$ is generated via the transition from $J/\psi$ by the electric field.
\end{abstract}

\maketitle
\section{Introduction}    

Deconfined quark-gluon plasma (QGP) is expected to form in relativistic heavy-ion collisions due to high energy density and high temperature. Relativistic Heavy-Ion Collider (RHIC) at Brookhaven National Laboratory (BNL) performs experiments in Au -Au collisions at center-of-mass energy $\sqrt{s_{NN}} = 200\,{\rm GeV}$ per nucleon pair. Large Hadron Collider (LHC) at CERN performs experiments in Pb - Pb collisions at center-of-mass energy $\sqrt{s_{NN}} = 2.76\,{\rm TeV}$. A lot of signals indicating the existence of QGP have been observed and studied in details in the past decades\cite{PhysRevC.78.034915,PhysRevLett.106.192301,IN3,PhysRevLett.89.244102,IN5,PhysRevC.72.064901}. Heavy quarkonia, due to their large mass, have been proposed as one of the ideal probes for the early stages of heavy ion collisions\cite{Matsui:1986dk}. Charmonium mass spectrum has been well studied with the parameterized Cornell potential, where relevant parameters can be fixed by the mass of low-lying charmonium states in vacuum\cite{PhysRevD.17.3090,PhysRevD.21.203,PhysRevD.50.2297,PhysRevLett.34.369}. At finite temperature, lattice QCD calculations suggest that the heavy quark potential inside quarkonium is partially screened by the deconfined medium\cite{Karsch:2003jg}.  Color screening effect sequentially melts the charmonium bound states at different temperatures.

The charmonium binding energy is usually taken as the difference between the charmonium mass and the open-charm threshold,
\begin{equation}
   \epsilon_B^0=2m_D-m_{\Psi},
   \label{b_e}
\end{equation}
with $m_D\simeq1.87\,$GeV. In vacuum the $D{\bar D}$ pair is usually considered as the open charm threshold for charmonium states. The low-lying charmonium states typically have binding energies on the order of several hundred MeV, e.g., $\epsilon_B^{J/\psi}=640\,$MeV.

In relativistic heavy-ion collisions, external electric field is produced by colliding heavy ions and is on the order of a few $m^2_{\pi}$ in the early stage of the collisions\cite{BZDAK2012171}. Note that $J/\psi$ mean radius is around $\langle r \rangle_{J/\psi}\sim0.5$\,fm. The electric potential energy between $c$ and $\bar c$, $eE\langle r\rangle $, is about several hundred MeV which is comparable with the binding energy of charmonium states.
Therefore, it is necessary to study the effects of external electric field on the evolution of charmonium states in the deconfined medium.
In this work, we focus on the effects of initial electric field on primordially produced charmonium.

The paper is organized as follows. In Sec.\,{\ref{fram}}, a brief introduction to the method we used in our calculation is presented. We explain the origin of screening effect of QGP and the source of the time-dependent electric field in the collision. In the end of this section we give the angular-momentum decomposed form of the Hamiltonian. Sec.\,{\ref{MSD}} shows numerical results with both electric field and QGP and a comparison is made with the results without electric field. Finally, conclusions and outlook are given in Sec.\,{\ref{Discussion}}.

\section{Framework}
\label{fram}
We adopt the non-relativistic Schr\"odinger equation to describe the evolution of the charmonium states in QGP\cite{Karsch:1987zw},
\begin{equation}\label{SEqn1}
\begin{split}
   &i{\partial \over \partial t}\psi(r,\theta)=\hat H \psi(r,\theta),\\
   &{\hat H}={1\over 2\mu}{\hat p}^2+V_{\bar q q}(r;T)+V_E(r,\theta),
     \end{split}
\end{equation}
here $r=|\boldsymbol{r}_{c}-\boldsymbol{r}_{\bar c}|$ is the length of the relative coordinate between $c$ and $\bar c$. $\psi(r,\theta)$ is the wave function of the charmonium and $\mu$ is the reduced mass. We take charm quark mass as 1.25\,GeV in our calculation. $V_{\bar q q}(r;T)$ is the potential between $c$ and $\bar c$ and it depends on the temperature $T$ of the medium after heavy-ion collisions. $V_E(r,\theta)$ is the potential of the electric field. We will give detailed explanations of these two potentials in the following parts.

\subsection{The potential between $c$ and $\bar c$ in QGP}\label{Review}
Heavy quarks move inside the charmonium with a speed $\langle v^2/c^2\rangle \sim 0.25$ for $J/\psi$\cite{Karsch:1987zw}. The charmonium can be described in non-relativistic Schr\"odinger approach with the Cornell potential which gives the mass spectrum of the charmonium states\cite{PhysRevD.17.3090,PhysRevD.21.203},
\begin{equation}\label{Eq1}
V_{\bar q q}(r;T=0)=-{4\over 3}{\alpha_s \over r}+\sigma r,
\end{equation}
with $\alpha_s \simeq 0.2$ and $\sigma \simeq 1\,\text{GeV/fm}$\cite{Satz_2006}. The first term originates from the one-gluon exchange interaction, and the second linear term reflects the confining interaction.

If a charmonium is immersed in QGP, the color force between $c$ and $\bar c$ is screened by the surrounding colored partons in a way similar to the electron plasma: the $c(\bar c)$ quark attracts partons with opposite color charges and forms the "Debye cloud". A phenomenological ansatz for the screened Cornell potential\cite{Karsch1988} is,
\begin{equation}\label{Eq3}
V_{\bar q q}(r;T)={\sigma\over\mu_D\left(T\right)}\left(1-\text{e}^{-\mu_D\left(T\right)r}\right)-{4\alpha_s\over 3r}\text{e}^{-\mu_D\left(T\right)r}.
\end{equation}

\begin{figure}[!htb]
   \begin{center}
      \includegraphics[width = 8.0cm]{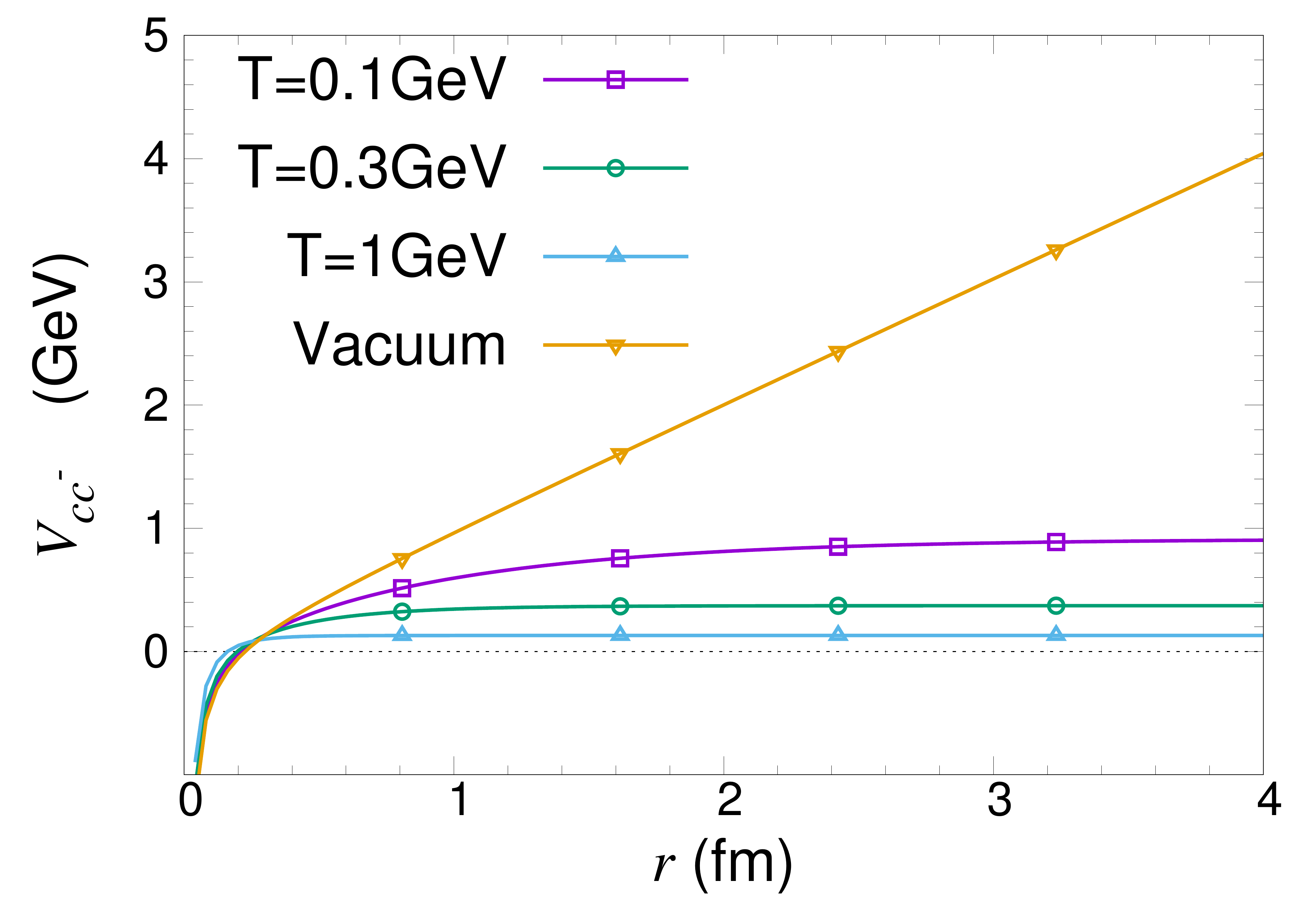}
      \includegraphics[width = 8.0cm]{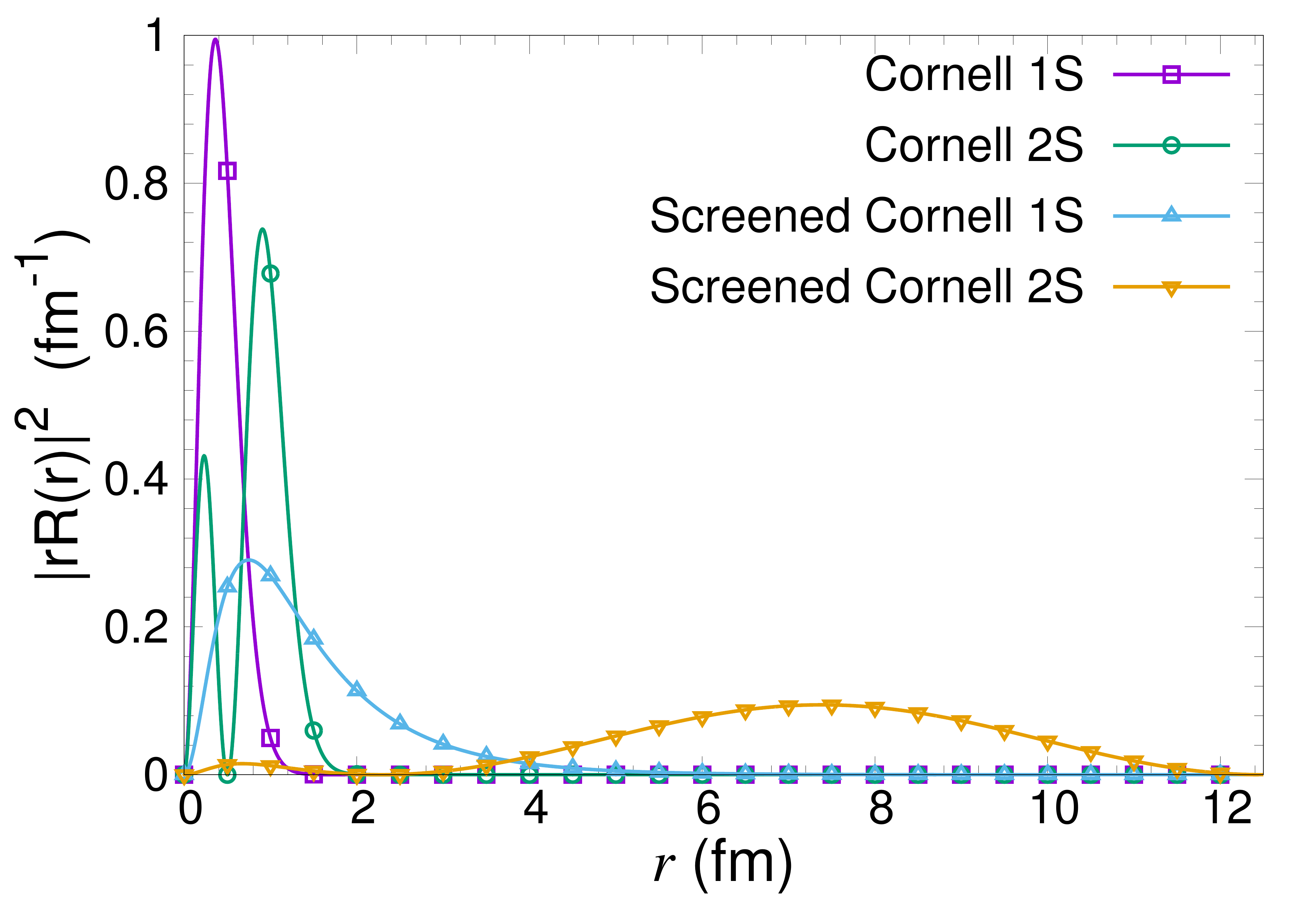}
     \caption{The left panel shows vacuum Cornell potential and the screened Cornell potential at different temperatures; the right panel shows the radial distribution of the 1S and 2S state of the vacuum Cornell potential and the screened Cornell potential.}\label{fig1}
   \end{center}
\end{figure}

Debye-screening lowers the charmonium binding energies, and contributes to the charmonium dissociation rate. According to thermal pQCD, the Debye mass is related to the temperature of the medium $T$ via
\begin{equation}\label{Eq4}
\mu_D^2\left(T\right)= g^2T^2\left({N_c\over 3}+{N_f\over 6}\right),
\end{equation}
here $g=1.5$ is the strong coupling constant. $N_c=3$ is the number of colors and $N_f=3$ is the number of flavors.
The vacuum Cornell potential and the screened Cornell potentials at different temperatures are shown in the left panel of FIG.\ref{fig1}. We can see that the screened Cornell potential at large $r$ decreases with temperature. The confining potential becomes very weak at temperature above critical temperature $T_c=170\,$MeV\cite{ct}. From FIG.\ref{fig1}, at sufficiently high temperature $c$ and $\bar c$ cannot form bound state any more because the color interaction between them is screened. Based on this mechanism $J/\psi$ suppression was first suggested in 1986 as a signature of QGP\cite{Matsui:1986dk}.
In this paper we ignore the spin degrees of freedom and thus the fine splittings are ignored. $J/\psi, \psi'$, and $\chi_c$ denote the eigenstates of the vacuum Cornell potential in Eq.(\ref{Eq1}) and throughout this paper they are used interchangeably with 1S, 2S, and 1P, respectively. 1S and 2S eigenstates of the vacuum Cornell potential and the screened Cornell potential are compared in the left panel of FIG.\ref{fig1}. We see that the eigenstates of the screened Cornell potential tends to dissolve.
\subsection{Electric field in QGP after heavy-ion collisions}\label{Elt}
\begin{figure}[!htb]
   \begin{center}
      \includegraphics[width = 8.0cm]{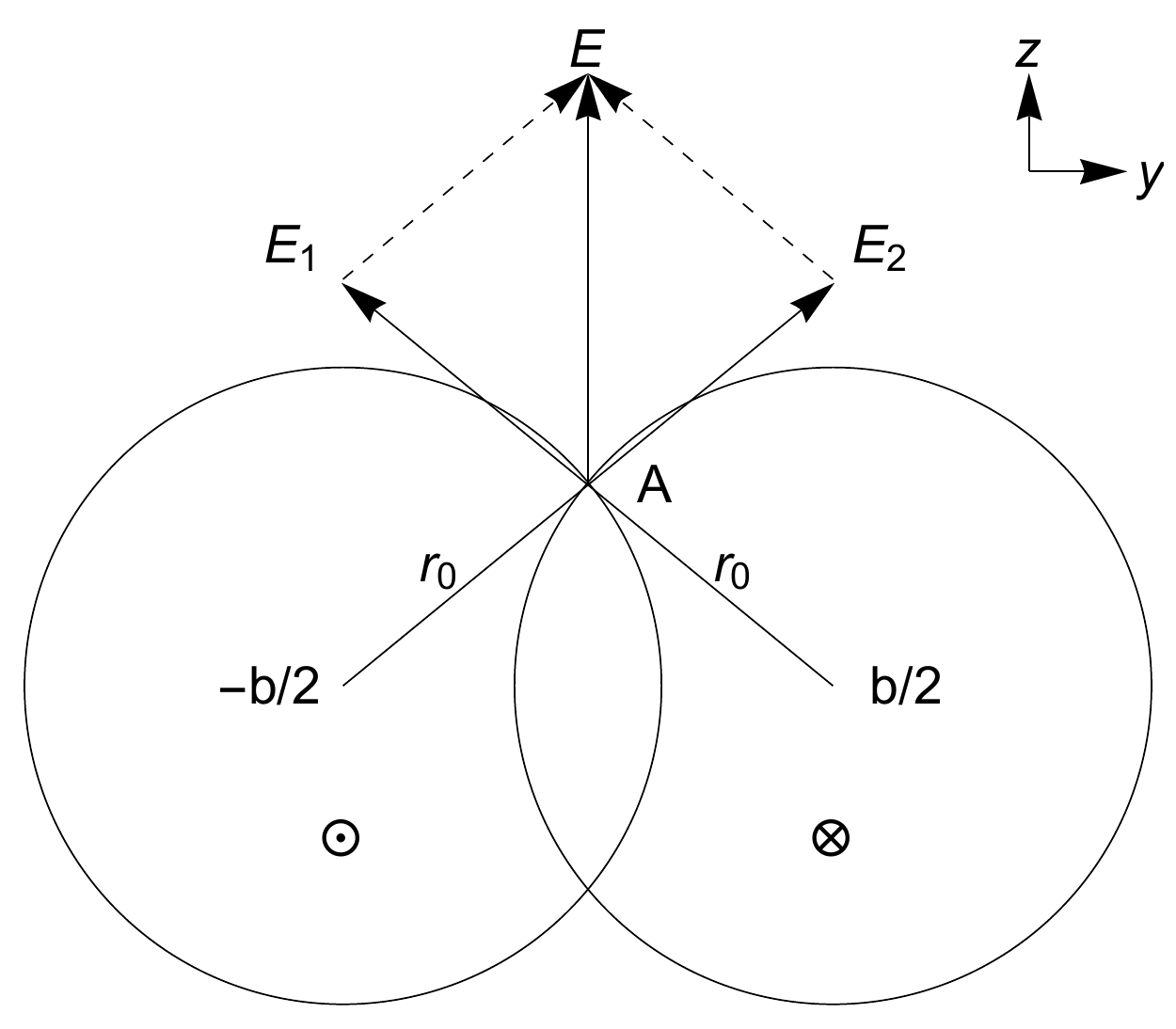}
      \includegraphics[width = 8.0cm]{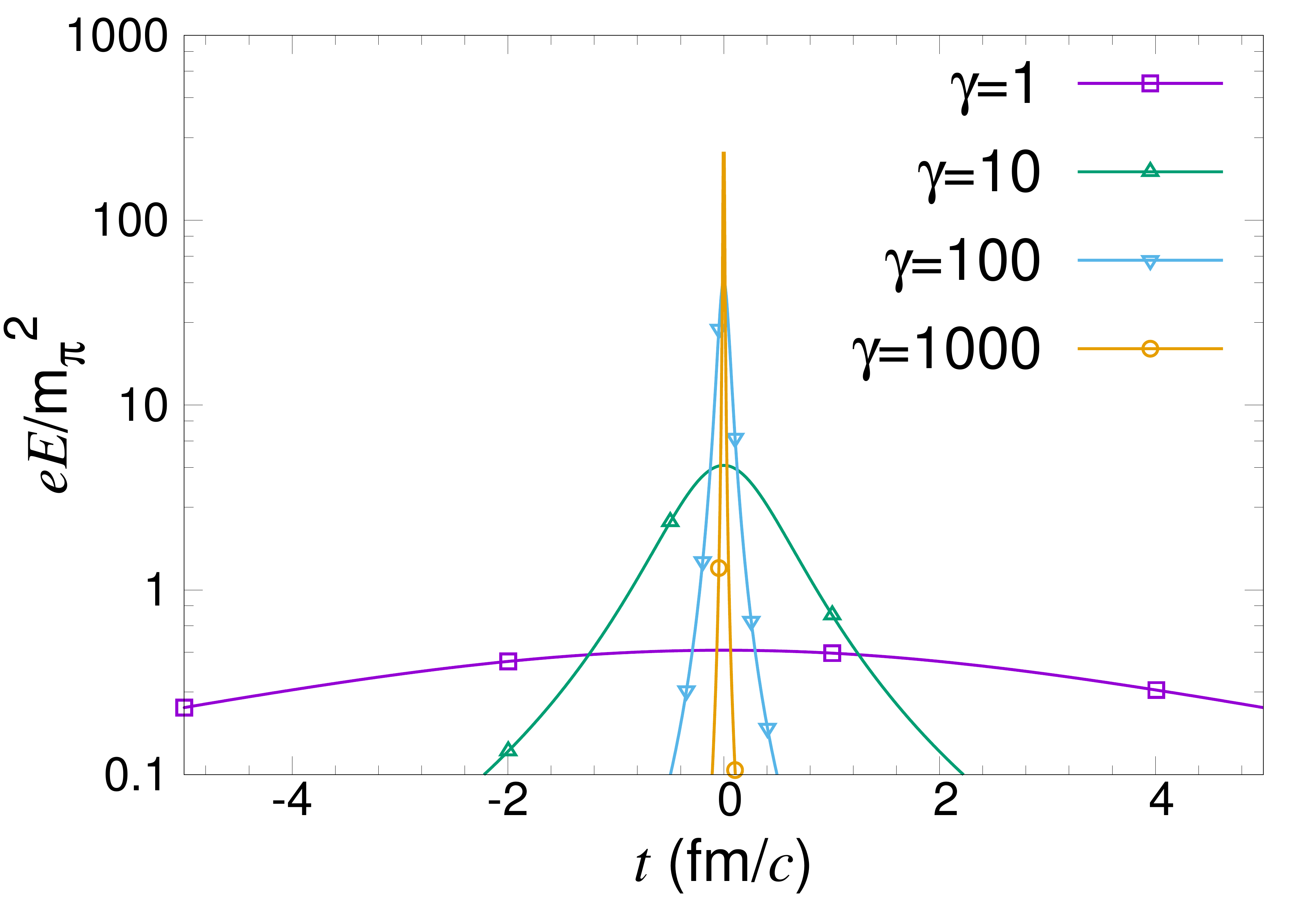}
\caption{The left panel: the electric field at point A is a combination of the electric fields generated by the two moving heavy ions. The right panel shows $eE(\gamma,t)$ in the unit of $m^2_{\pi}$ ~(with $m^2_{\pi}=0.018\,{\rm GeV}^{2}$).
}\label{fig2}
\end{center}
\end{figure}

Ultra-relativisic heavy-ion collisions can create a strong and time-dependent electric field with a peak magnitude $eE\sim m_{\pi}^2$\cite{PhysRevC.85.044907}. The electric field $E(t)$ in this work originates from two moving heavy ions with impact parameter $b$ and is schematically shown in the left panel of FIG.\ref{fig2}.

Electric field generated by moving charges after heavy ion collisions is position-dependent. In order to see the order of the effect of the electric field on the evolution of the charmonium states, we take a representative point A located at the surface of the intersection of the two heavy ions as shown in the left panel of FIG.\ref{fig2}. The heavy ions are moving in the direction perpendicular to the plane of the page and the heavy ion at right side is moving inward while the other one is moving outward. The electric field at point A generated by the two moving heavy-ions is given by Eq.(\ref{Efield}). It is in $z$ direction and is the combination of the electric fields generated by the two moving heavy ions. In our calculation the electric field is taken as a uniform electric field identical to the electric field at point A. We defer the study of non-uniform electric field to a future paper. The electric field strength at point A generated by one of the two moving heavy ions with a radius $r_0$ and Lorentz factor $\gamma ={1\over \sqrt{1-v^2}}$ is\cite{9780471309321}:
\begin{equation}
E={\gamma Zer_0\over (r_0^2+\gamma^2v^2t^2)^{3/2}},
\end{equation}
which gives the total electric field strength at point A as:
\begin{equation}
\begin{split}
eE=&{\gamma Ze^2r_0\over (r_0^2+\gamma^2v^2t^2)^{3/2}}\cdot {2\sqrt{r_0^2-\left(b\over2\right)^2}\over r_0}\\
=&{4\pi\alpha\gamma Z \sqrt{4r_0^2-b^2}\over\left(r_0^2+\gamma^2v^2t^2\right)^{3/2}},
\end{split}
\label{Efield}
\end{equation}
here $\alpha={1\over137}$ is the electromagnetic coupling constant. $Z$ is the proton number. $A$ is the relative atomic mass and $r_0$ is the radius of heavy ions\cite{heyde_2018}:
\begin{equation}\label{Re}
   r_0=1.1A^{1\over 3}({\rm fm}).
\end{equation}

The right panel of FIG.\ref{fig2} shows the time-dependence of the $eE(\gamma,t)$ generated by two gold nuclei with impact parameter $b=10\,$fm at different $\gamma$'s. $t=0$ is defined as the moment of the collision.

As shown in the right panel of FIG.\ref{fig2} the electric field vanishes quickly after the collision. We define the lifetime of the electric field as the time it takes for the electric field to decrease from its peak value to 10\% of the peak value. The peak magnitude of $eE$ increases with $\gamma$, while the electric field's lifetime decreases rapidly with $\gamma$.

\subsection{Hamiltonian}
The wave function of the charmonium states can be represented as:
\begin{equation}\label{Eqn}
\psi(r,\theta,\phi)=\sum_{lm}R_{lm}(r)Y_l^m(\theta,\phi),
\end{equation}
where $Y_l^m(\theta,\phi)$ are spherical harmonics. In this case, considering that the electric field does not change the magnetic quantum number $m$, we choose the states in our calculation with $m=0$. We consider the charmonium  states with mass up to $4\,\text{GeV}$ (in vacuum) which include specifically 1S ($J/\psi$), 2S ($\psi'$), 1P ($\chi_c$), 2P and 1D. Their respective masses from Eq.(\ref{Eq1}) are $m_{J/\psi}=3.09\,$GeV, $m_{\psi'}=3.70\,$GeV, $m_{\chi_c}=3.48\,$GeV, $m_{2P}=3.98\,$GeV, and $m_{1D}=3.78\,$GeV.

For the radial wave function, we define $U(r)=rR(r)$. The radial Hamiltonian in Eq.(\ref{SEqn1}) can then be written as:
\begin{equation}\label{Eq7}
\begin{aligned}
   &{\cal H}= -{1\over 2\mu}{\partial^2 \over \partial r^2}+{V}_{\bar q q}(r) +{l\left(l+1\right)\over 2 \mu r^2}-E{r}\cos\theta.
\end{aligned}
\end{equation}

By multiplying each side of the Sch\"odinger equation by $\sum_l|Y_l^0(\theta)\rangle \langle Y^0_l(\theta)|$ and integrating $\theta$ out we obtain the radial Sch\"odinger equation as:
\begin{equation}\label{SEqn}
i{\partial \over \partial t}U_l(r)=\int \text{d}r'\sum_{l'}{\cal H}_{ll'}(r,r') U_{l'}(r'),
\end{equation}
here ${l}=0, 1, 2, \cdots$ are the eigenvalues of angular momentum of the charmonium states and $U_l(r)$ is the radial wave function of the state with angular momentum $l$. Then the Hamiltonian
${\cal H}_{ll'}(r,r')$
in basis of spherical harmonics takes the following form:
\begin{equation}\label{Eq7theta}
\begin{aligned}
   &{\cal H}_{0}(r)=-{1\over 2\mu}{\partial^2 \over \partial r^2}+{V}_{\bar q q}(r),\\
   &{\cal H}_{00}(r)={\cal H}_{0}(r),\\
   &{\cal H}_{01}(r)={\cal H}_{10}(r)=-{\sqrt{3}\over 3}Er,\\
   &{\cal H}_{11}(r)={\cal H}_{0}(r)+{2\over 2 \mu r^2},\\
   &{\cal H}_{20}(r)={\cal H}_{02}(r)=0,\\
   &{\cal H}_{12}(r)=-{\sqrt{15}\over 30}Er,\\
   &{\cal H}_{22}(r)={\cal H}_{0}(r) +{6\over 2 \mu r^2}.
\end{aligned}
\end{equation}

We use natural units $\hbar=c=1$.
From the Hamiltonian we can see that the uniform electric field can only induce transitions between states that differ in angular momentum by 1, which is consistent with the selection rule for electric dipole transitions.

\section{Numerical results}\label{MSD}
In solving the time-dependent Schr\"odinger equation, we use the MSD2 method\cite{PhysRevE.49.4684}.
In order to see the evolution of $J/\psi$, we make contour plots of $|\psi(r,\theta,\phi)|^2=\sum_l|R_l(r)|^2|Y_l^0(\theta,\phi)|^2$ at azimuthal angle $\phi=0$ and $\phi=\pi$ (y-z plane) by taking into account the azimuthal symmetry. In the following sections we will compare with the evolution of radial distribution of $c(\bar c)$ with and without the external electric field. Since at RHIC or LHC energy, the lifetime of the electric field is within 0.2\,fm$/c$, and the lifetime of QGP is typically several fm$/c$, in this work we consider the evolution in the first 2\,fm$/c$ when the electric field and QGP have strongest impact on the production of charmonia.

\subsection{Time evolution of $J/\psi$ and $\chi_c$ in QGP at constant temperature without electric field}
In this section we consider the charmonium states dissociation process at the constant temperature of $1.5\,T_c$, a typical temperature of QGP generated in heavy-ion collisions.

\begin{figure}[!htb]
   \begin{center}
      \includegraphics[width = 8.0cm]{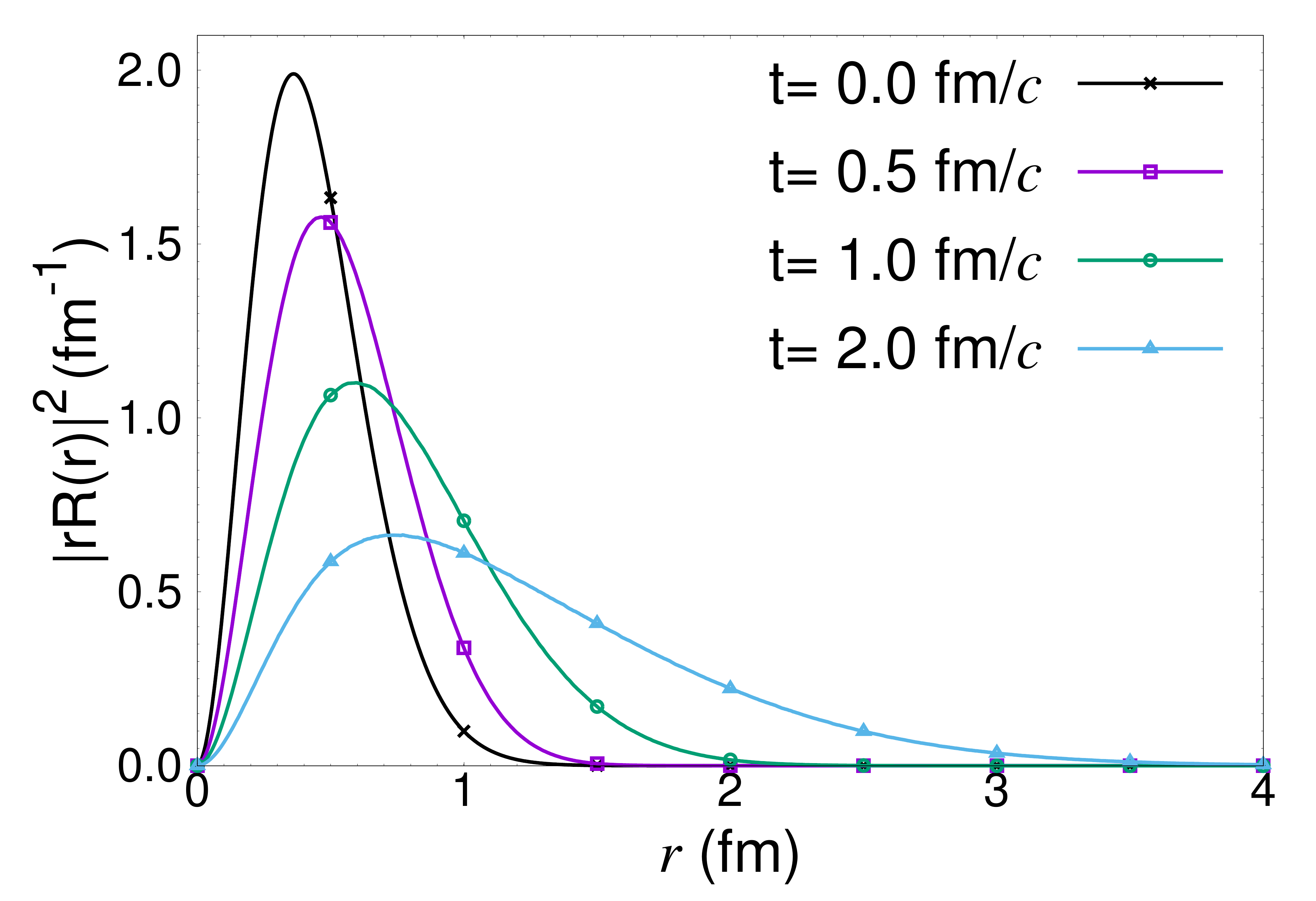}
      \includegraphics[width = 8.0cm]{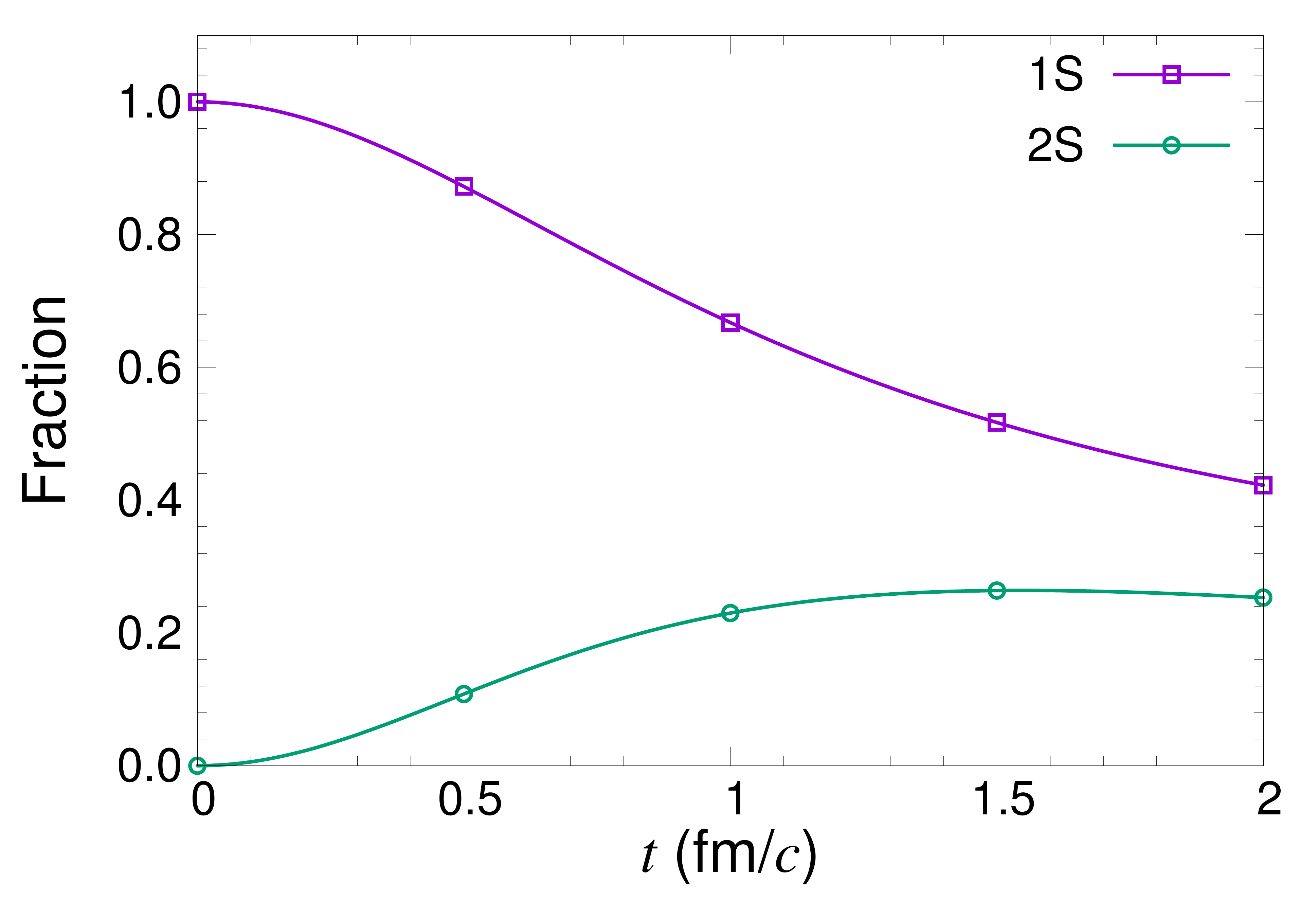}
     \caption{
 The initial state is $J/\psi$ (in vacuum) and $T=1.5T_c$. The radial distribution $|rR(r,t)|^2$ of $J/\psi$ is plotted in the left panel. The fractions are plotted in the right panel.
}\label{fig4a}
\end{center}
\end{figure}

FIG.\ref{fig4a} shows the dissociation process of $J/\psi$.
The fractions are defined as the possibility of charmonium states projected onto eigenstates of the vacuum Cornell potential.

In the left panel of FIG.\ref{fig4a}, the wave function of charmonium broadens with time which suggestes that $J/\psi$ is being dissociated by the colored partons in QGP. The fractions in FIG.\ref{fig4a} show the dissociation effect and the contour plots FIG.\ref{fig4} show that the evolved wave function of charmonium is still with spherical symmetry.
\begin{figure*}[!htb]
   \begin{center}
      \includegraphics[width = 6.6cm]{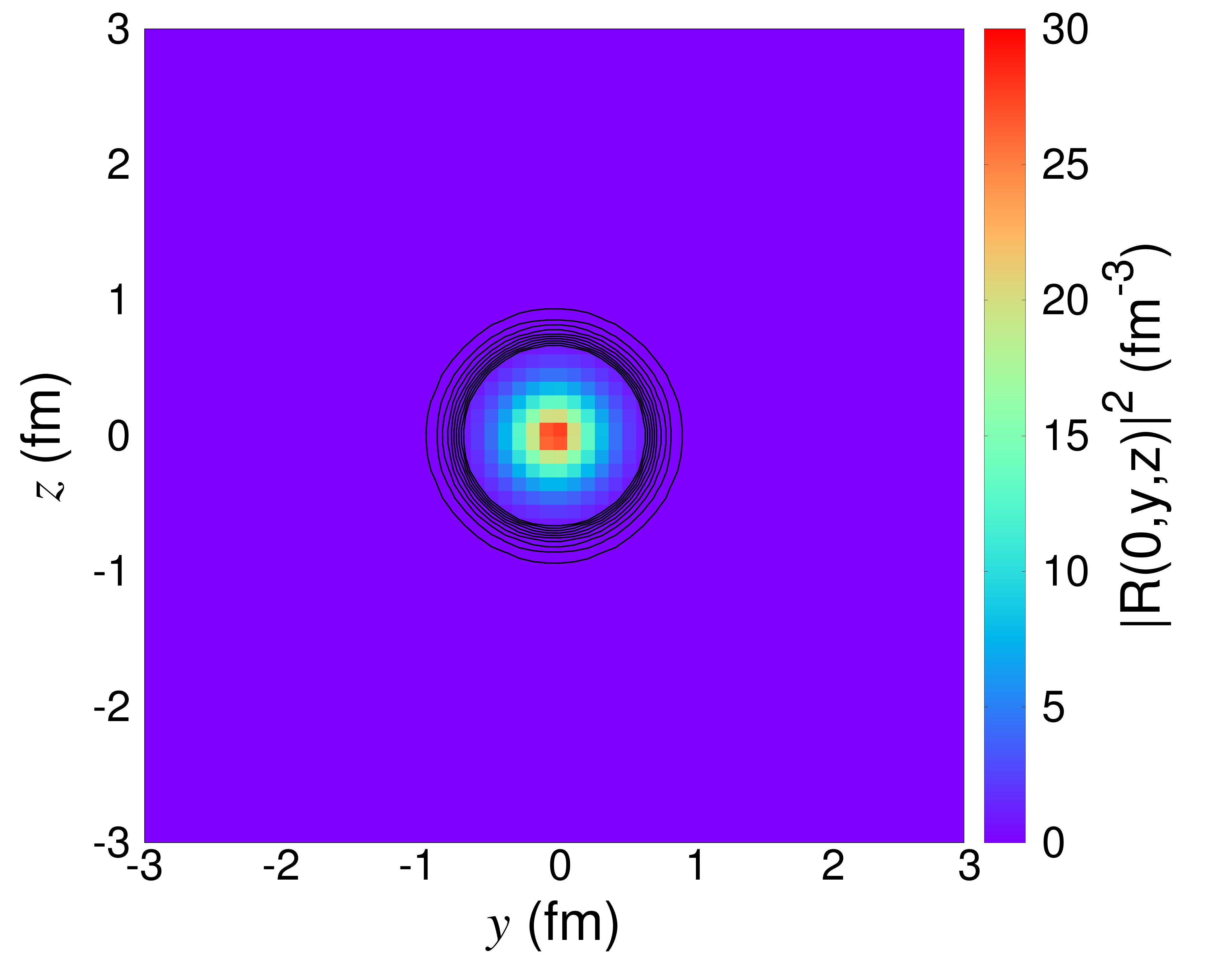}
      \includegraphics[width = 6.6cm]{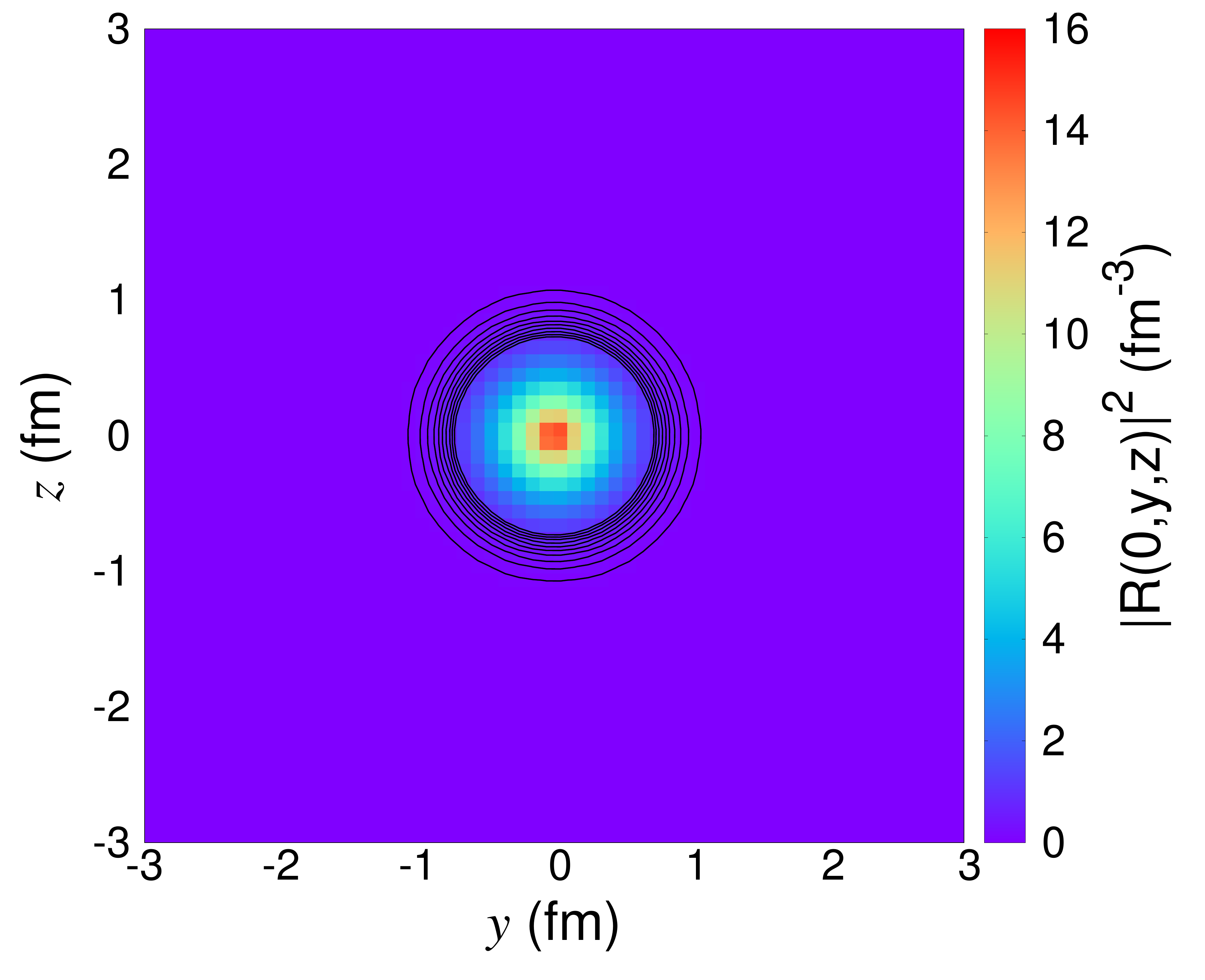}
\end{center}
\end{figure*}
\begin{figure}[!htb]
   \begin{center}
      \includegraphics[width = 6.6cm]{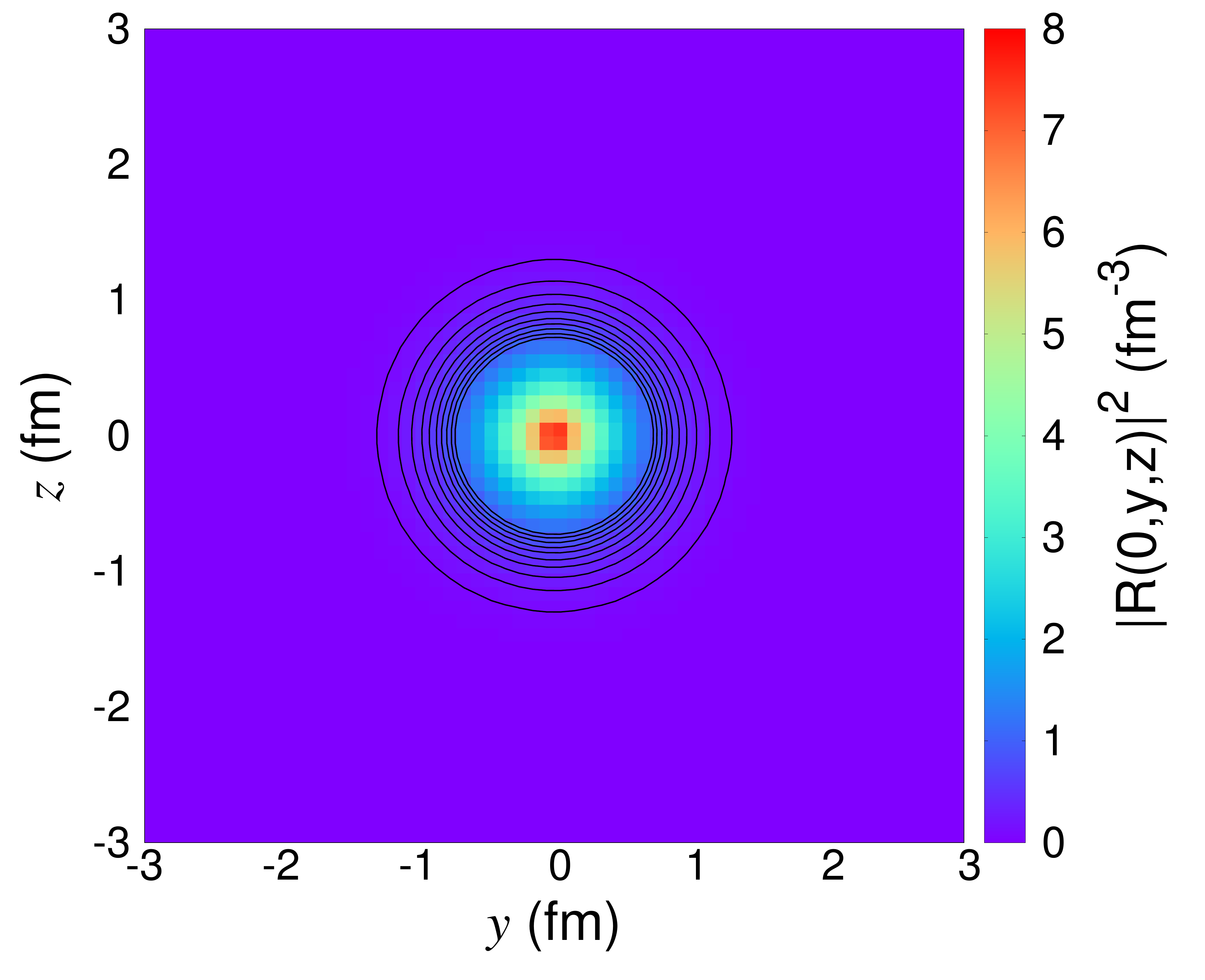}
      \includegraphics[width = 6.6cm]{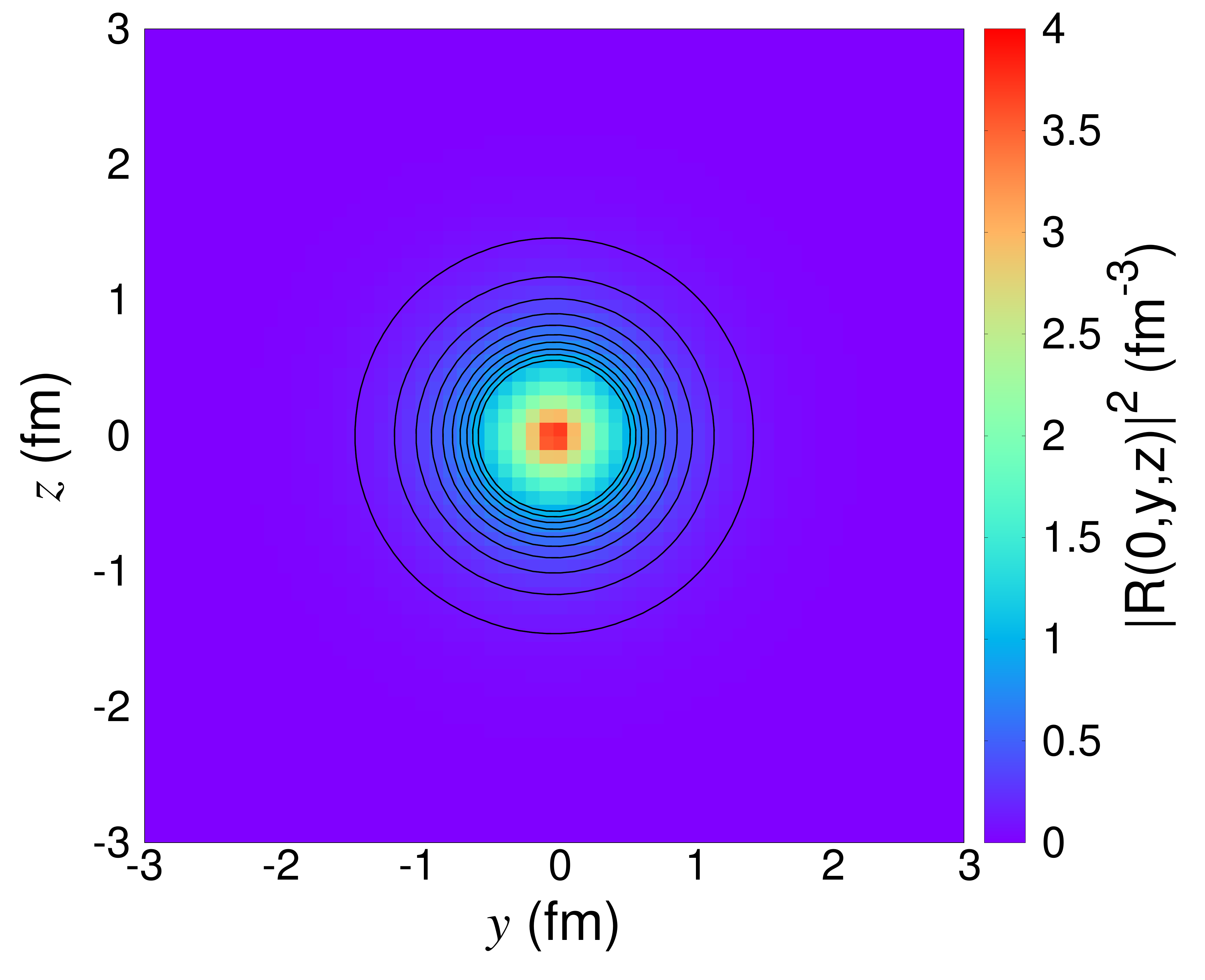}
     \caption{
 The radial distribution of $c(\bar c)$ at $t=0$, $t=0.5\,\text{fm}/c$, $t=1\,\text{fm}/c$ and $t=2\,\text{fm}/c$.
}
\label{fig4}
\end{center}
\end{figure}

\begin{figure}[!htb]
   \begin{center}
      \includegraphics[width = 8.0cm]{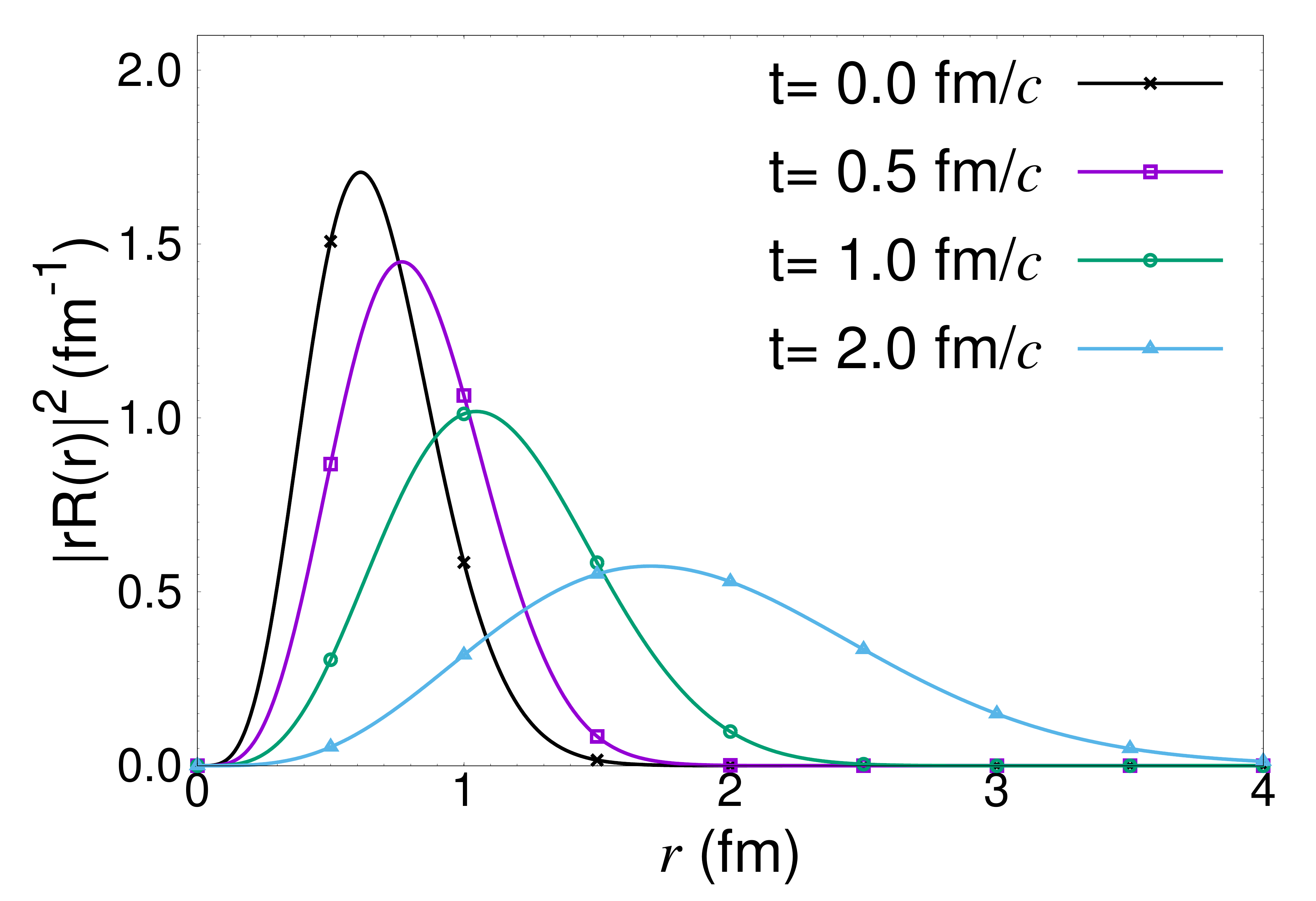}
      \includegraphics[width = 8.0cm]{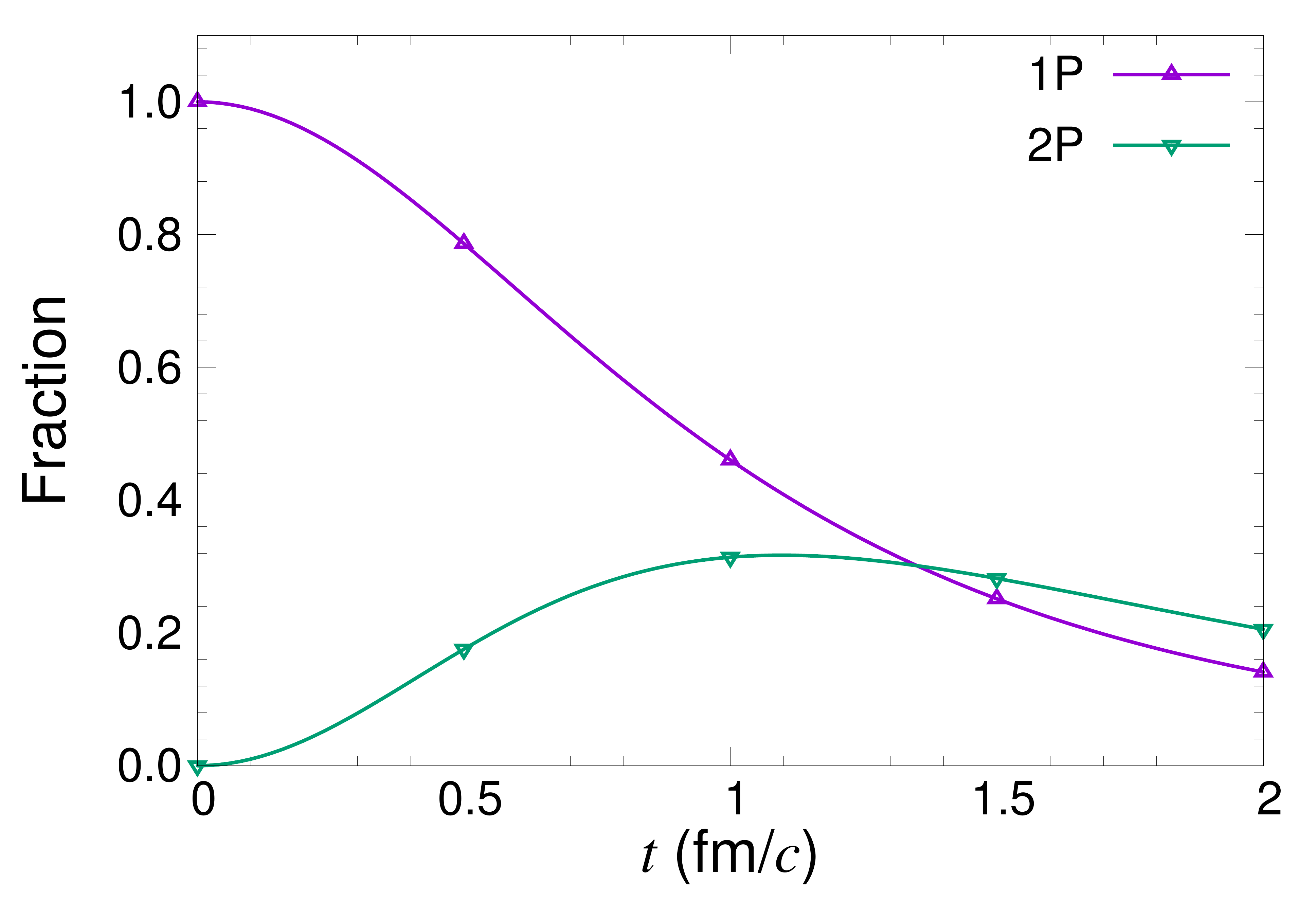}
     \caption{
  The initial state is $\chi_c$ (in vacuum) and $T=1.5T_c$. The radial distribution $|rR(r,t)|^2$ of $\chi_c$ is plotted in the left panel. The fractions are plotted in the right panel.
}\label{fig5a}
\end{center}
\end{figure}
\begin{figure*}[!htb]
   \begin{center}
      \includegraphics[width = 6.6cm]{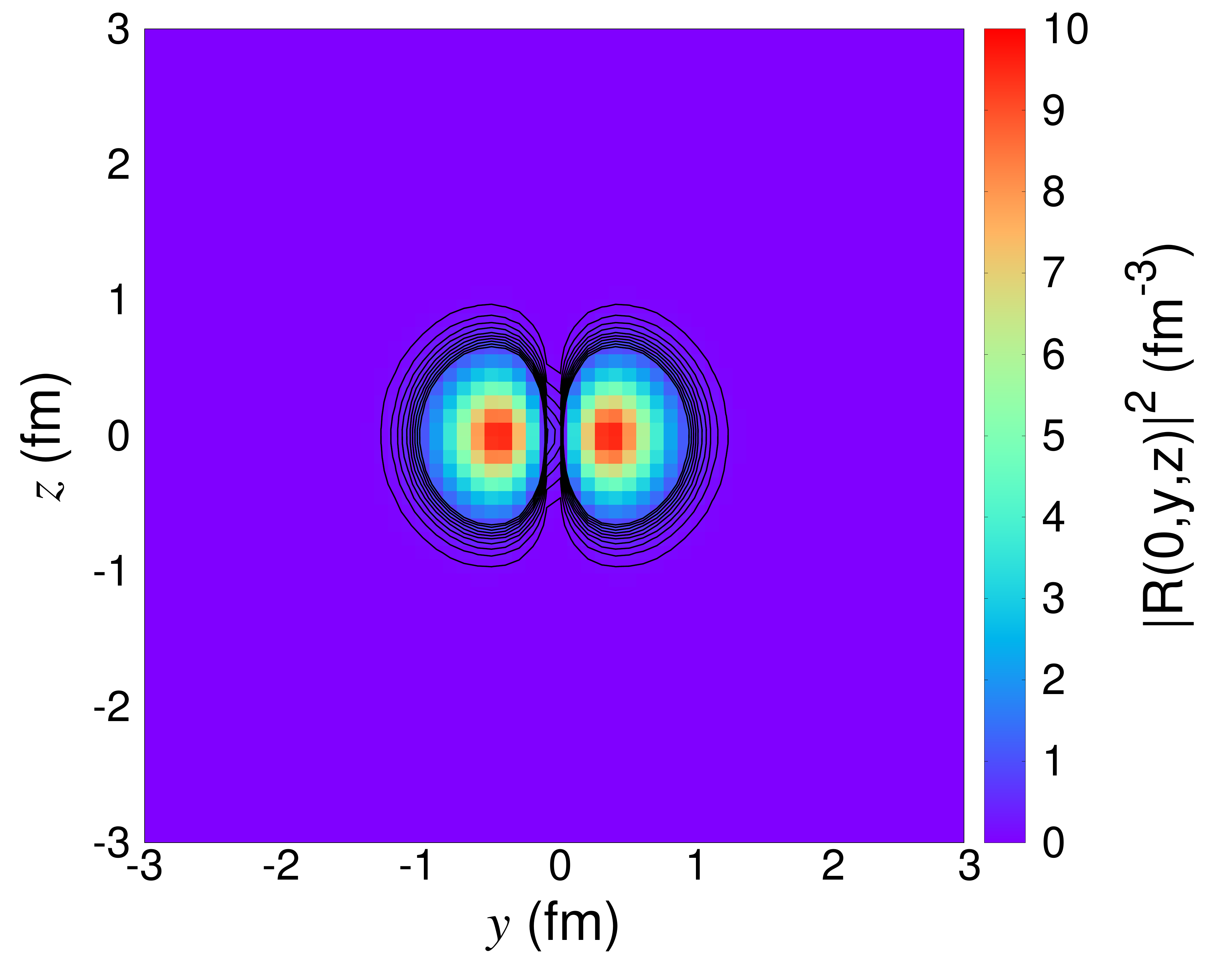}
      \includegraphics[width = 6.6cm]{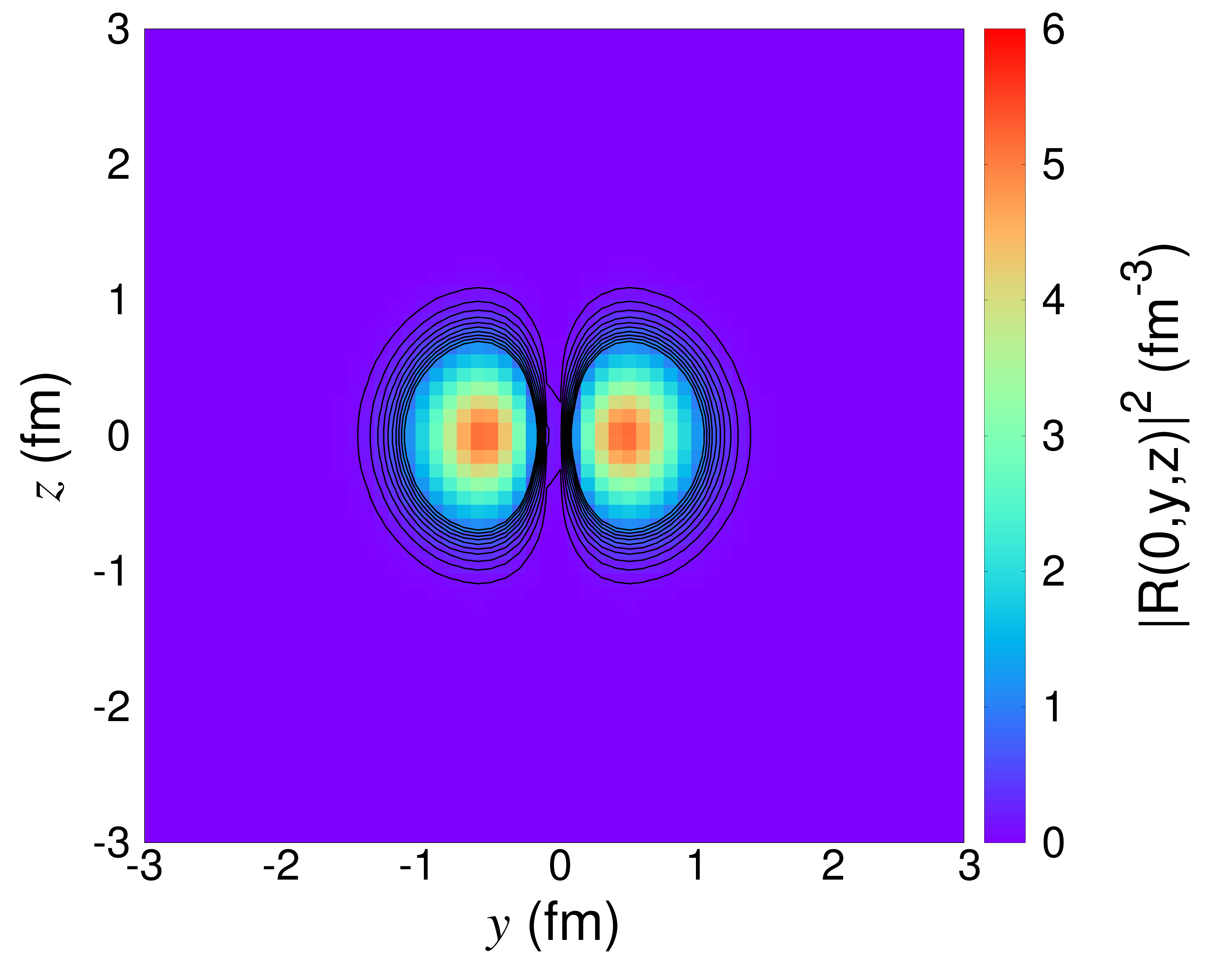}
\end{center}
\end{figure*}
\begin{figure}[!htb]
   \begin{center}
      \includegraphics[width = 6.6cm]{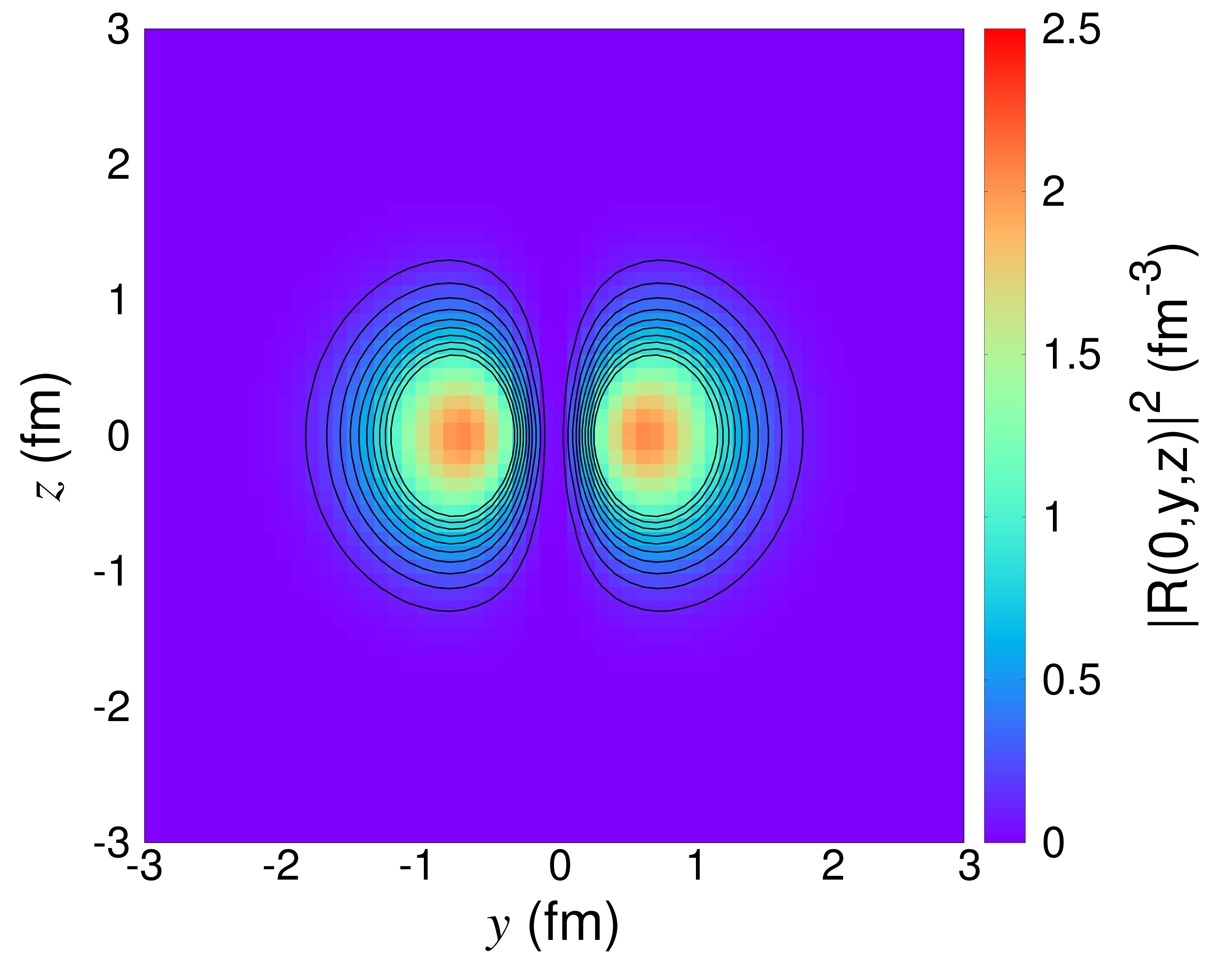}
      \includegraphics[width = 6.6cm]{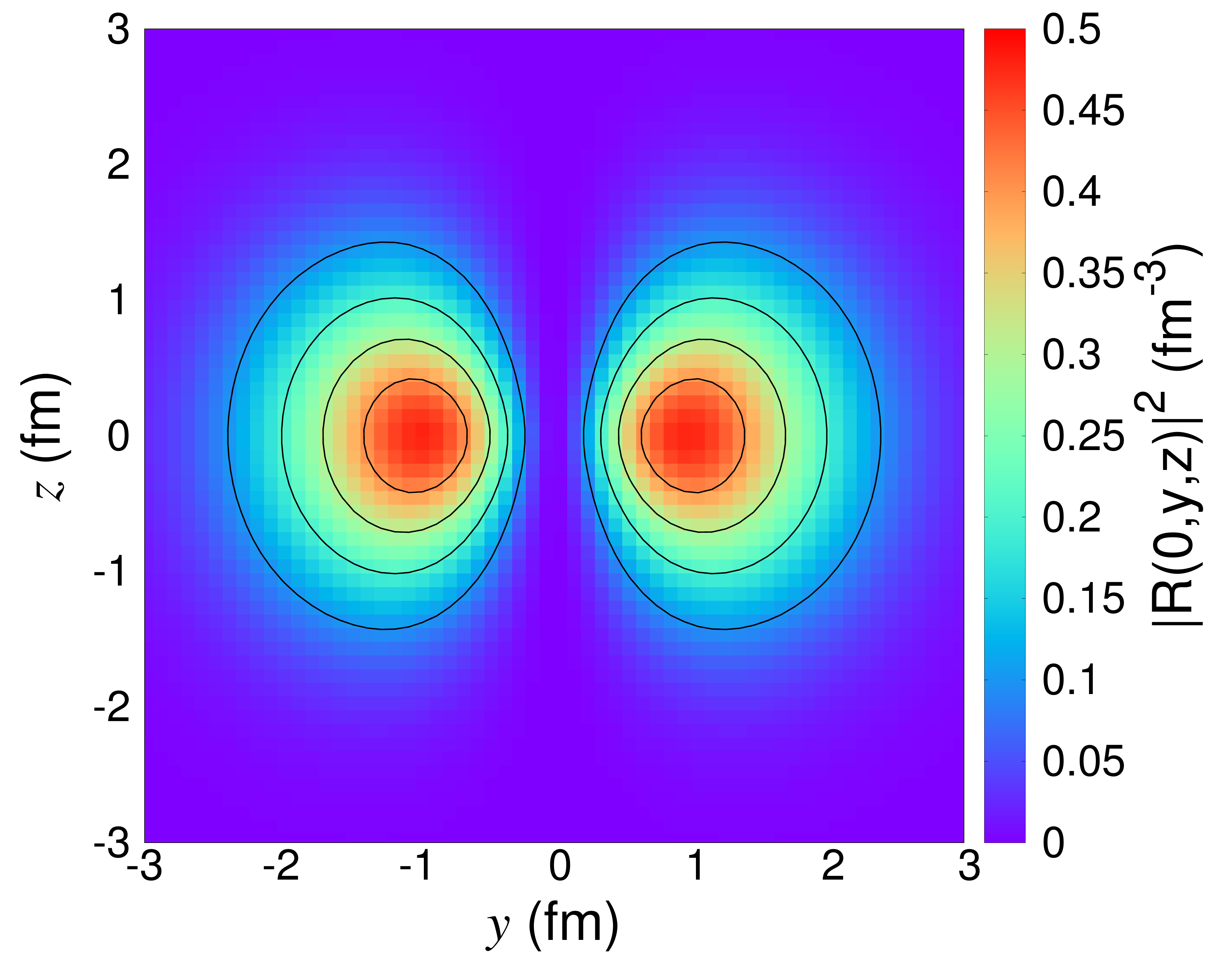}
     \caption{
 The radial distribution of $c(\bar c)$ at $t=0$, $t=0.5\,\text{fm}/c$, $t=1\,\text{fm}/c$ and $t=2\,\text{fm}/c$.
}
\label{fig5}
\end{center}
\end{figure}

In FIG.\ref{fig5a}-\ref{fig5} the initial state is taken to be $\chi_c$, a P-wave state of the vacuum Cornell potential.
Since the Hamiltonian without electric field is spherically symmetric, the transitions between states with different angular momenta are forbidden. Comparing with FIG.\ref{fig4a} and FIG.\ref{fig5a}, we see that $\chi_c$ is dissociated faster than $J/\psi$.
\subsection{Time evolution of $J/\psi$ and $\chi_c$ in QGP at decreasing Temperature without electric field} 
\begin{figure}[!htb]
   \begin{center}
      \includegraphics[width = 8.0cm]{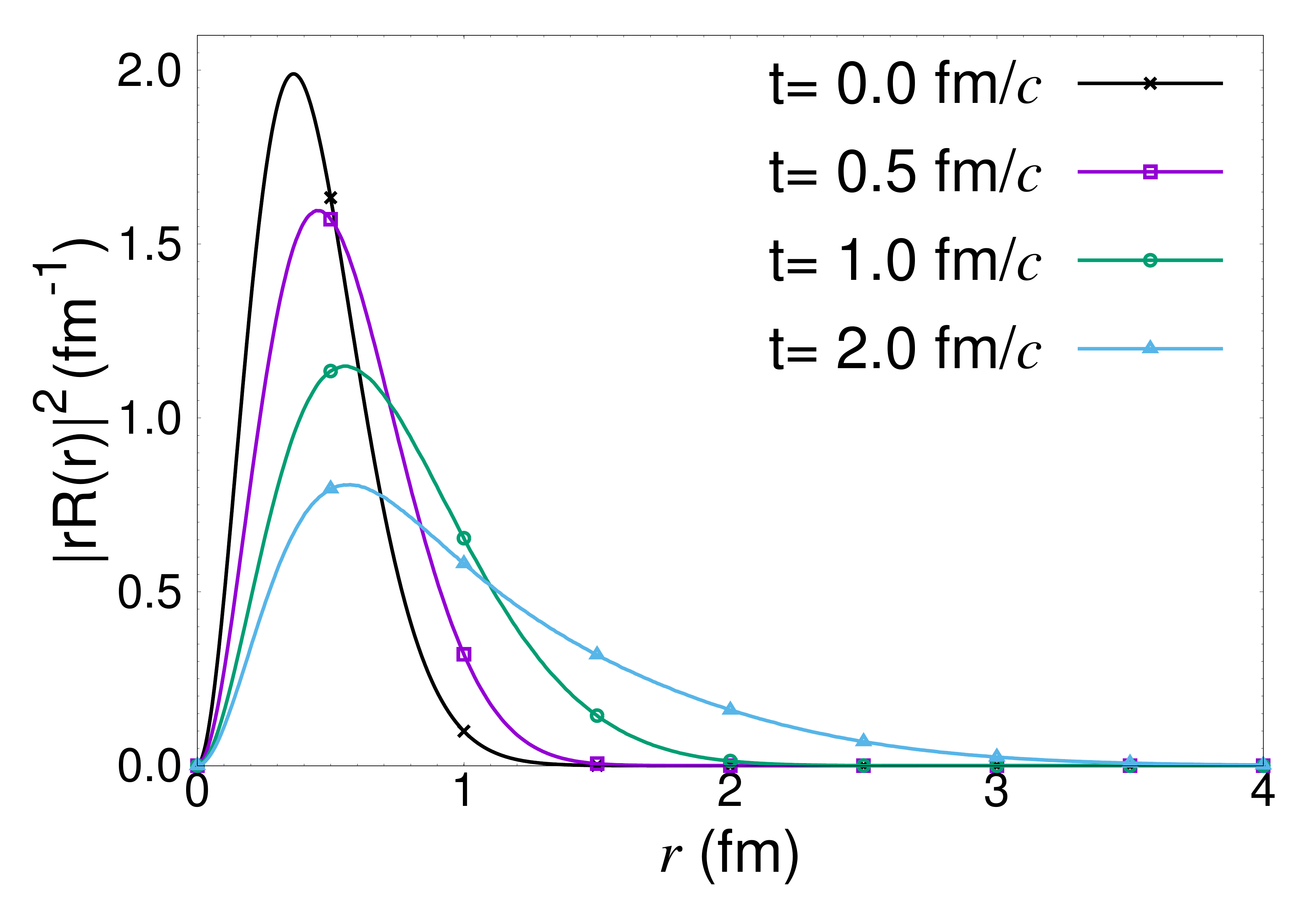}
      \includegraphics[width = 8.0cm]{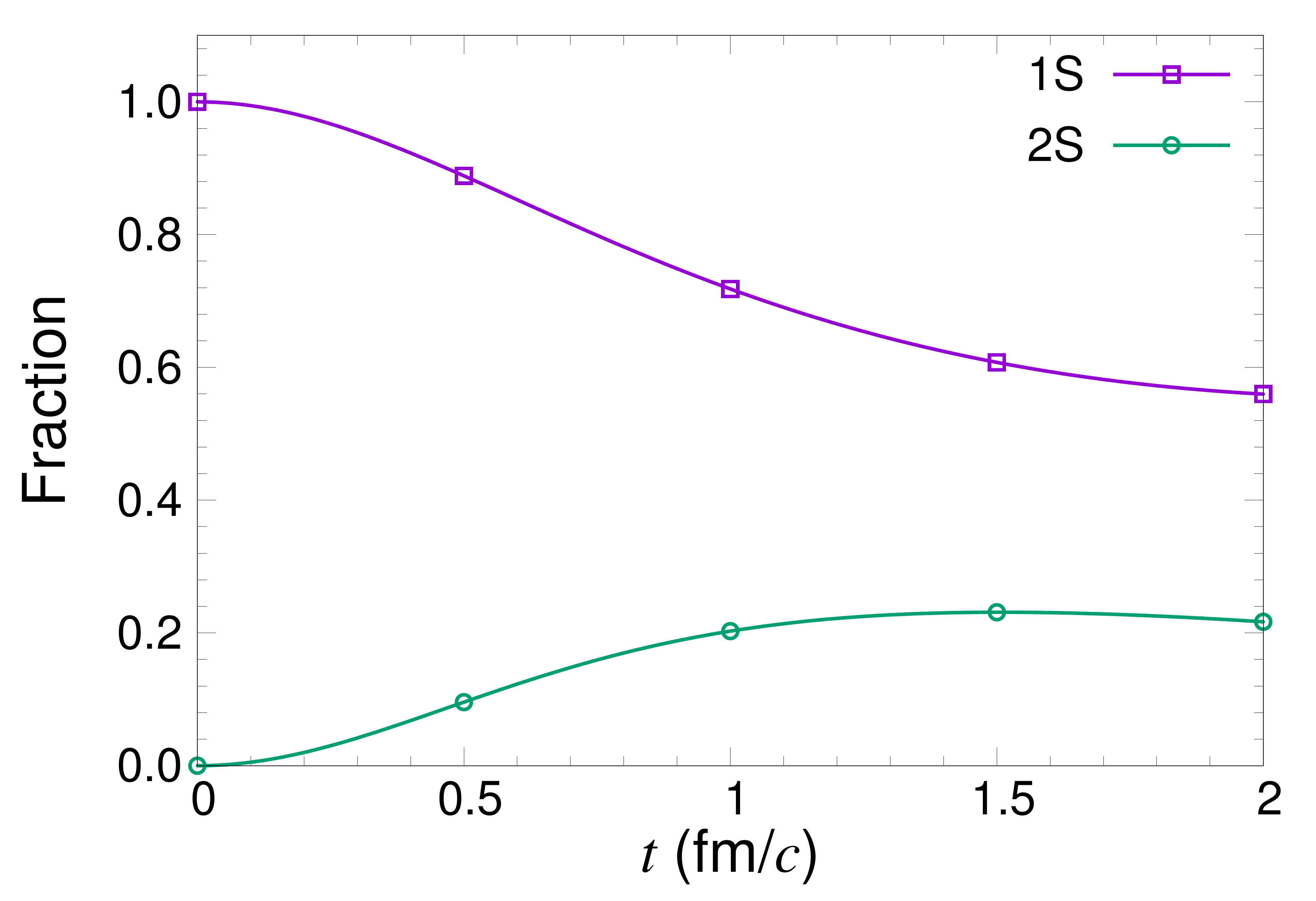}
     \caption{
The initial state is $J/\psi$ (in vacuum). And the temperature in the screened Cornell potential decreases linearly in time from $1.5\,T_c$ to $T_c$ in 2\,fm$/c$. The radial distribution $|rR(r,t)|^2$ of $J/\psi$ is plotted in the left panel. The fractions are plotted in the right panel. 
}\label{fig6a}
\end{center}
\end{figure}

\begin{figure}[!htb]
   \begin{center}
      \includegraphics[width = 6.5cm]{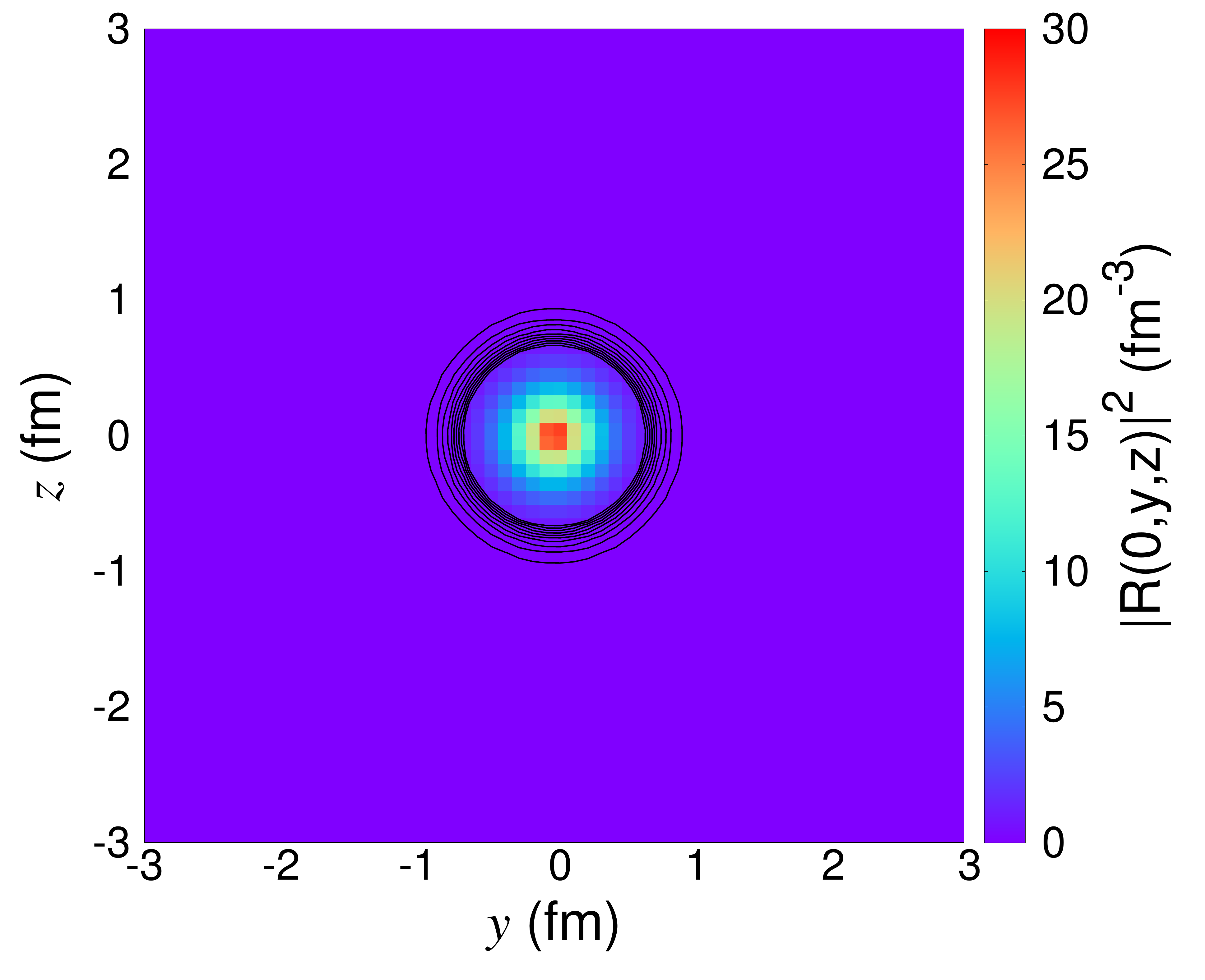}
      \includegraphics[width = 6.5cm]{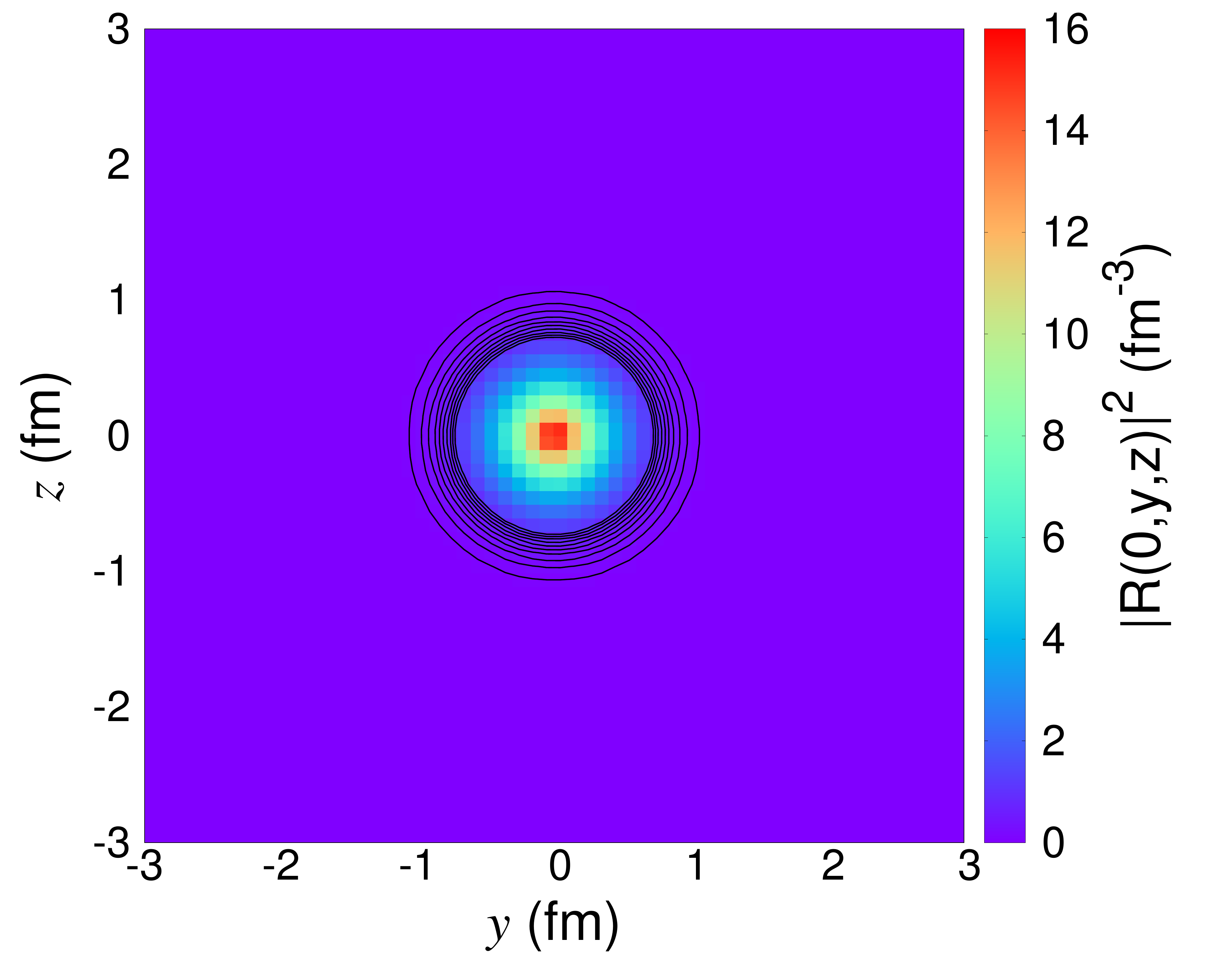}
      \includegraphics[width = 6.5cm]{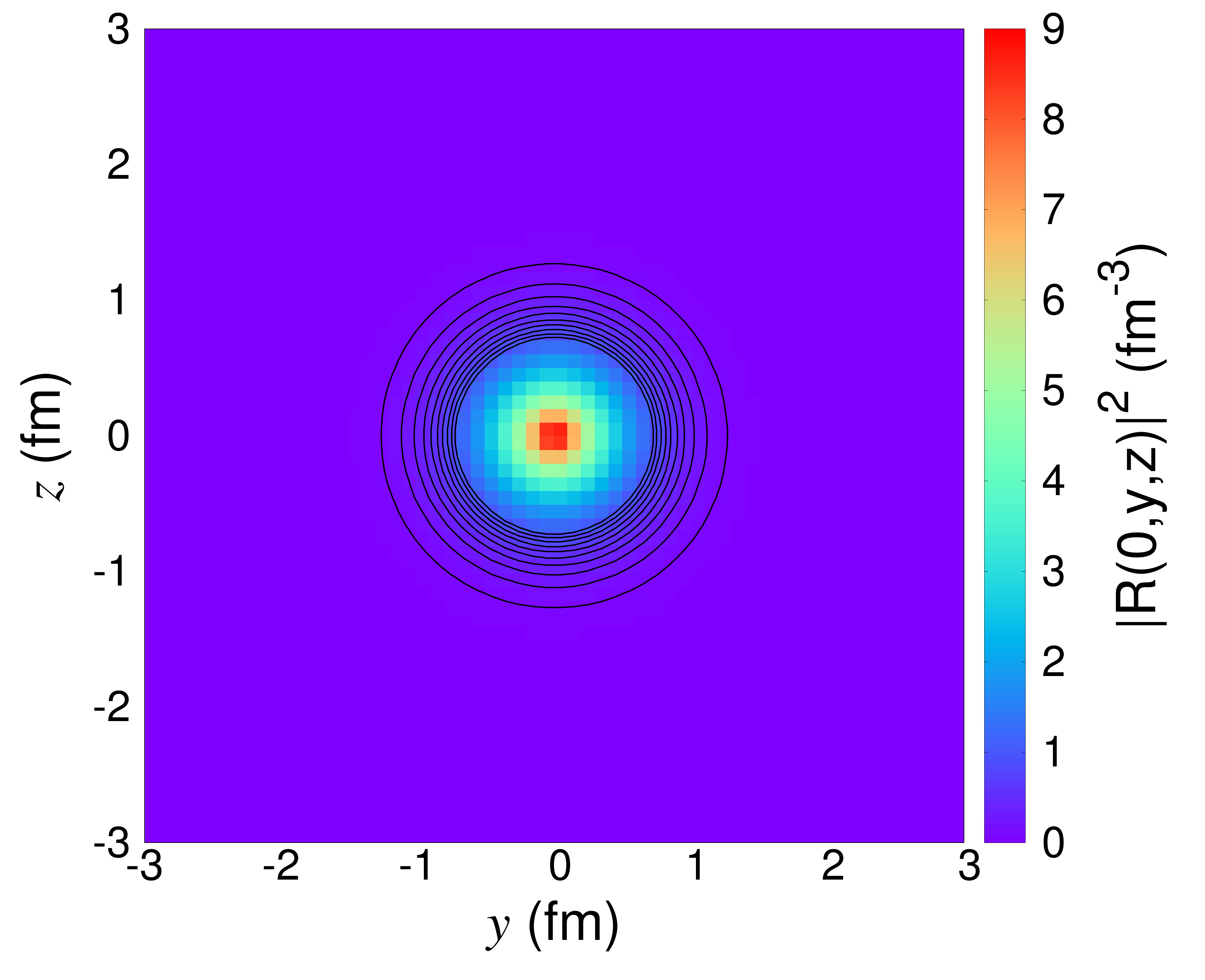}
      \includegraphics[width = 6.5cm]{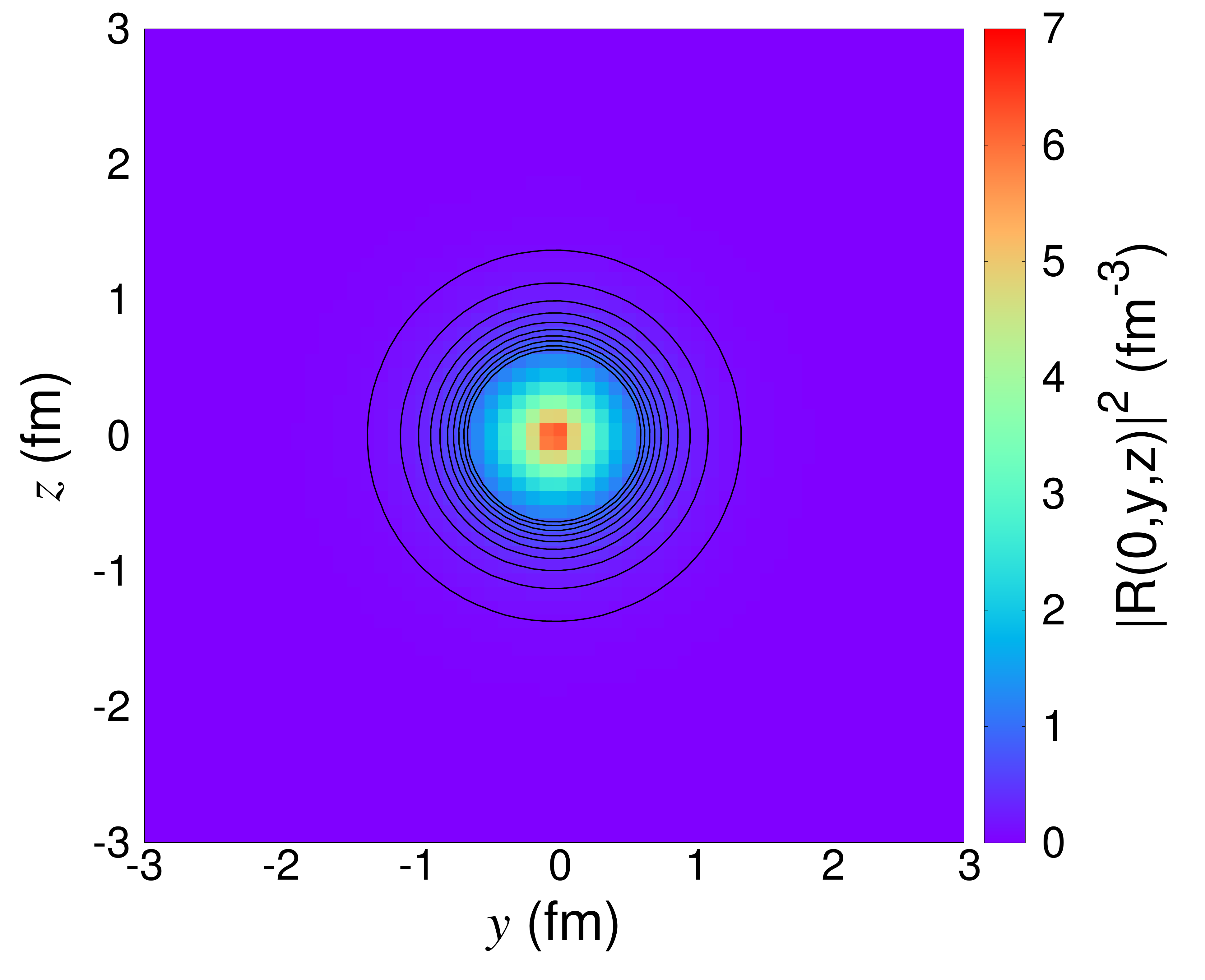}
     \caption{
The radial distribution of $c(\bar c)$ at $t=0$, $t=0.5\,\text{fm}/c$, $t=1\,\text{fm}/c$ and $t=2\,\text{fm}/c$. 
}
\label{fig6}
\end{center}
\end{figure}

As QGP expands after the collision, the temperature of the hot medium decreases. For simplicity we model the evolution of temperature as a linearly decreasing process in which temperature decreases from $1.5T_c$ to $T_c$ in 2fm$/c$.
FIG.\ref{fig6a}-\ref{fig6} shows the evolution of $J/\psi$ in the cooling system. Comparing with FIG.\ref{fig4a}, we see that the rate of dissociation is slower than that at the constant temperature of $1.5\,T_c$.

\begin{figure}[!htb]
   \begin{center}
      \includegraphics[width = 8.0cm]{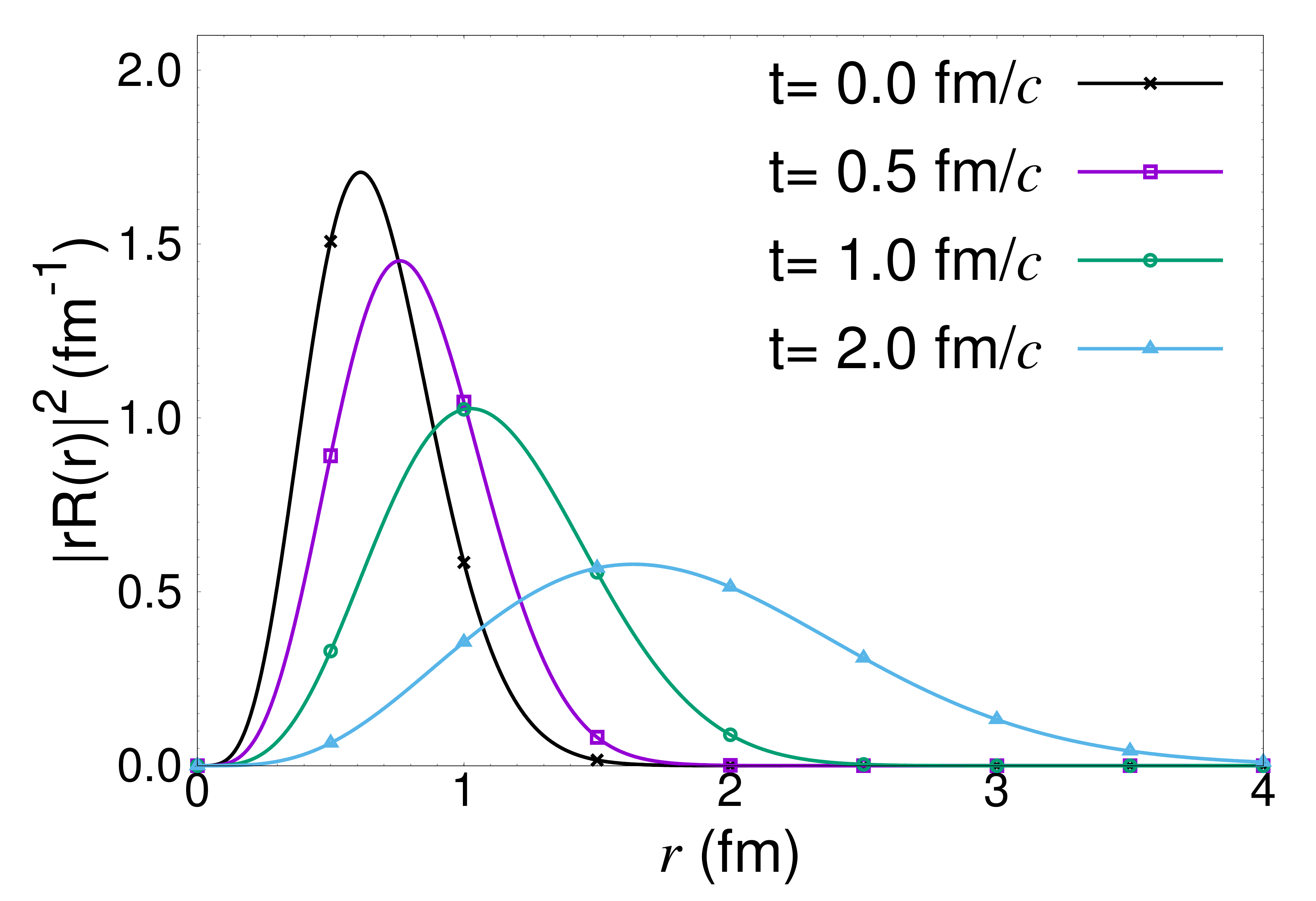}
      \includegraphics[width = 8.0cm]{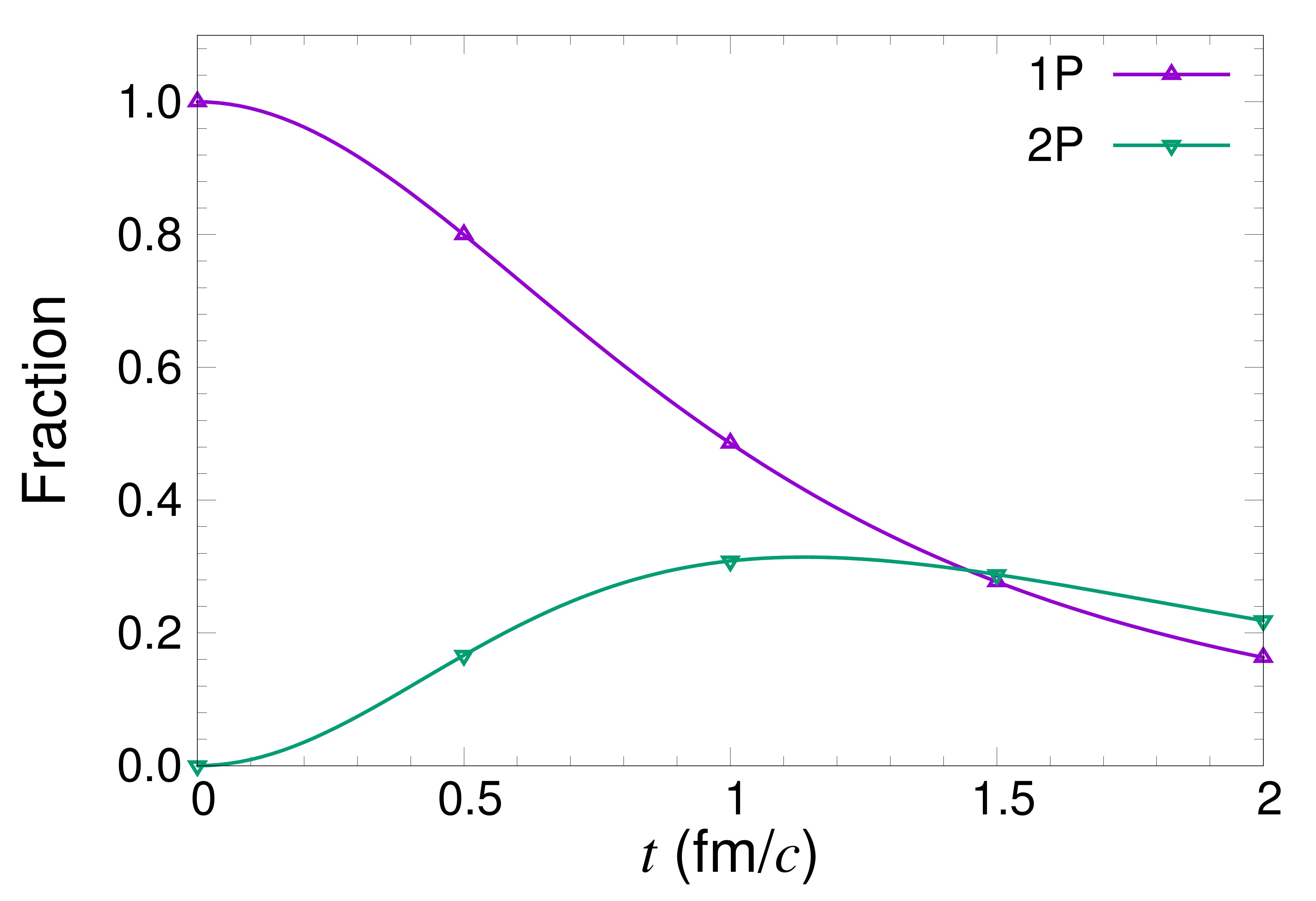}
     \caption{
 The initial state is $\chi_c$ (in vacuum). The temperature in the screened Cornell potential decreases linearly from $1.5\,T_c$ to $T_c$ uniformly in 2\,fm$/c$. The radial part $|rR(r,t)|^2$ of $\chi_c$ is plotted in the left panel. The fractions are plotted in right panel.
}\label{fig7a}
\end{center}
\end{figure}

\begin{figure}[!htb]
   \begin{center}
      \includegraphics[width = 6.5cm]{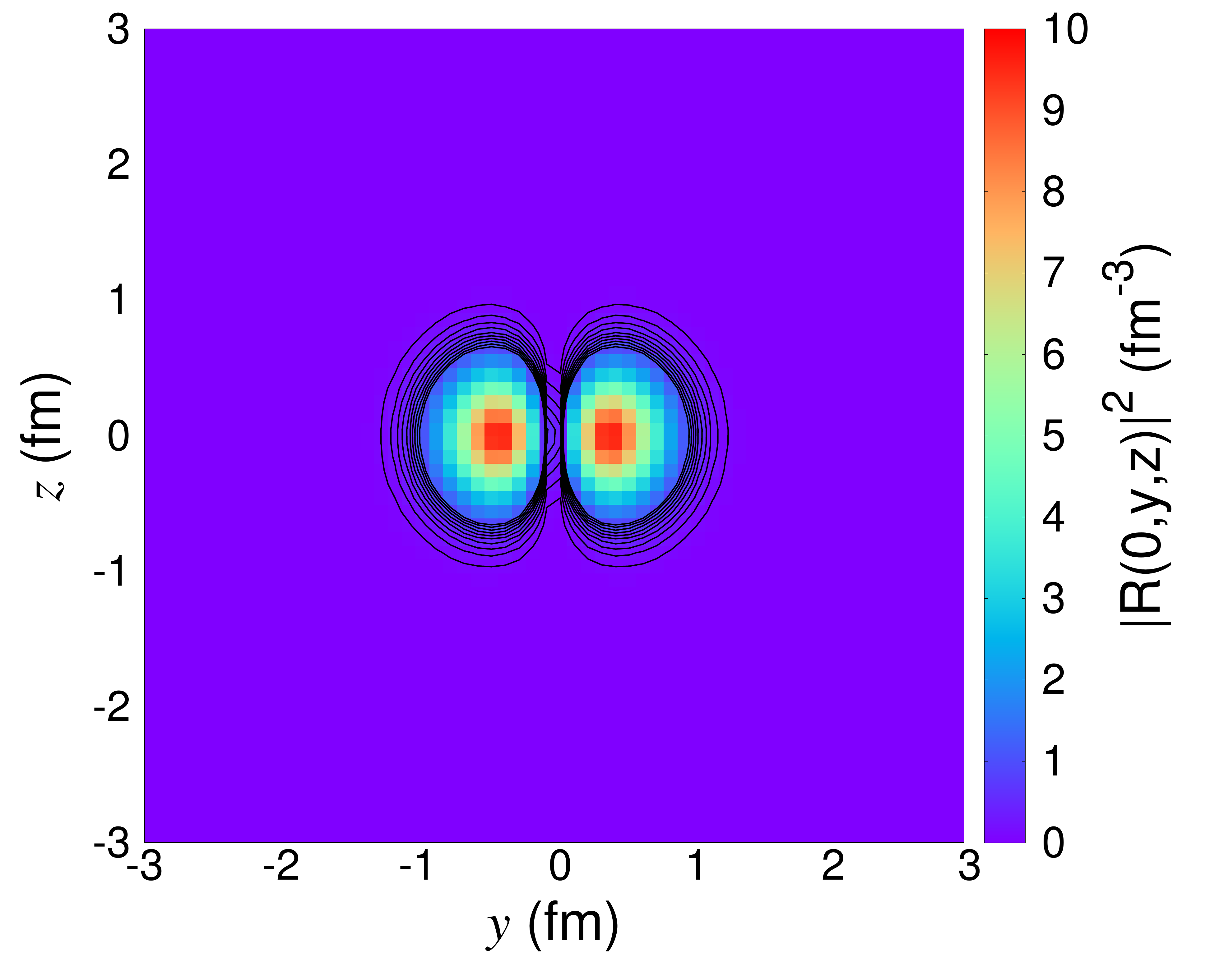}
      \includegraphics[width = 6.5cm]{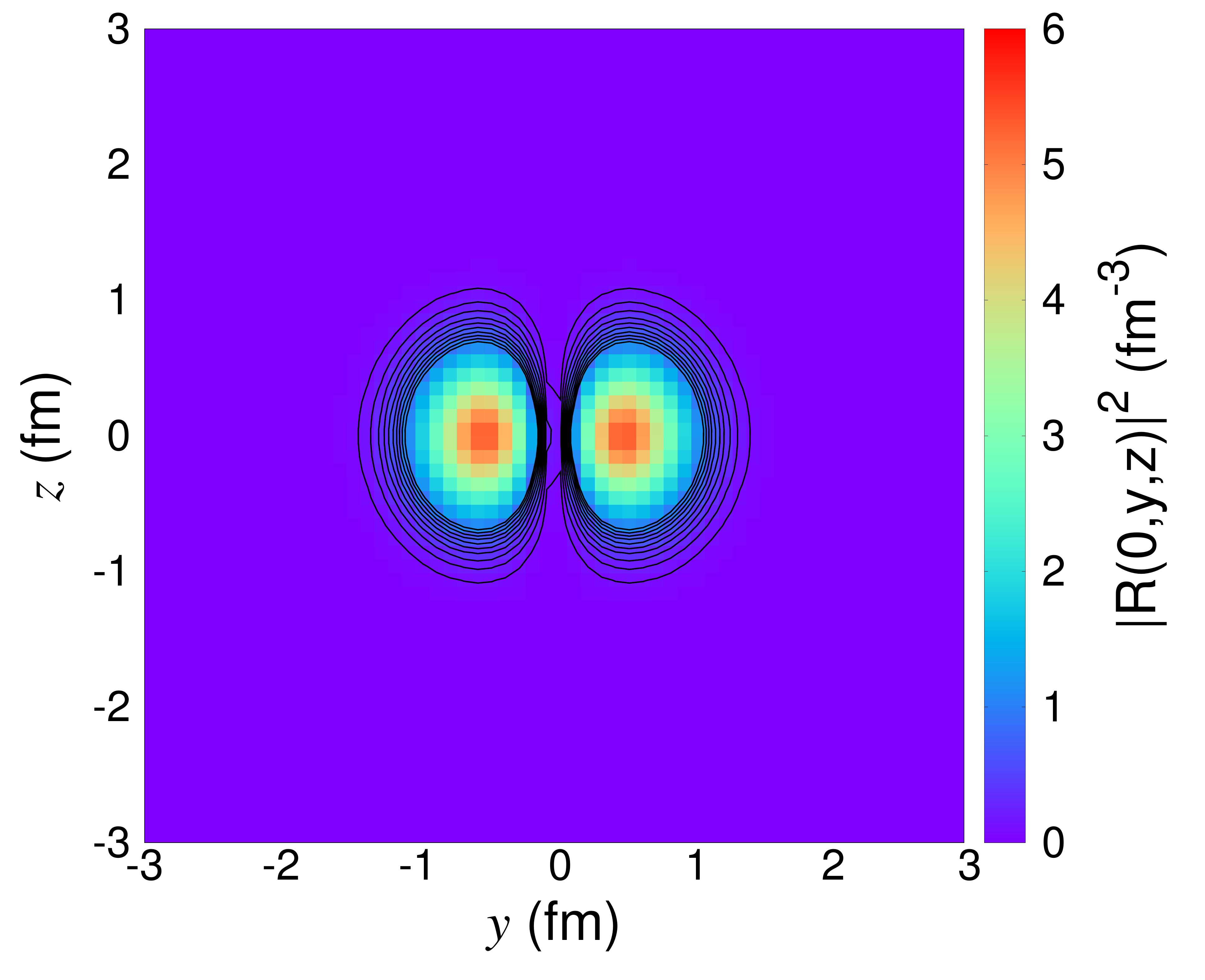}
      \includegraphics[width = 6.5cm]{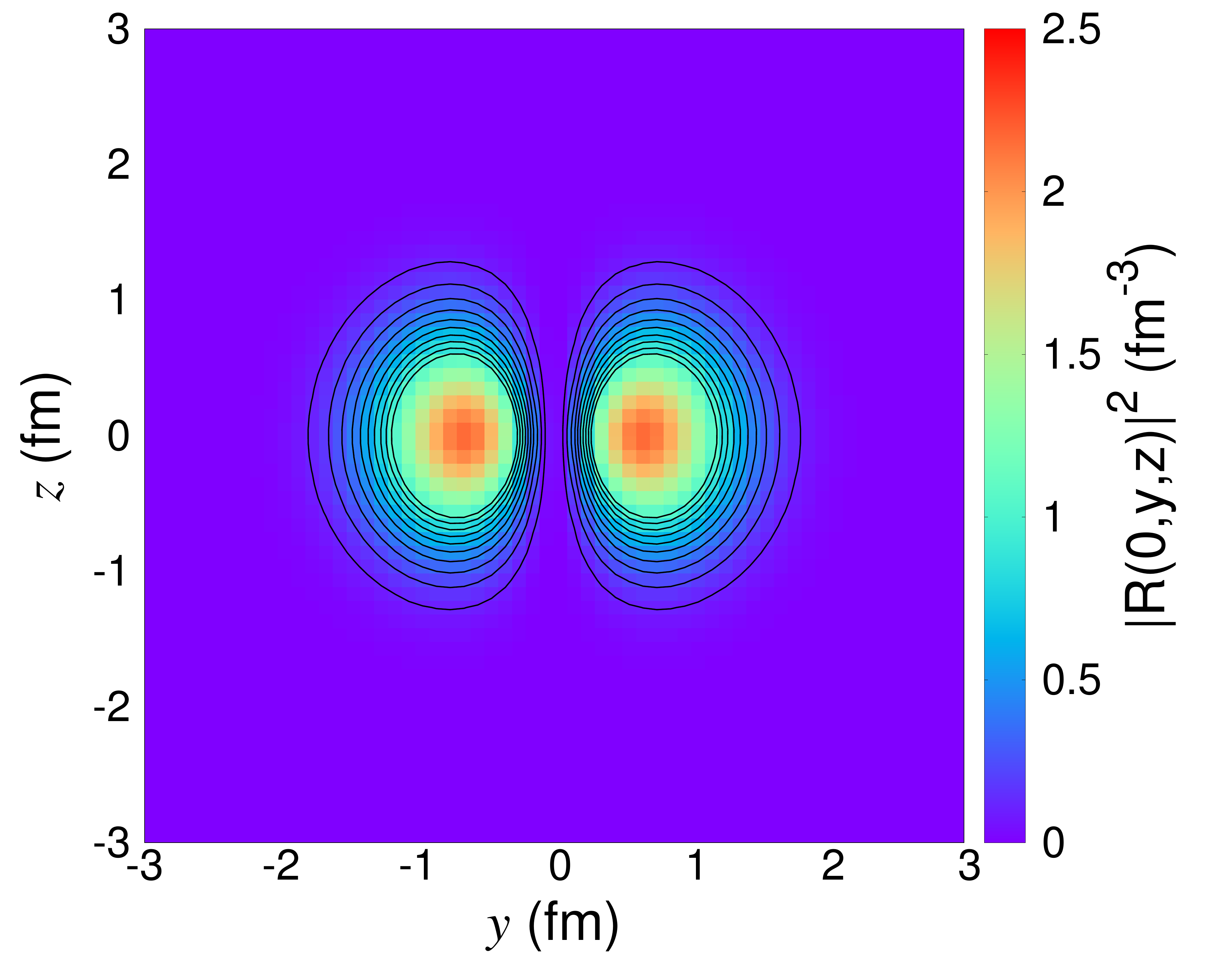}
      \includegraphics[width = 6.5cm]{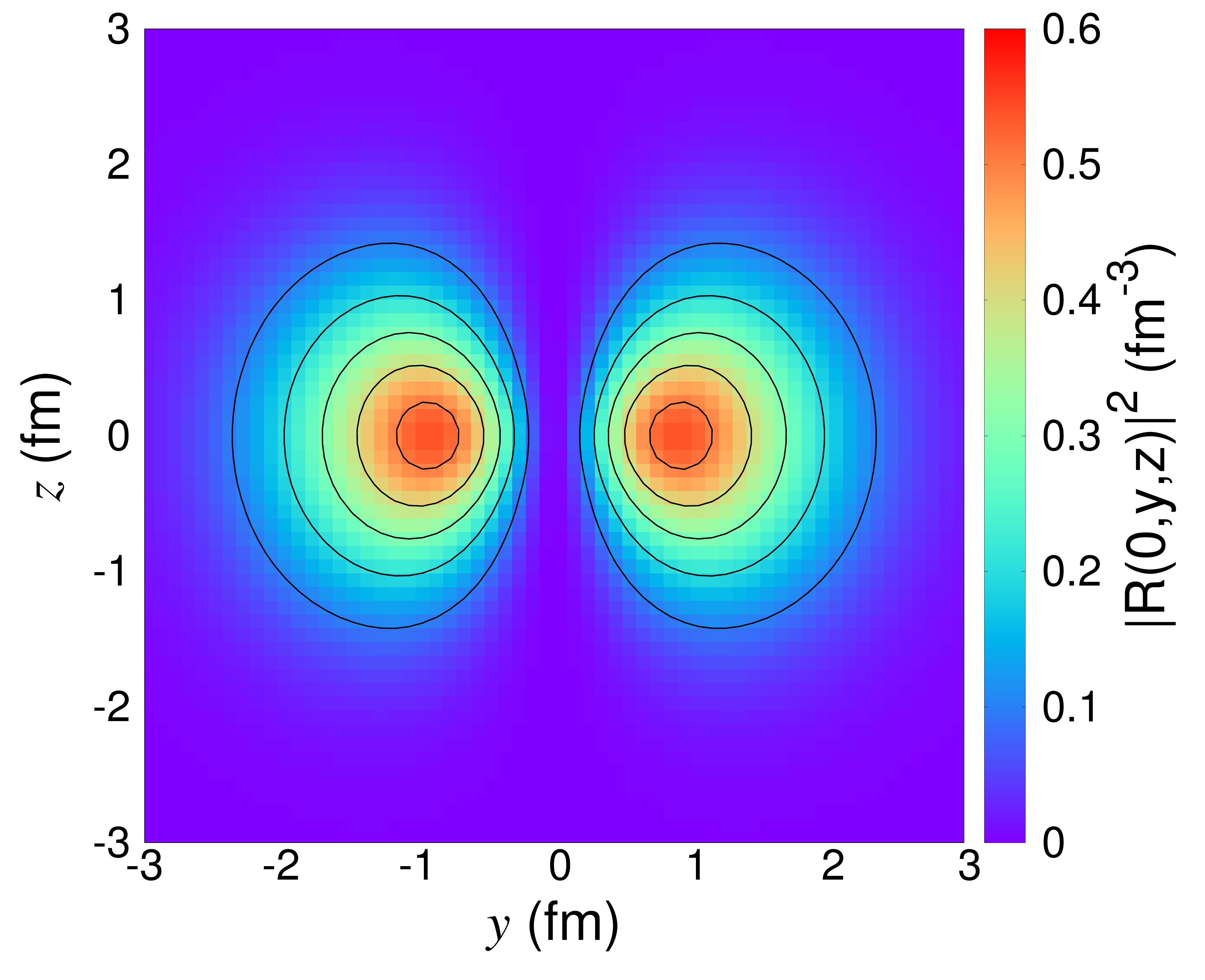}
     \caption{
The radial distribution of $c(\bar c)$ at $t=0$, $t=0.5\,\text{fm}/c$, $t=1\,\text{fm}/c$ and $t=2\,\text{fm}/c$.
}
\label{fig7}
\end{center}
\end{figure}

FIG.\ref{fig7a}-\ref{fig7} shows the dissociation process of $\chi_c$ in QGP with decreasing temperature. We can see that time-dependent temperature has a greater impact on the fractions of S wave states than those of P-wave states.
\newpage
\subsection{Time evolution of $J/\psi$ and $\chi_c$ in QGP at decreasing Temperature with electric field}
In the following parts we present the evolution of charmonium states in the electric field introduced in {Sec.\ref{Elt}}. Meanwhile the temperature is assumed to drop linearly in time from 1.5\,$T_c$ to $T_c$ in 2\,fm$/c$ in this process.

\subsubsection{Au-Au collision with $\gamma\simeq 100$}
We first consider Au-Au collisions in RHIC with center-of-mass energy $\sqrt{s_{NN}}=200\,$GeV per nucleon pair with  $\gamma\simeq 100$. The lifetime of electric field is about 0.2fm$/c$ as shown in the right panel of FIG.\ref{fig2}.

\begin{figure}[!htb]
   \begin{center}
      \includegraphics[width = 8.0cm]{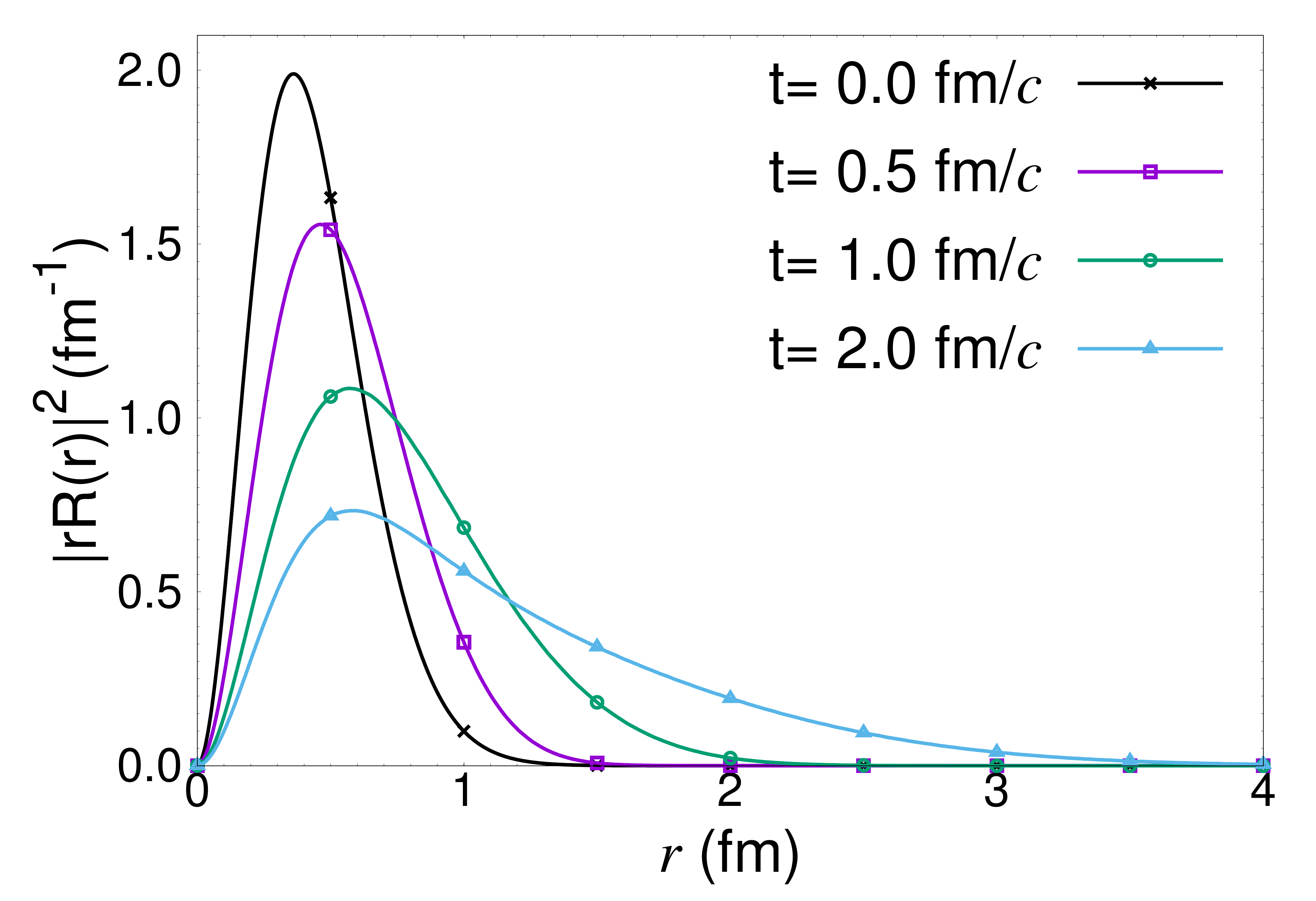}
      \includegraphics[width = 8.0cm]{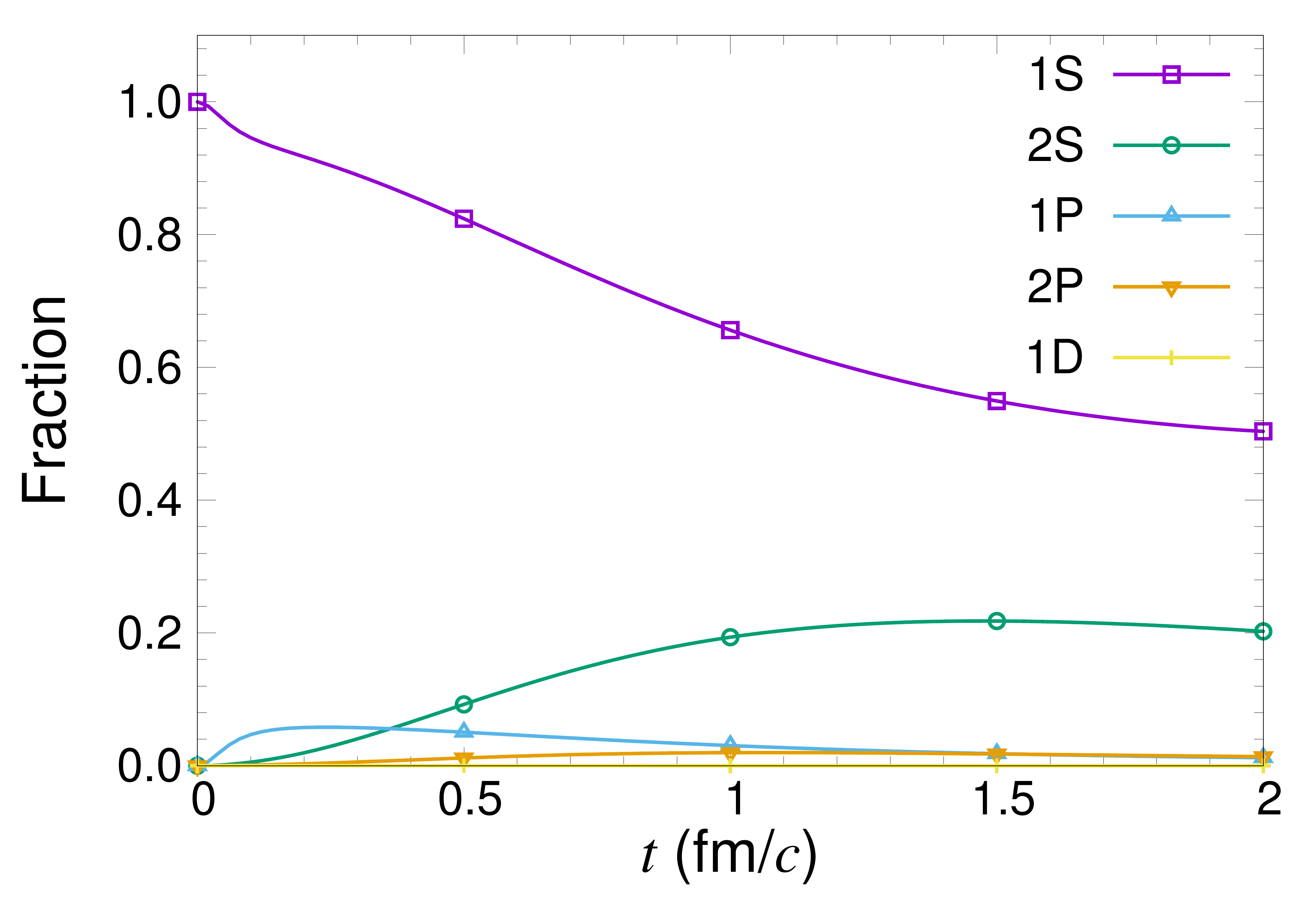}
     \caption{
The initial state is $J/\psi$ (in vacuum). The radial distribution $|rR(r,t)|^2$ of $J/\psi$ is plotted in the left panel. The fractions are plotted in the right panel. 
}\label{fig8a}
\end{center}
\end{figure}

\begin{figure}[!htb]
   \begin{center}
      \includegraphics[width = 6.5cm]{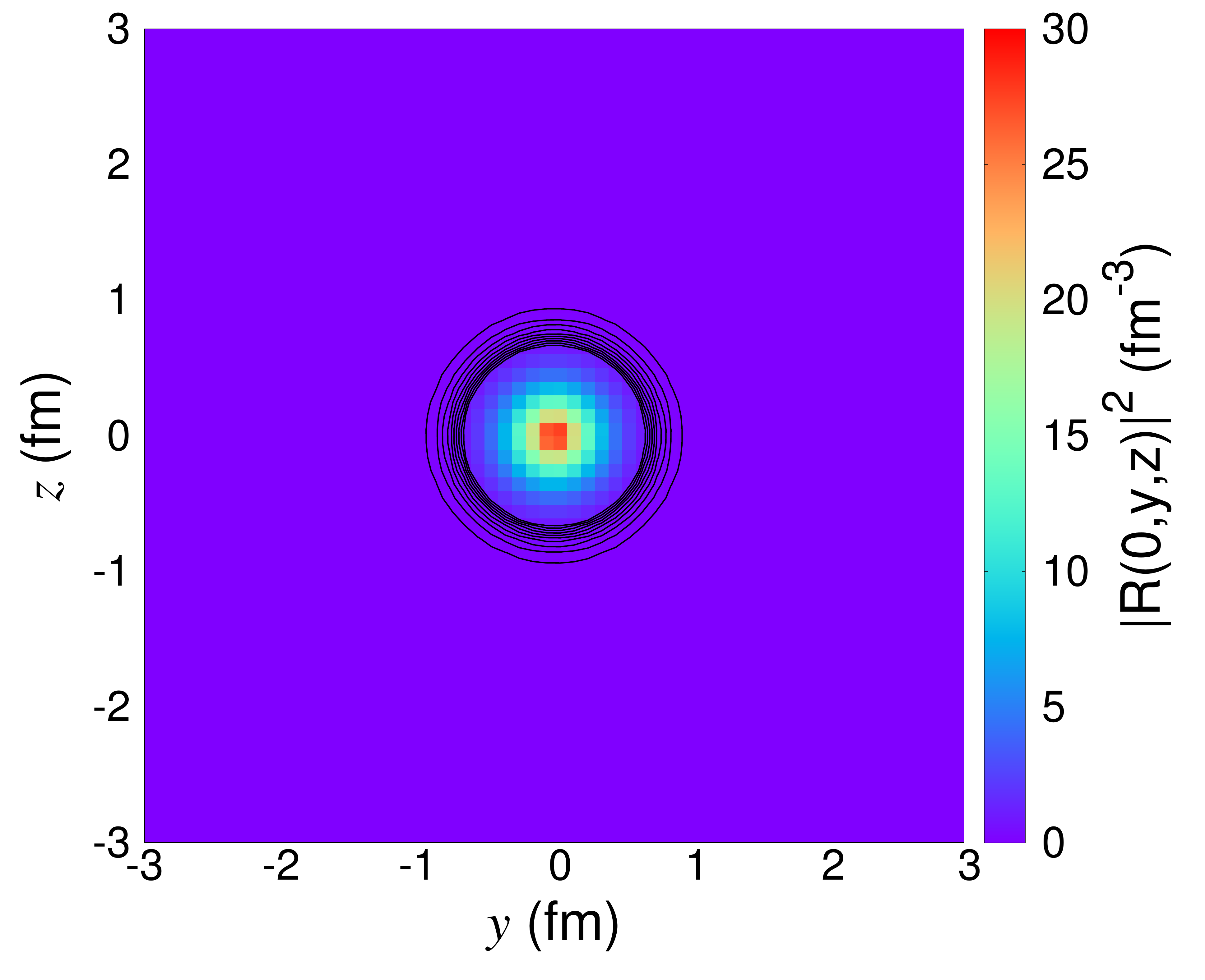}
      \includegraphics[width = 6.5cm]{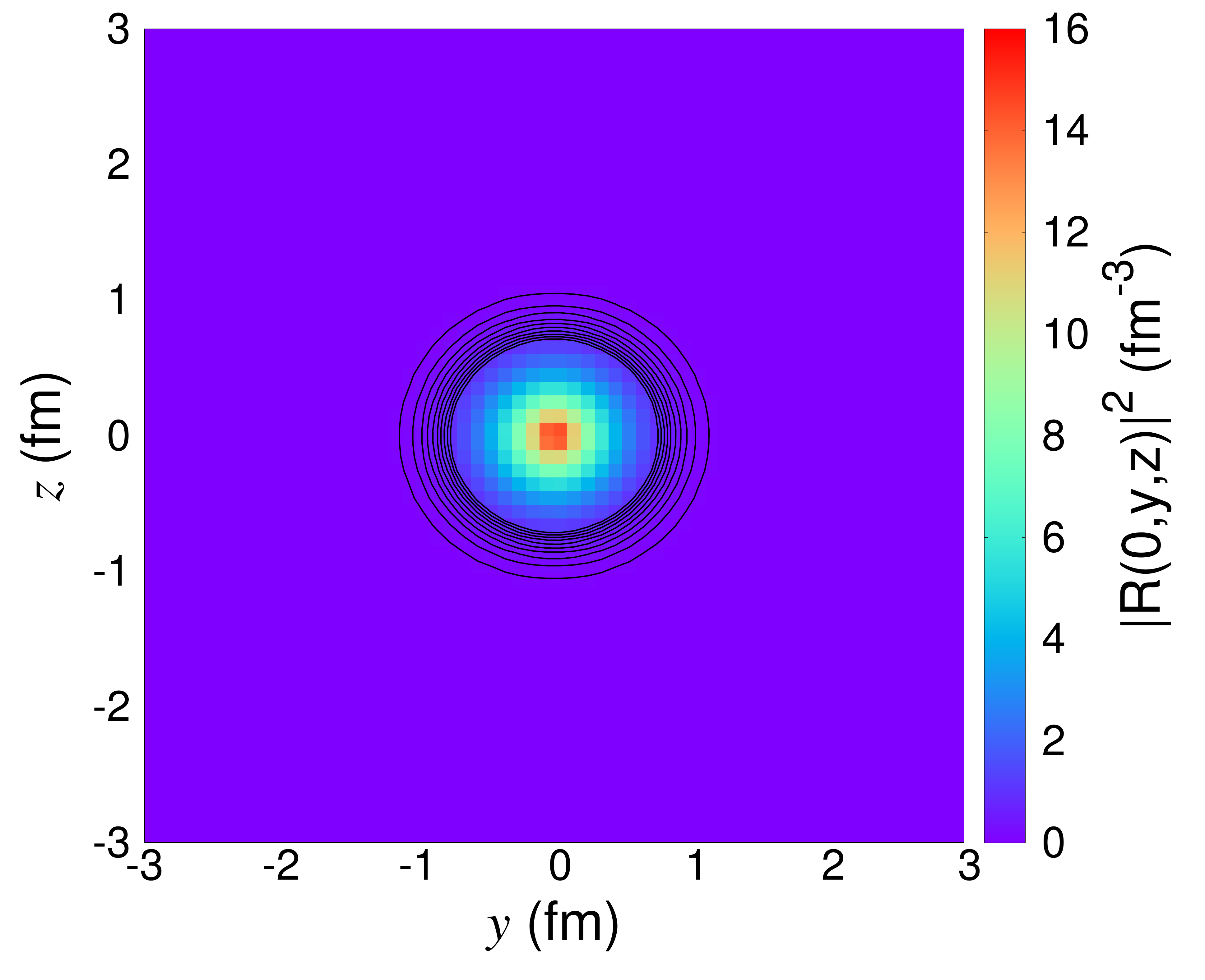}
      \includegraphics[width = 6.5cm]{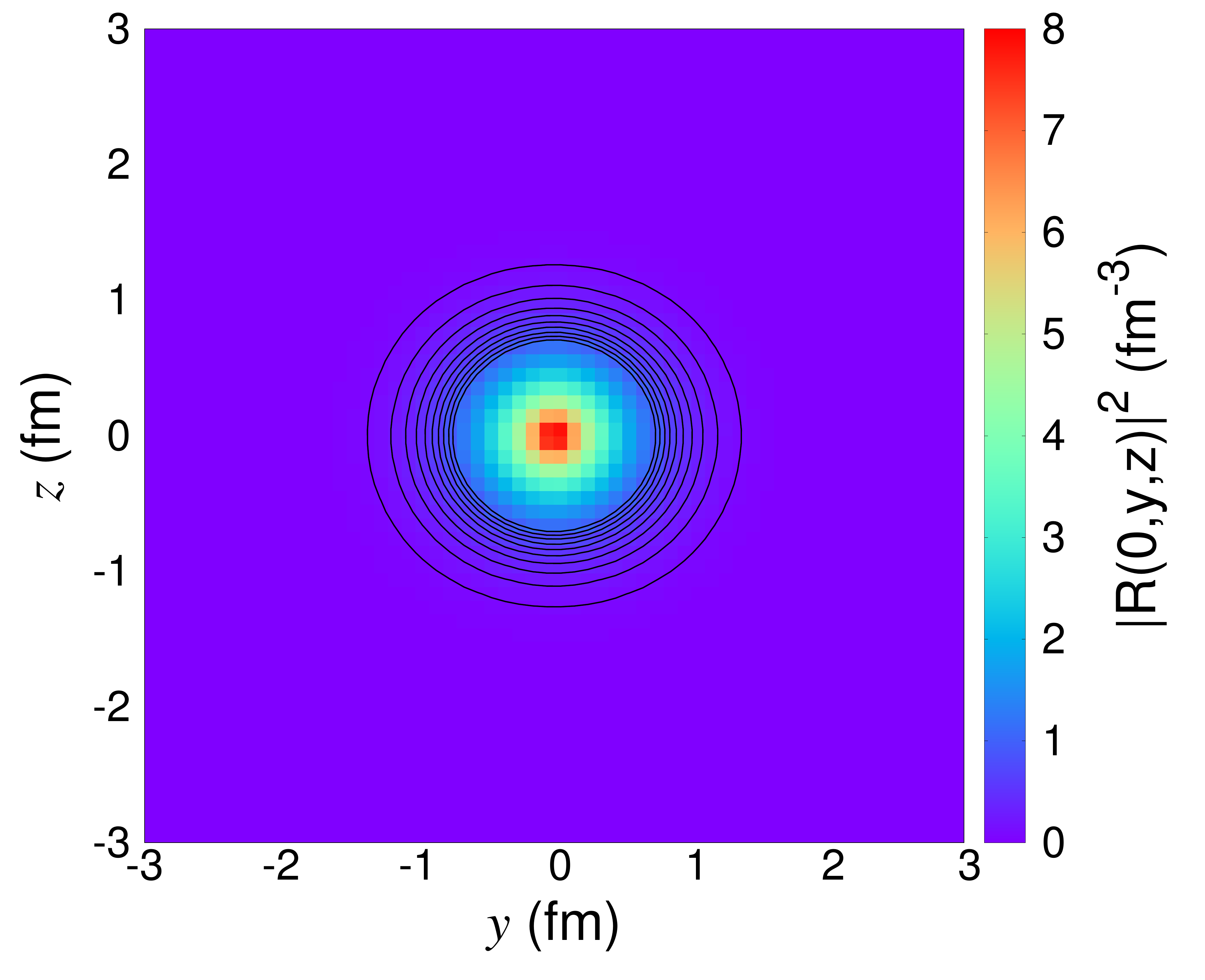}
      \includegraphics[width = 6.5cm]{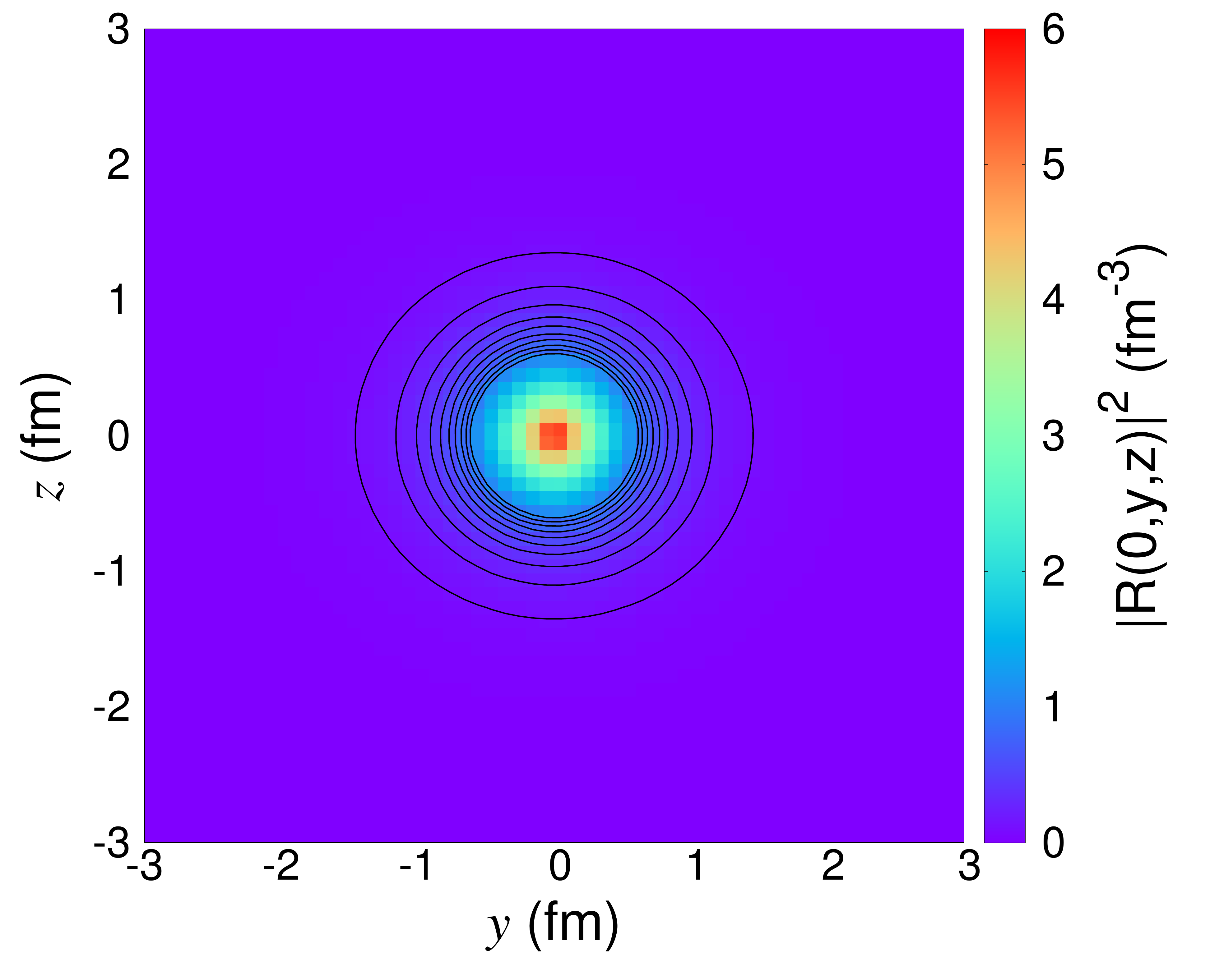}
     \caption{
The radial distribution of $c(\bar c)$ at $t=0$, $t=0.5\,\text{fm}/c$, $t=1\,\text{fm}/c$ and $t=2\,\text{fm}/c$.
}
\label{fig8}
\end{center}
\end{figure}

FIG.\ref{fig8a}-\ref{fig8} shows the evolution of the charmonium system. For simplicity, the initial state is assumed to be $J/\psi$ (in vacuum). The system is with both cooling QGP and the time-dependent electric field. As shown in FIG.\ref{fig8a},  The fraction of $J/\psi$ drops to 0.5 in 2\,fm$/c$ which is faster than the case without the electric field, see FIG.\ref{fig6a}. In particular, the 1S fraction in FIG.\ref{fig8a} drops rapidly in the first 0.2\,fm$/c$ ($\sim$the lifetime of the electric field at $\gamma \simeq 100$). We note that the transition from 1S to 2S induced by the electric field is suppressed due to the selection rules for the electric dipole transitions. We also note that $\chi_c$ is generated due to the strong electric field. As a result the radial distribution of $c(\bar c)$ deviates from spherical symmetry.

\begin{figure}[!htb]
   \begin{center}
      \includegraphics[width = 8.0cm]{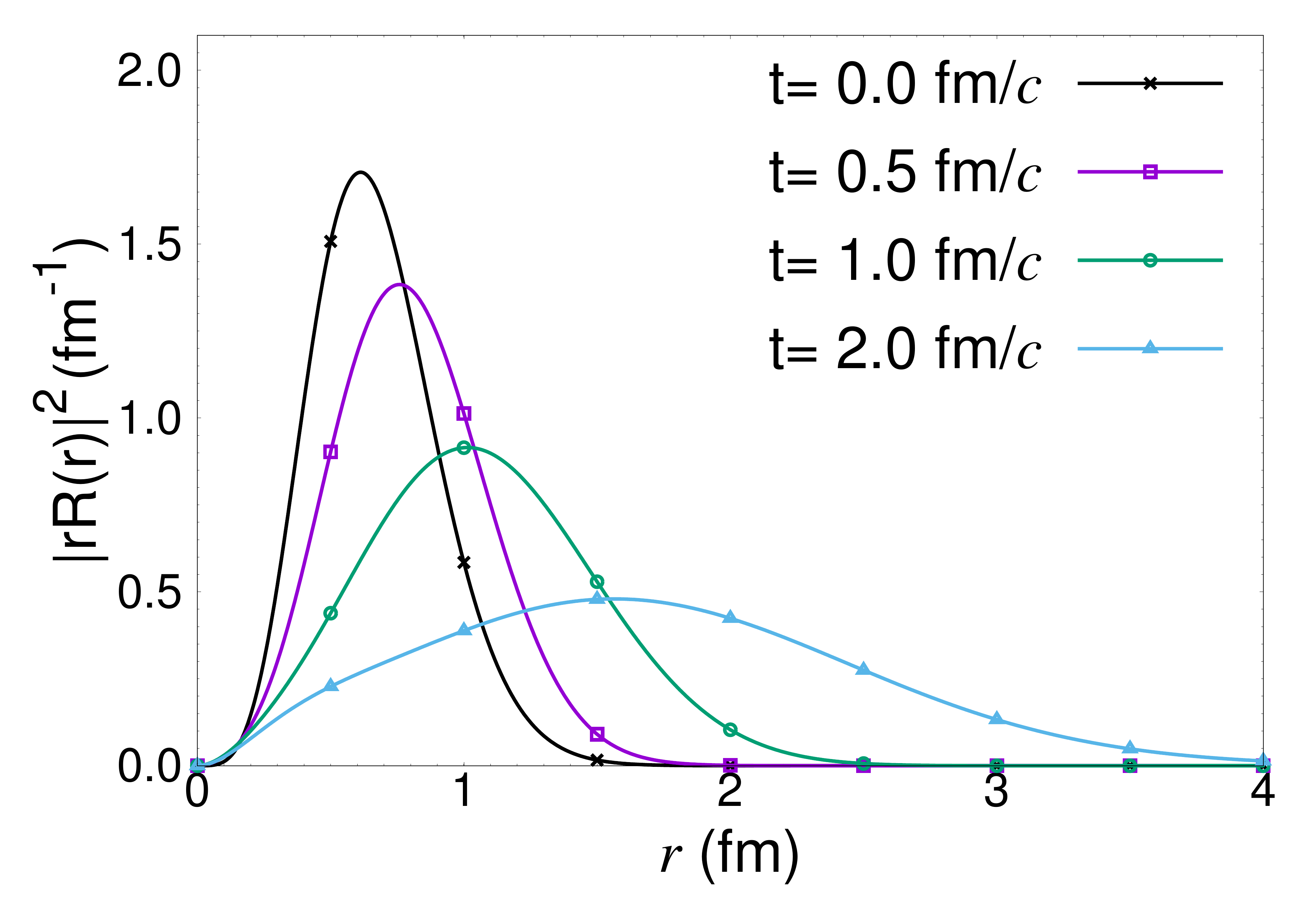}
      \includegraphics[width = 8.0cm]{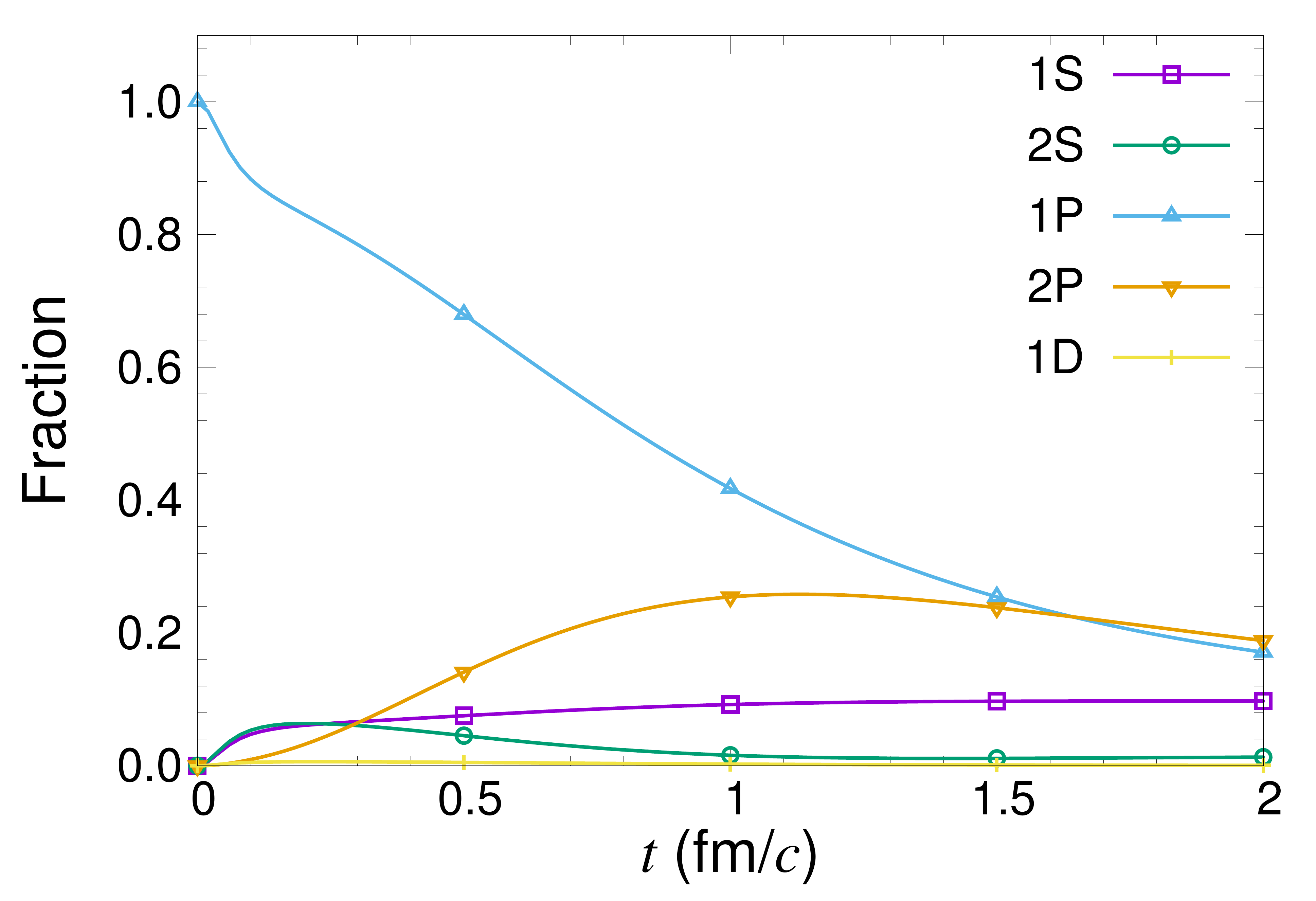}
     \caption{
The initial state is  $\chi_c$ (in vacuum). The radial distribution $|rR(r,t)|^2$ of $\chi_c$ is plotted in the left panel of the first row. The fractions are plotted in the right panel of the first row.
}\label{fig9a}
\end{center}
\end{figure}

\begin{figure}[!htb]
   \begin{center}
      \includegraphics[width = 6.5cm]{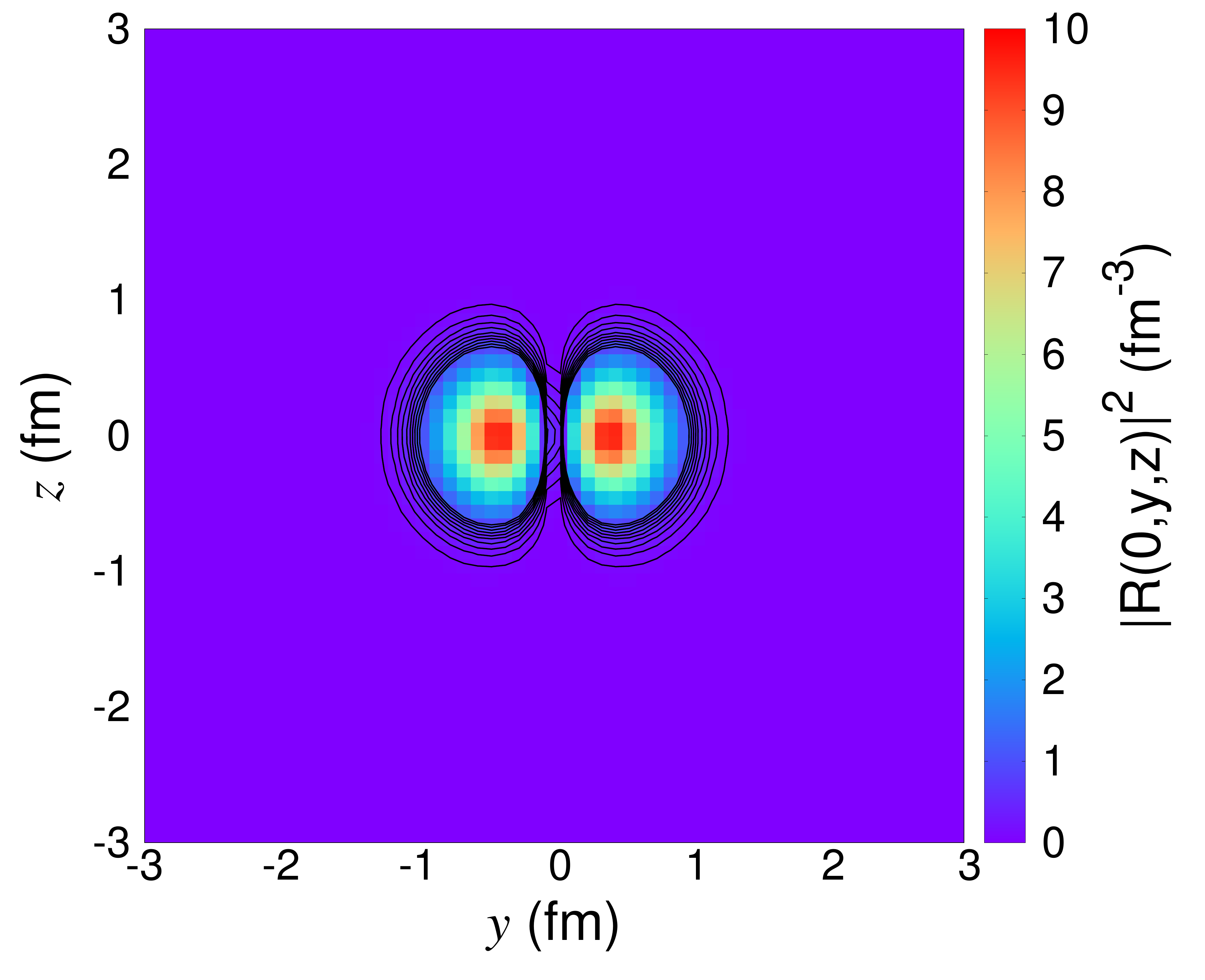}
      \includegraphics[width = 6.5cm]{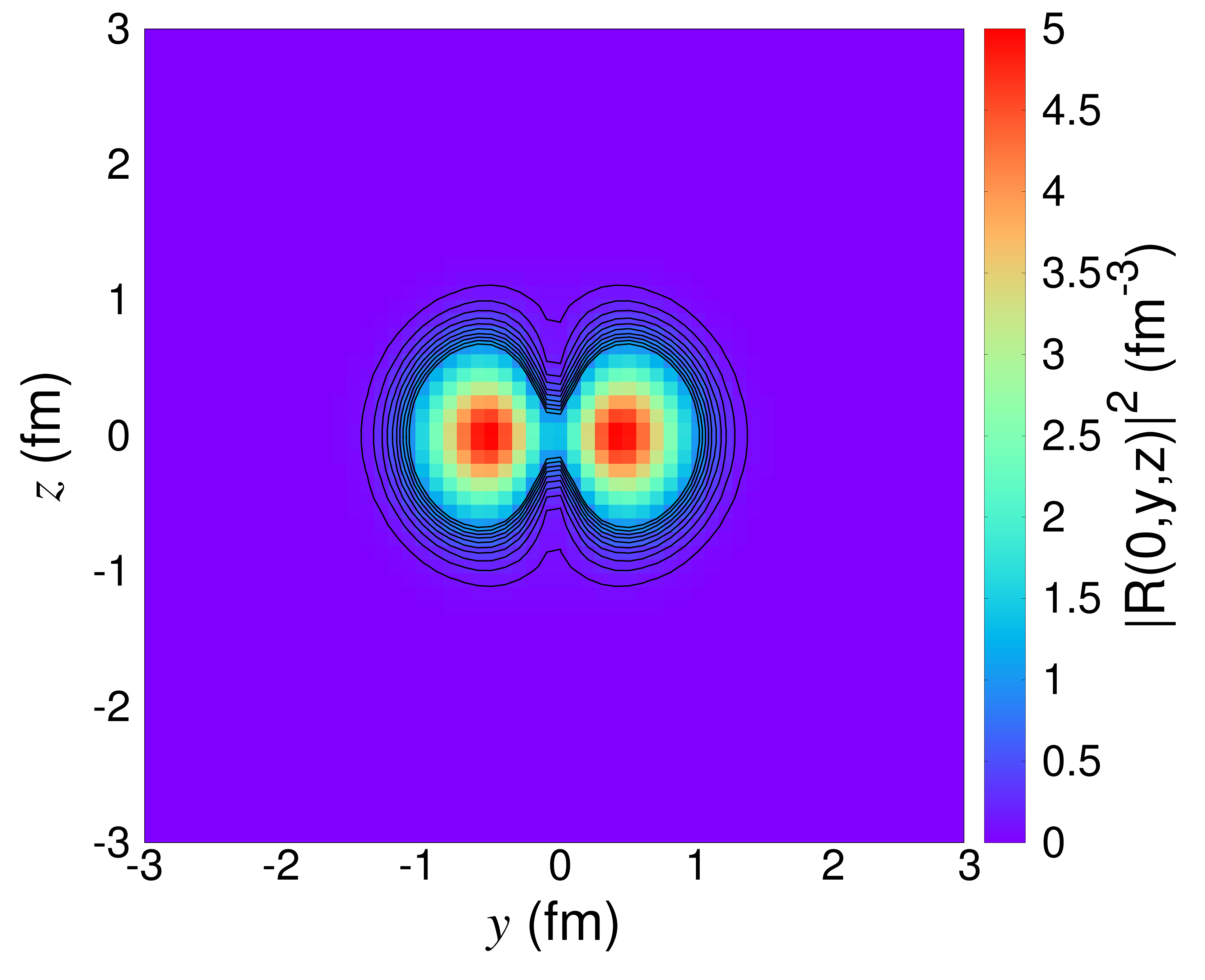}
      \includegraphics[width = 6.5cm]{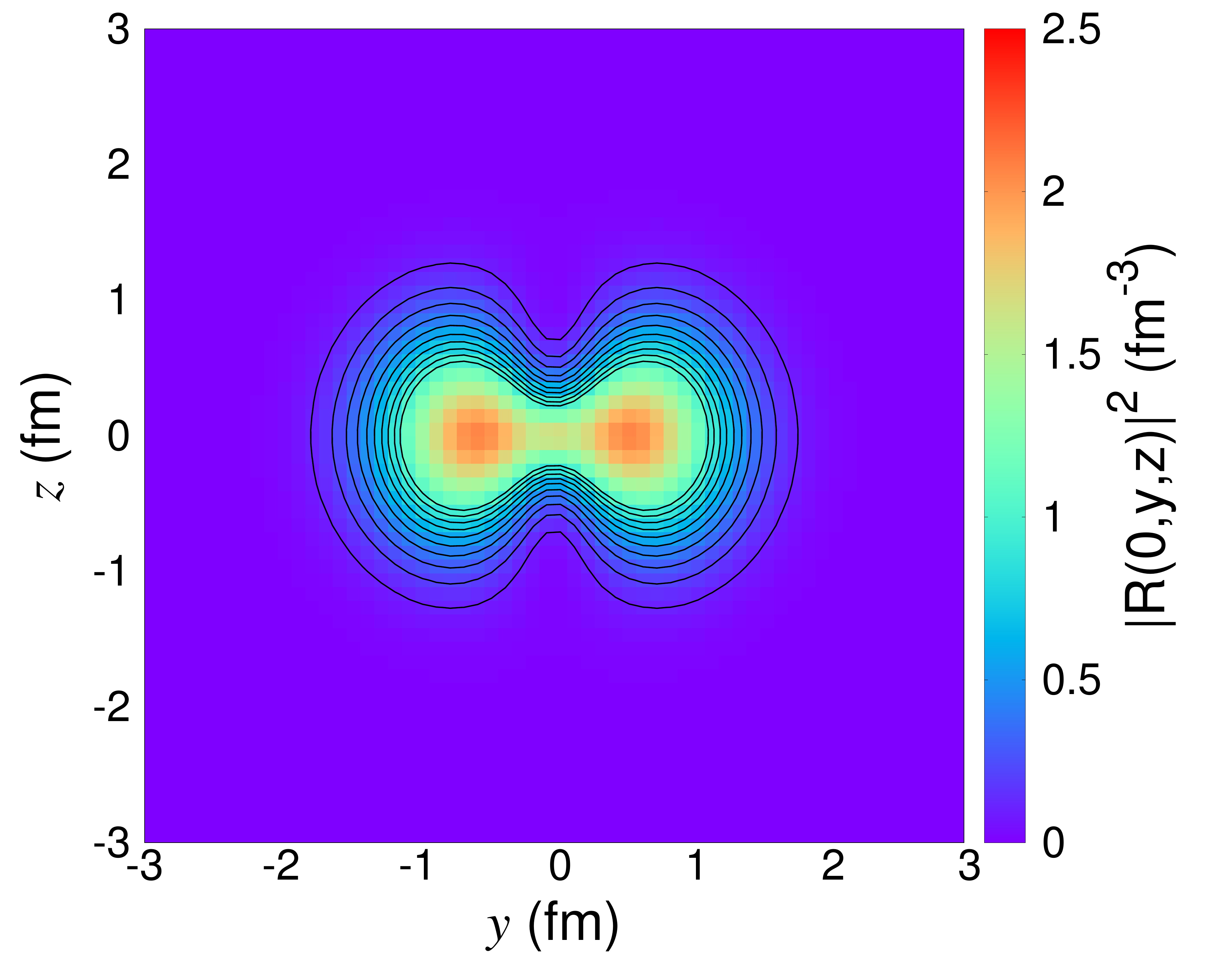}
      \includegraphics[width = 6.5cm]{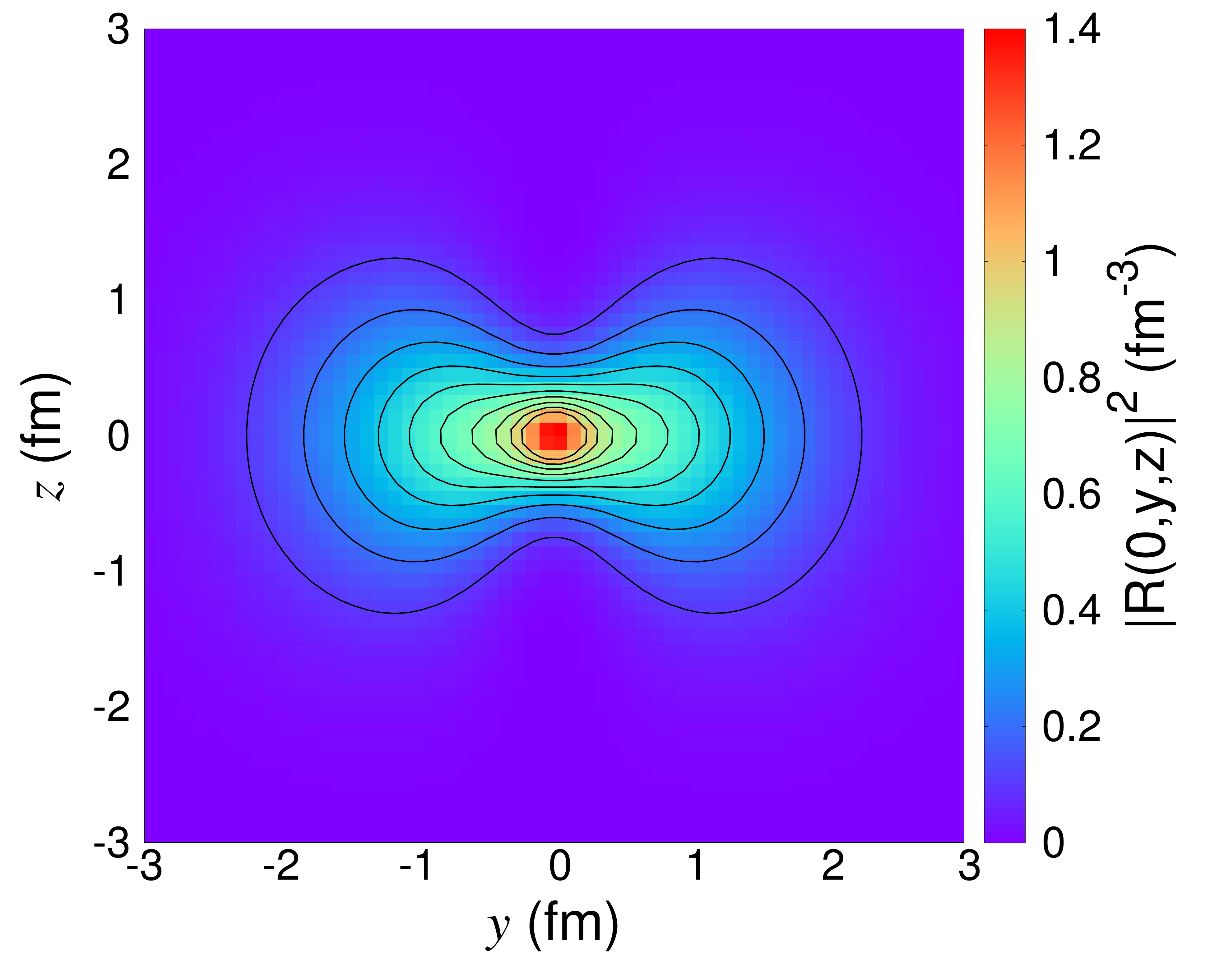}
     \caption{
The radial distribution of $c(\bar c)$ at $t=0$, $t=0.5\,\text{fm}/c$, $t=1\,\text{fm}/c$ and $t=2\,\text{fm}/c$. 
}
\label{fig9}
\end{center}
\end{figure}

In FIG.\ref{fig9a}-\ref{fig9}, the initial state of the charmonium is taken to be $\chi_c$ (in vacuum).
Comparing to the case without the electric field (see FIG.\ref{fig7a}), electric field induces transitions from 1P to 1S and 2S states. The fractions of the S-wave states and D-wave states are non-zero at $t$=2\,fm$/c$, which are from the transitions of the $\chi_c$ state. The contour plots indicate the S-wave components in the wave function because only S-wave states have non-zero probability at the origin. We note that the 1D state is slightly produced in the first 1\,fm$/c$ and is dissociated subsequently.

\subsubsection{Pb-Pb collision with $\gamma\simeq 1000$}
In our calculation we take $\gamma= 1000$ for simplicity. 
\begin{figure}[!htb]
   \begin{center}
      \includegraphics[width = 8.0cm]{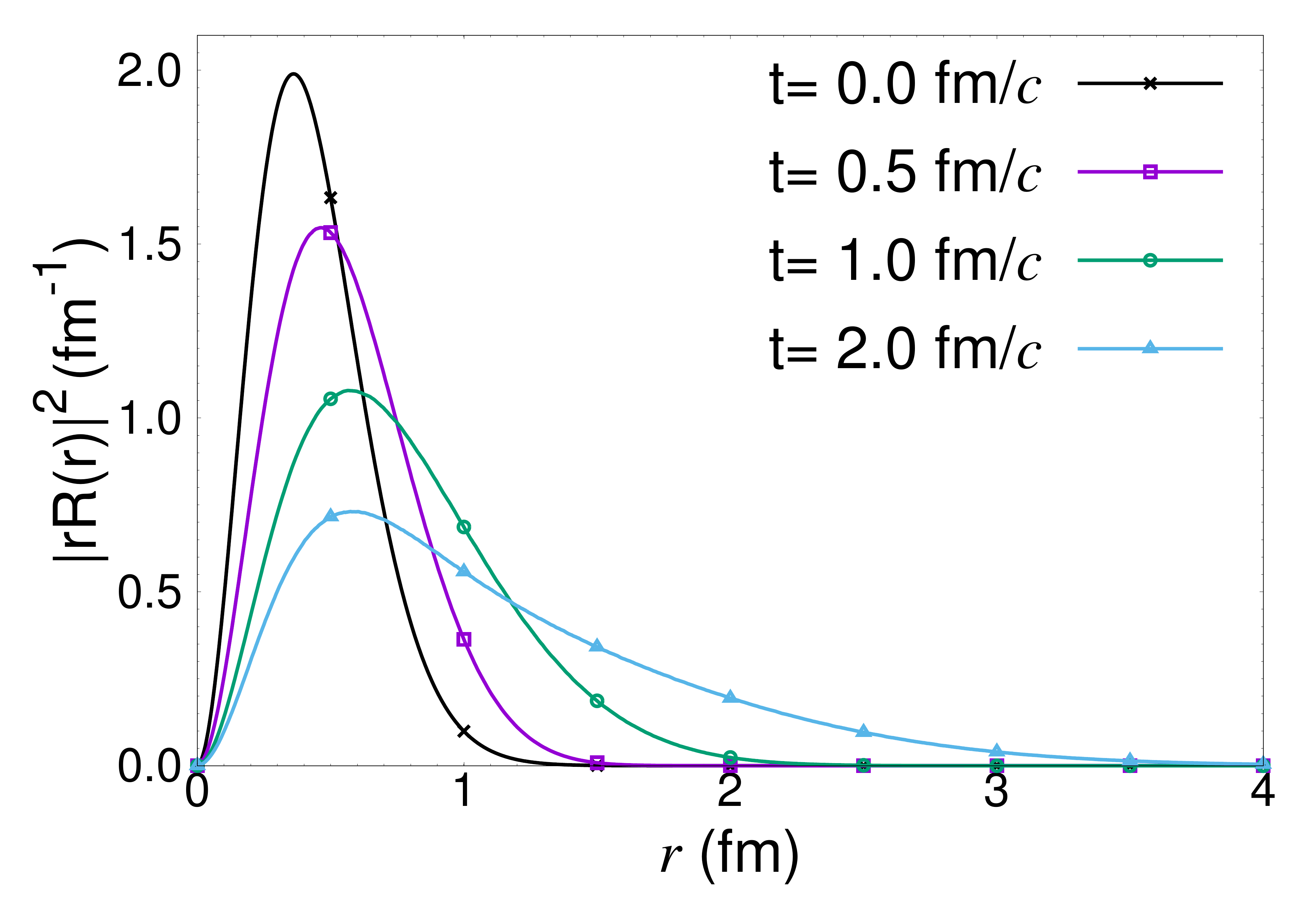}
      \includegraphics[width = 8.0cm]{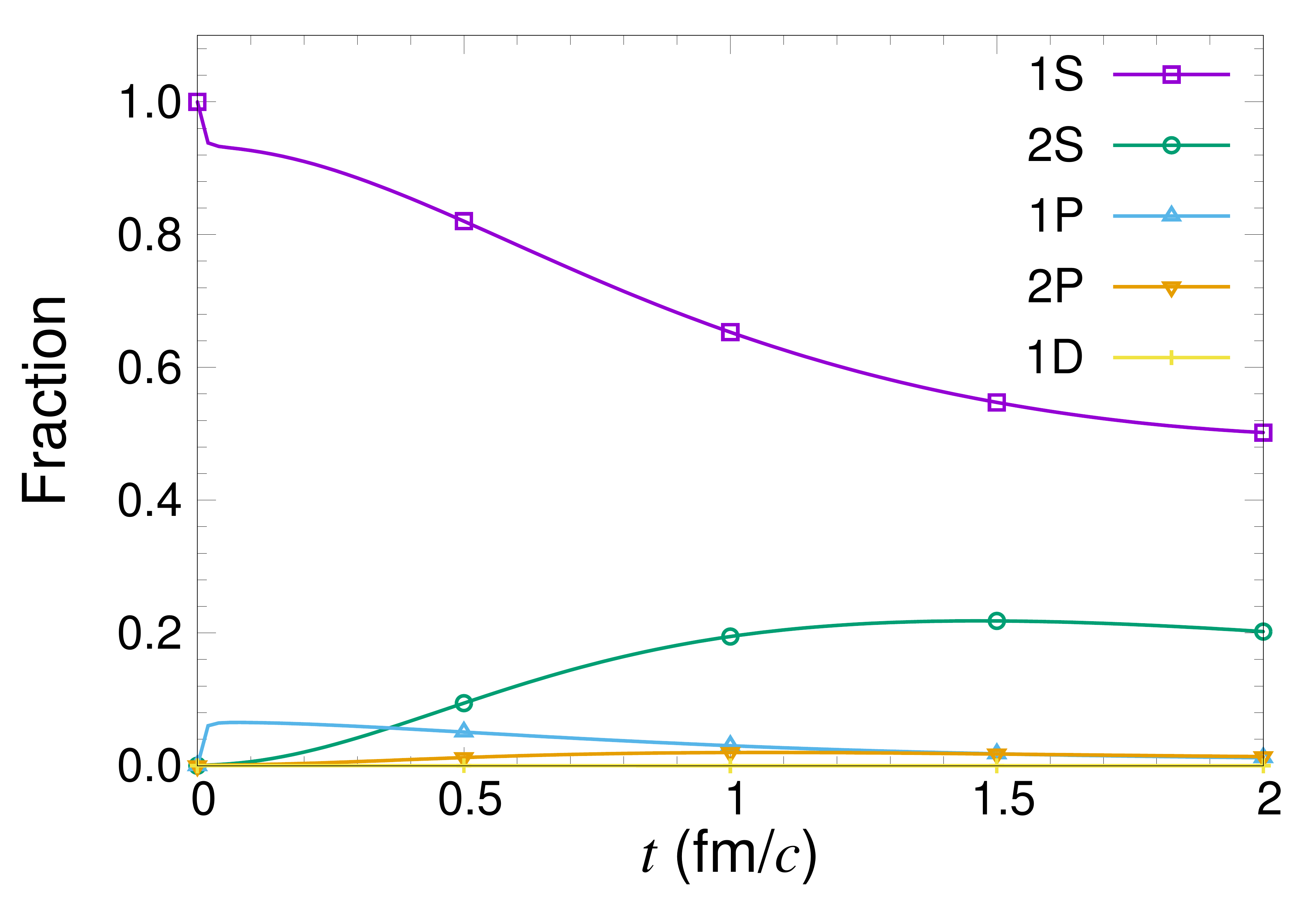}
     \caption{The initial state is $J/\psi$ (in vacuum). The radial distribution $|rR(r,t)|^2$ is plotted in the left panel. The fractions are plotted in right panel.
     }\label{fig10a}
\end{center}
\end{figure}
\begin{figure}[!htb]
   \begin{center}
      \includegraphics[width = 6.5cm]{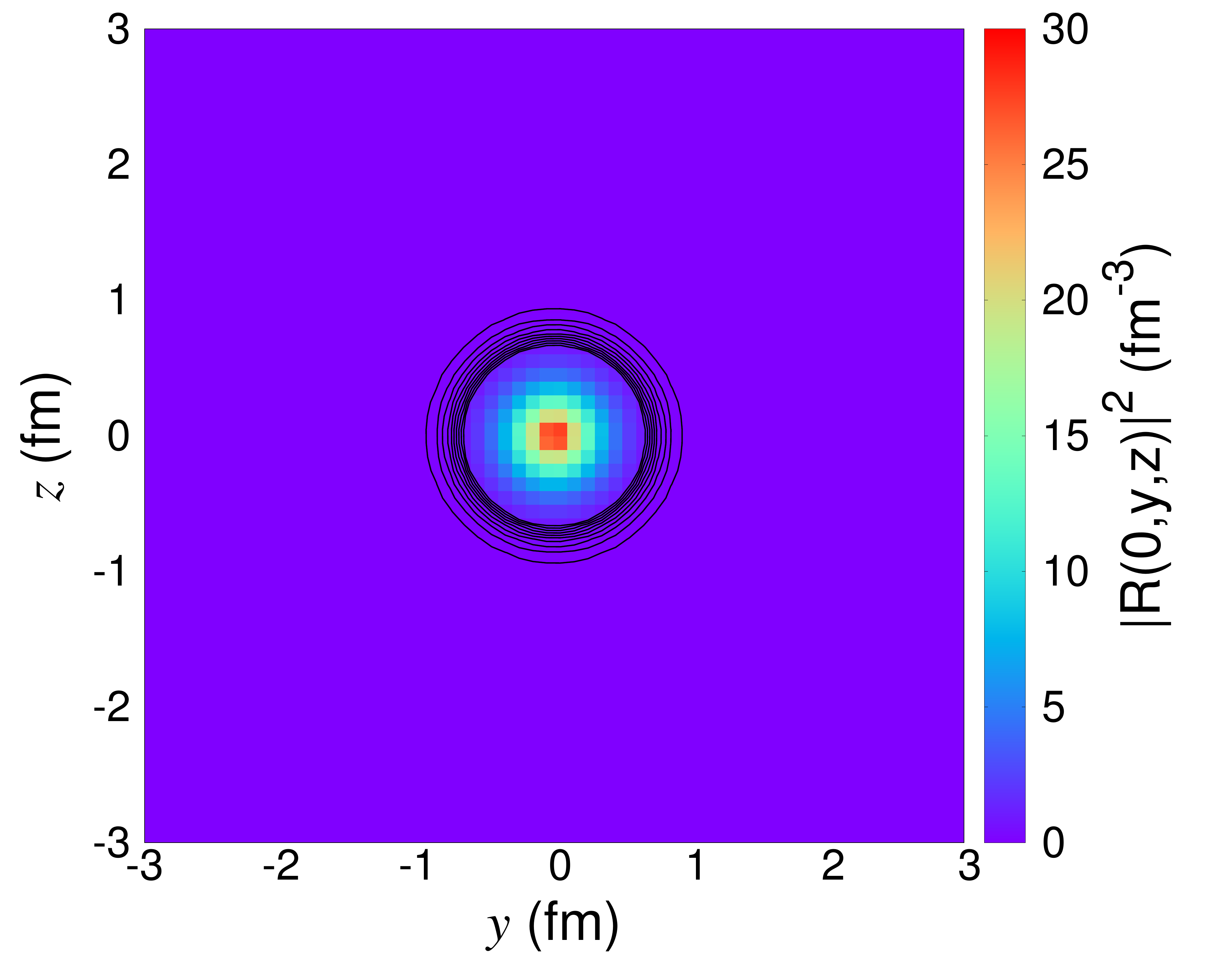}
      \includegraphics[width = 6.5cm]{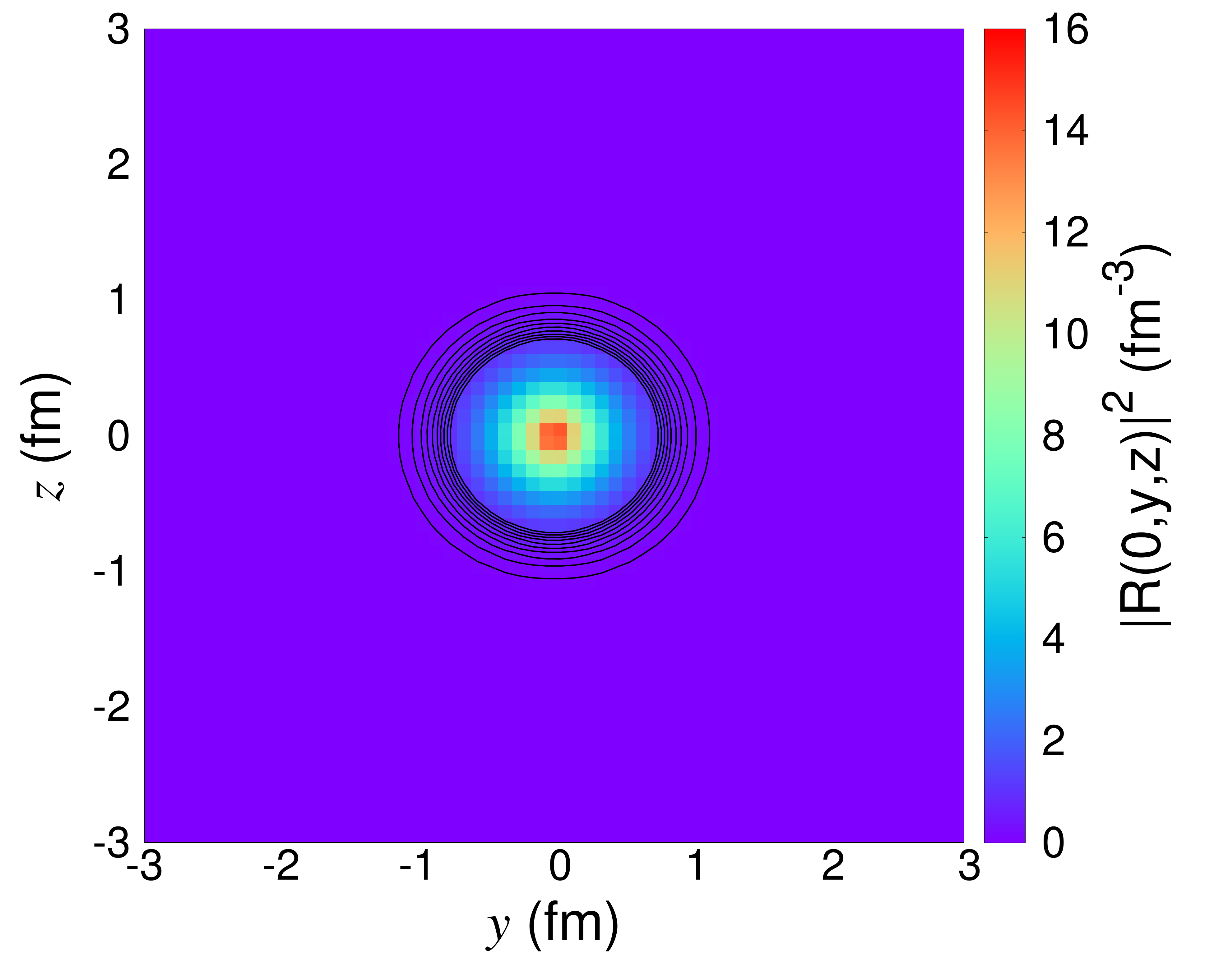}
      \includegraphics[width = 6.5cm]{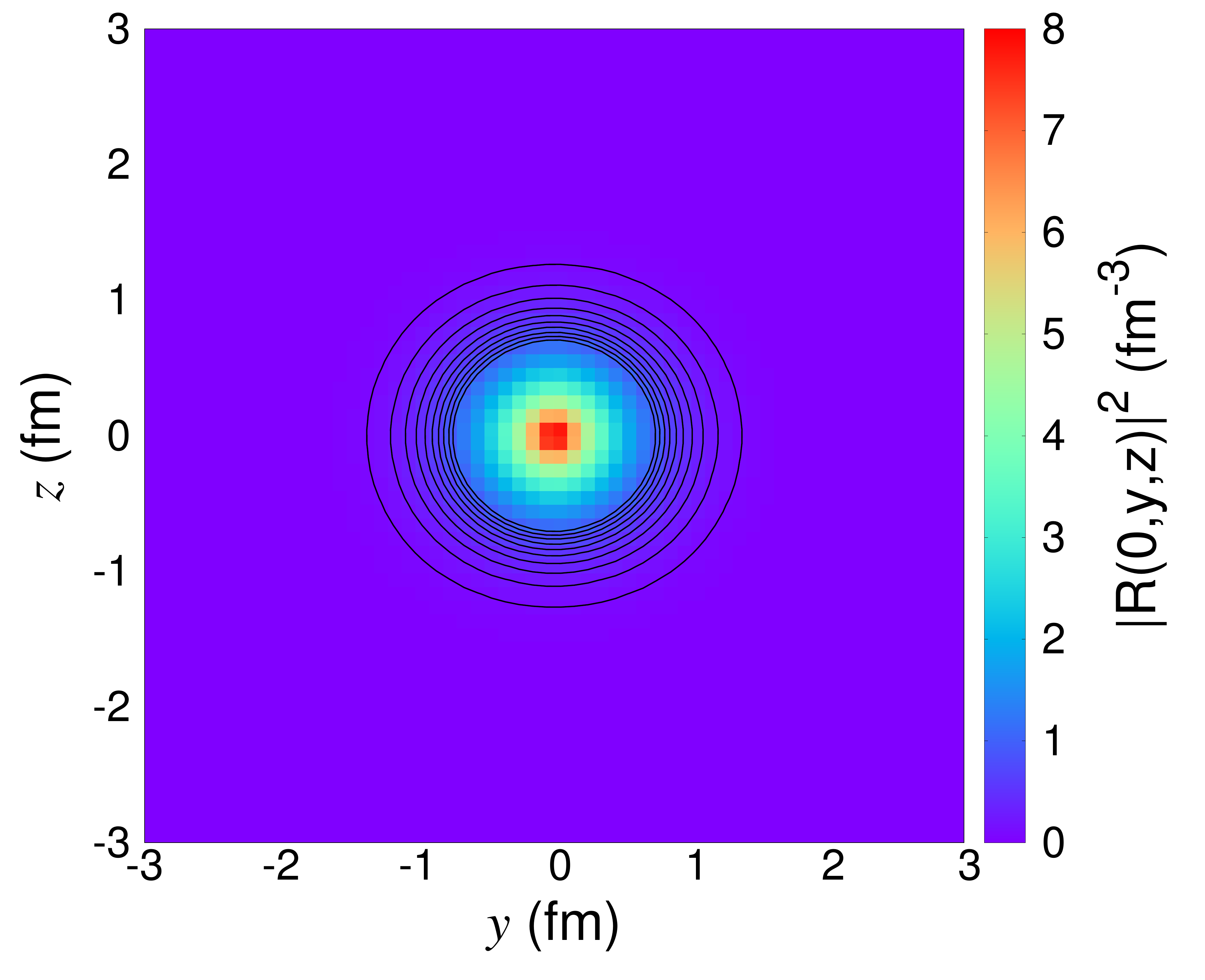}
      \includegraphics[width = 6.5cm]{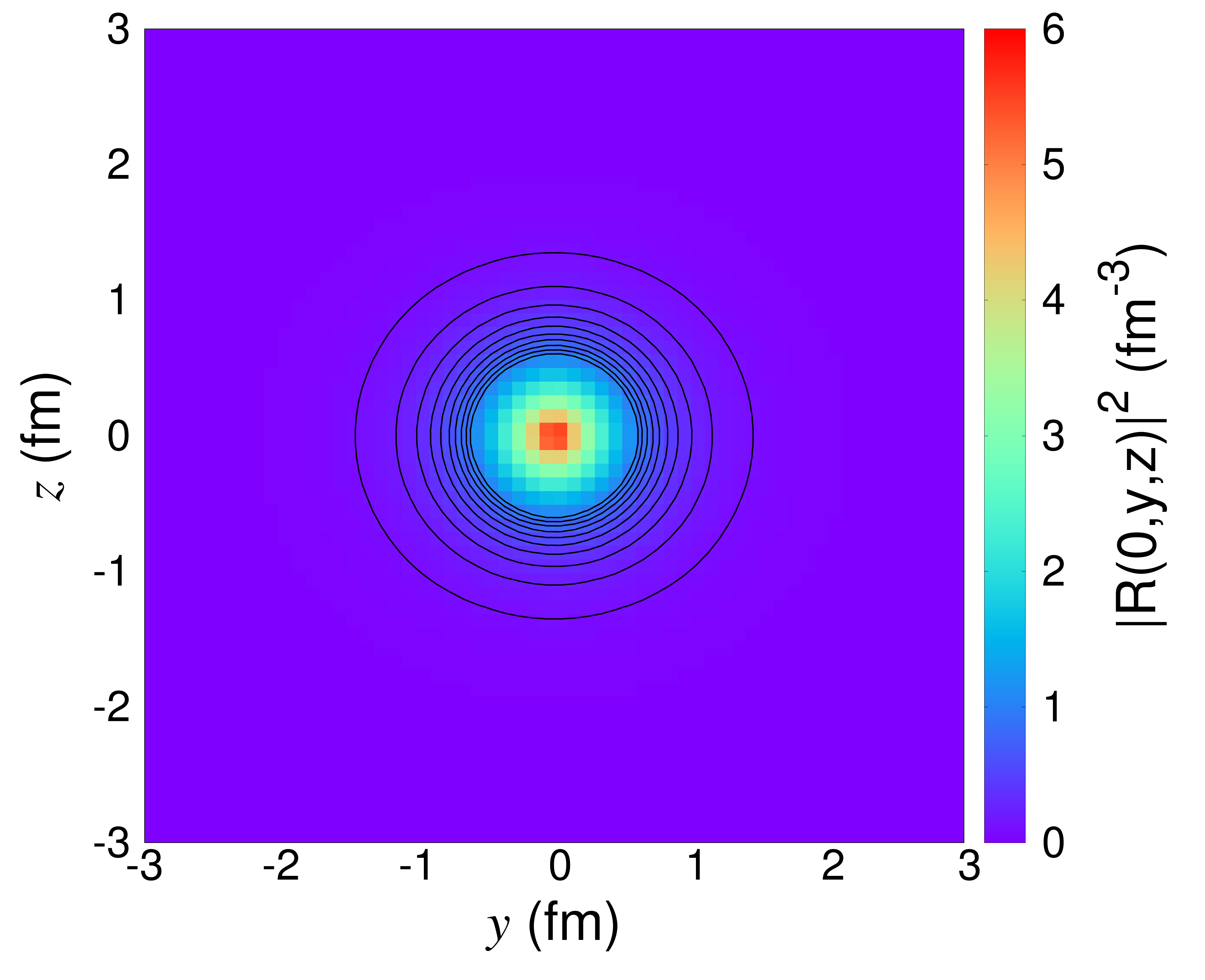}
     \caption{
 The radial distribution of $c(\bar c)$ at $t=0$, $t=0.5\,\text{fm}/c$, $t=1\,\text{fm}/c$ and $t=2\,\text{fm}/c$.
}
\label{fig10}
\end{center}
\end{figure}

FIG.\ref{fig10a}-\ref{fig10} shows the evolution of the charmonium system from the initial state of $J/\psi$ (in vacuum). The system is with both cooling QGP and the time-dependent electric field.
In FIG.\ref{fig10a}, in the first 0.02fm$/c$, the strong electric field causes a significant drop in the fraction of 1S state and a rapid increase in the fraction of 1P state. The duration in which 1S fraction drops fastest is approximately the lifetime of the electric field at $\gamma = 1000$ as shown in FIG.\ref{fig2}. And the contour plots show that the radial distribution of $c(\bar c)$ deforms from spherical symmetry significantly due to the strong electric field.
\begin{figure}[!htb]
   \begin{center}
      \includegraphics[width = 8.0cm]{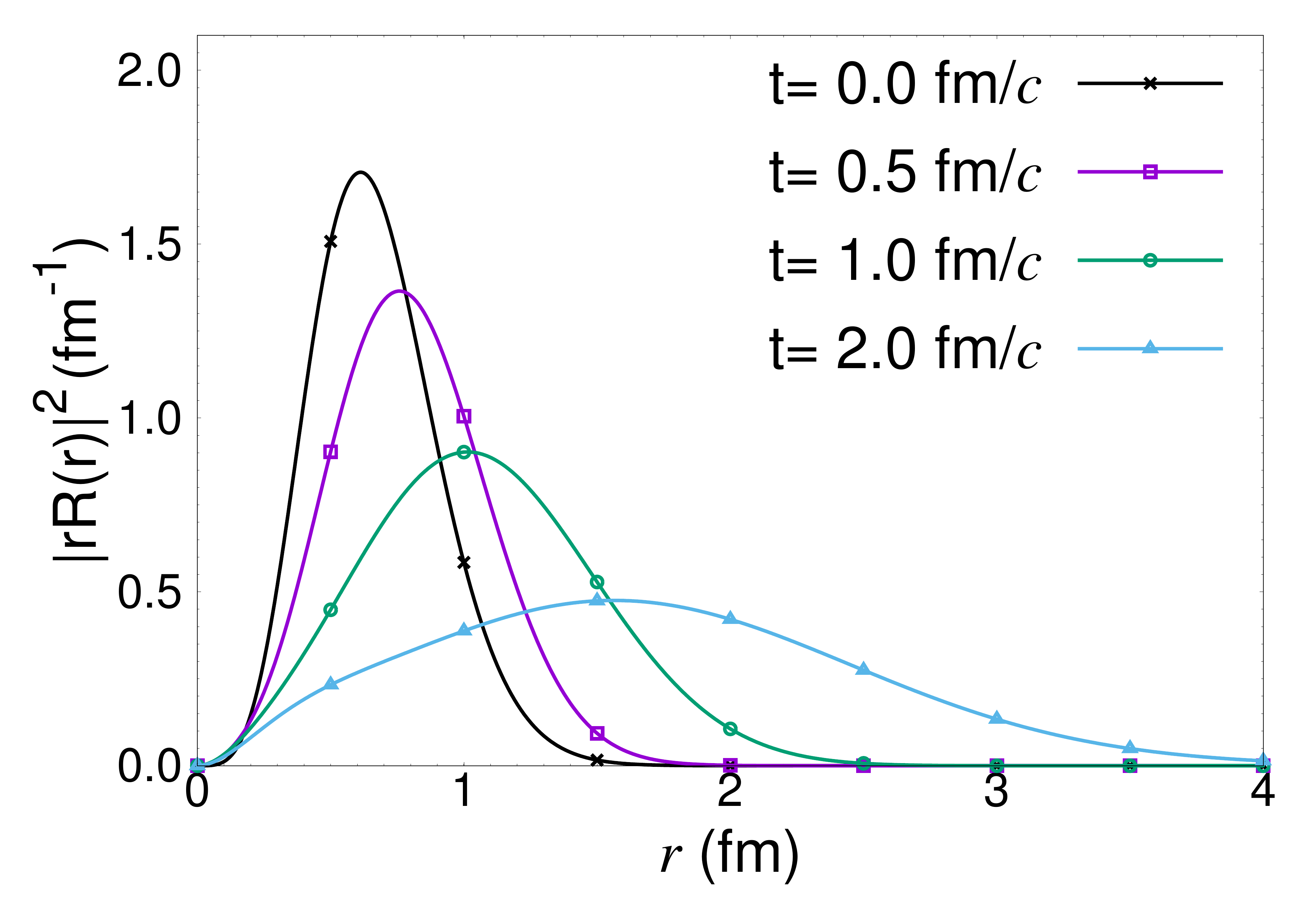}
      \includegraphics[width = 8.0cm]{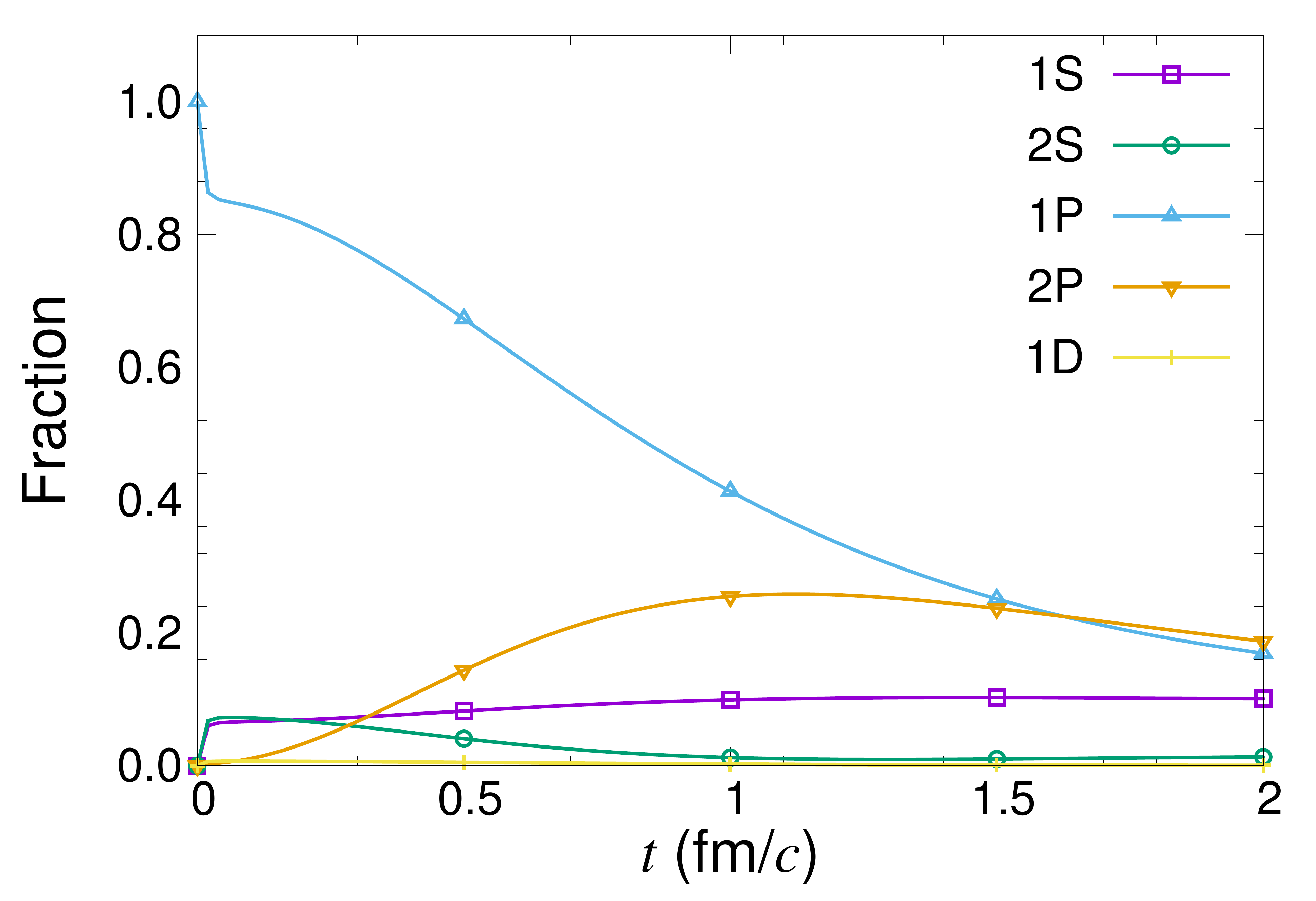}
\end{center}
\end{figure}
\begin{figure}[!htb]
   \begin{center}
      \includegraphics[width = 6.5cm]{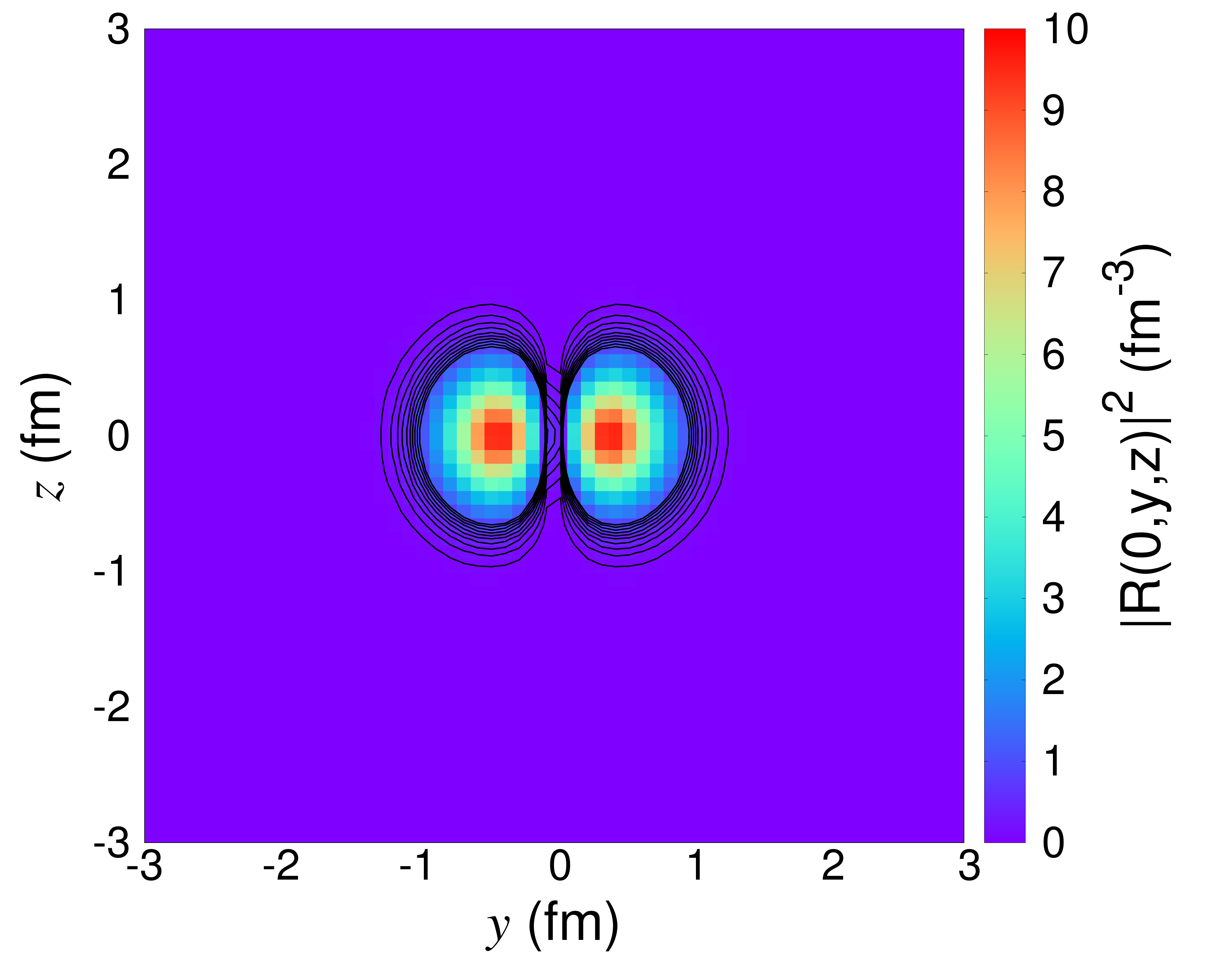}
      \includegraphics[width = 6.5cm]{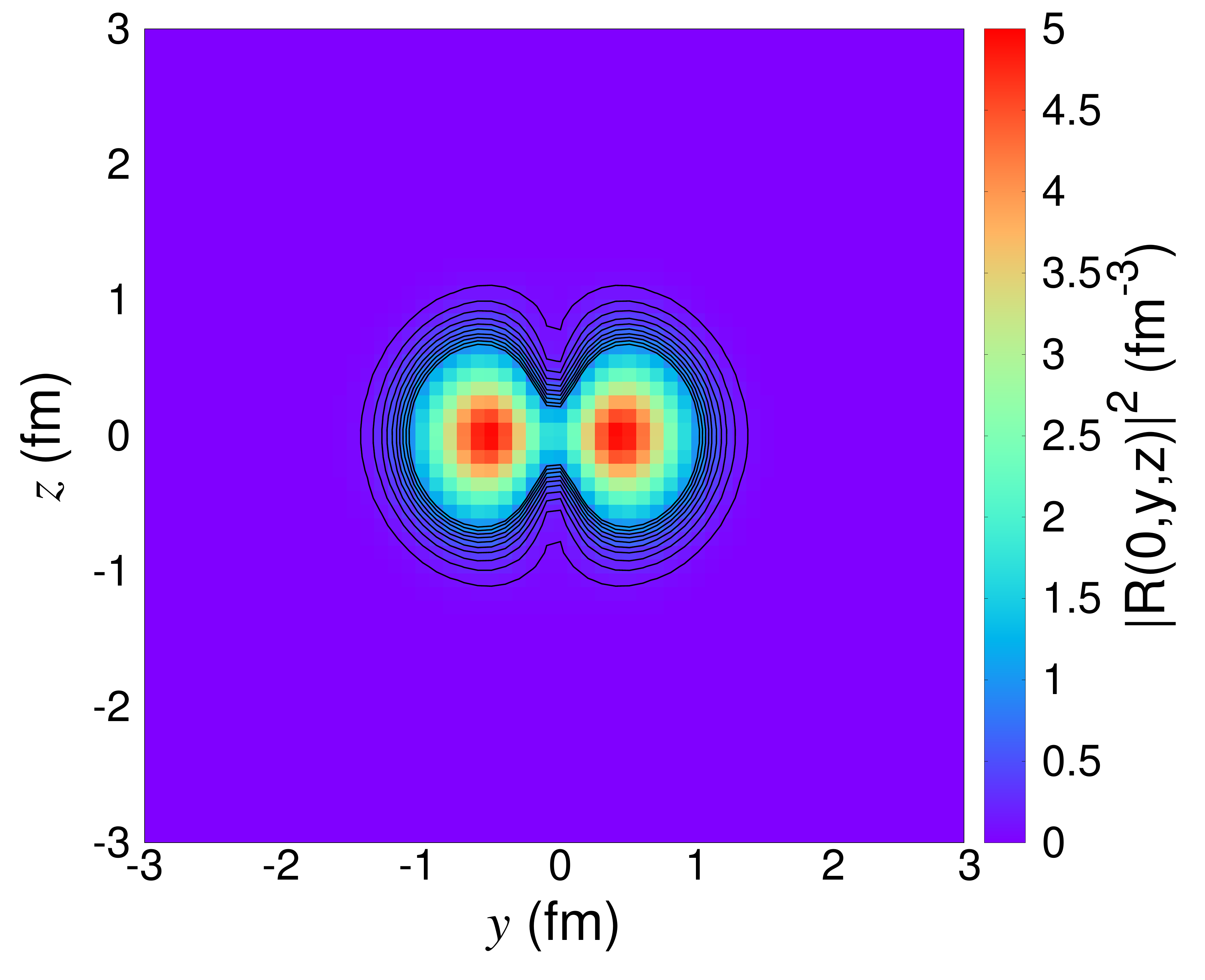}
      \includegraphics[width = 6.5cm]{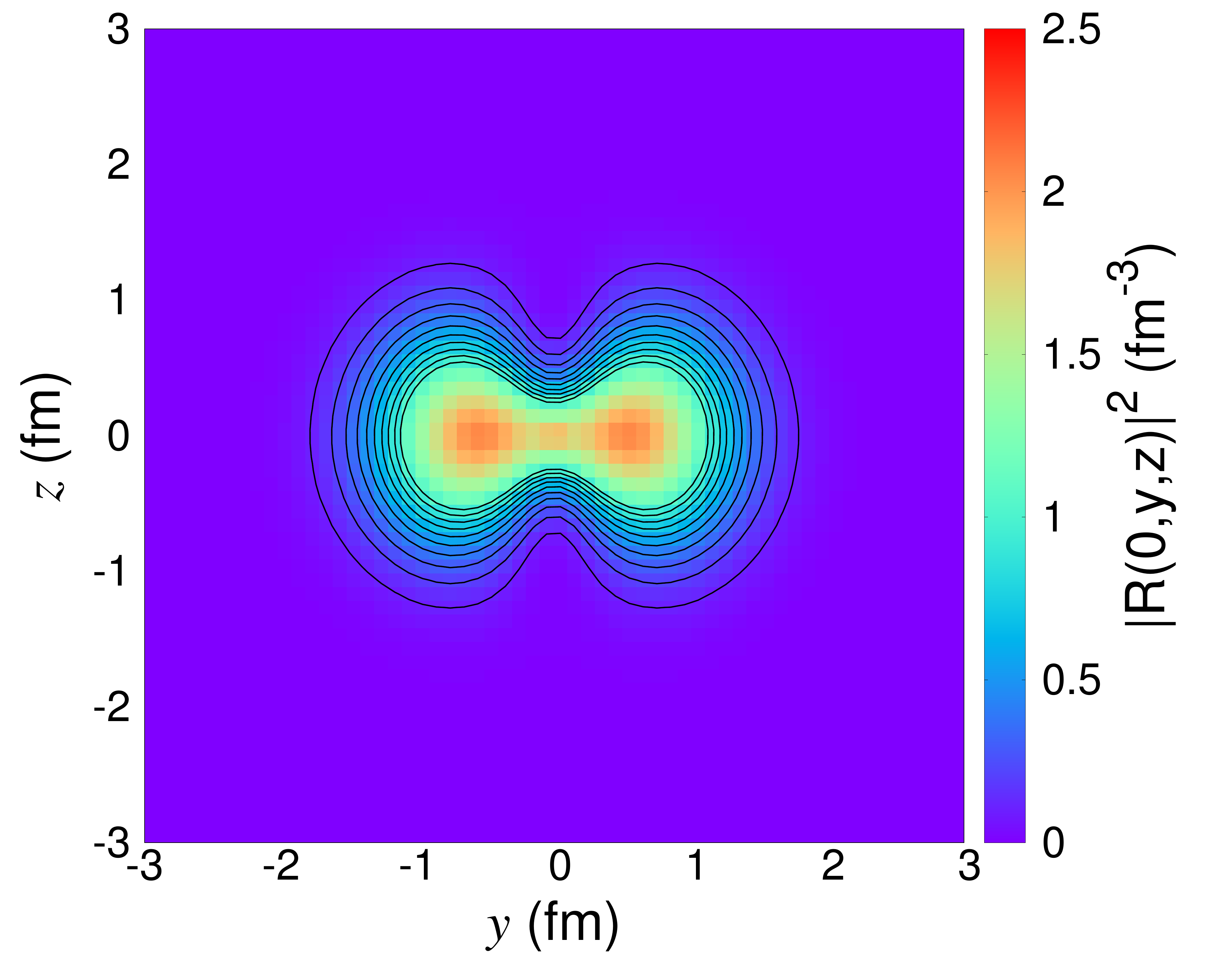}
      \includegraphics[width = 6.5cm]{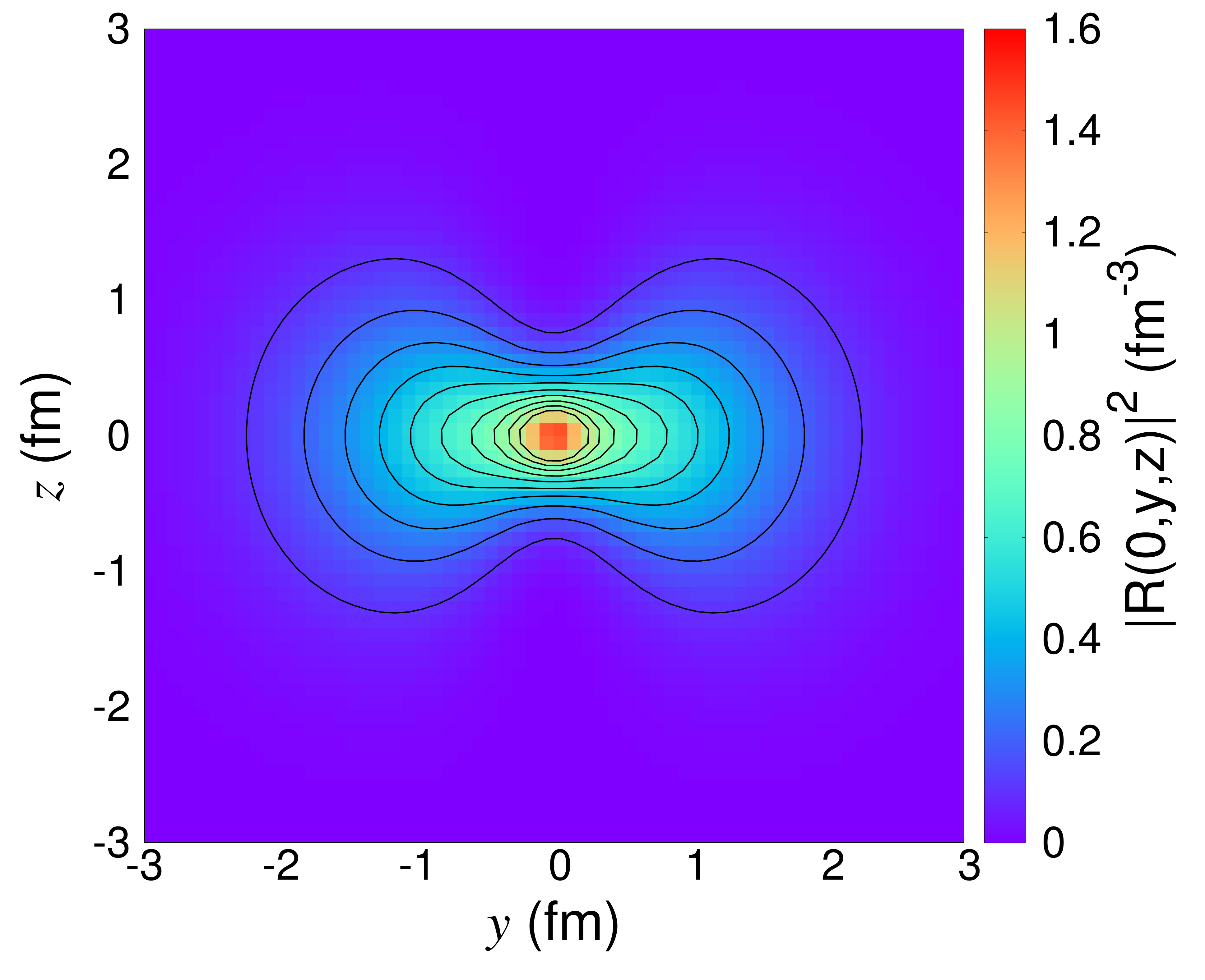}
     \caption{
The initial state is $\chi_c$ (in vacuum). The radial distribution $|rR(r,t)|^2$ of $\chi_c$ is plotted in left panel of the first row. The fractions are plotted in right panel of the first row. And the radial distribution of $c(\bar c)$ at $t=0$, $t=0.5\,\text{fm}/c$, $t=1\,\text{fm}/c$ and $t=2\,\text{fm}/c$.
}
\label{fig11}
\end{center}
\end{figure}

 FIG.\ref{fig11} shows the evolution of $\chi_c$ in the screened Cornell potential with decreasing temperature and the electric field generated in heavy-ion collisions with $\gamma\simeq 1000$.

Again, in the first 0.01fm$/c$, the electric field causes a significant drop in the fraction of 1P state and correspondingly a rapid increase in the fractions of 1S and 2S state. This is different from FIG.\ref{fig10a} where only 1P state is generated rapidly.

\section{Conclusions and outlook}\label{Discussion}

In relativistic heavy-ion collisions, extremely hot medium can be produced which is expected to be the deconfined phase of nuclear matter. Besides, strong electric field is produced when two nuclei collide with each other at nearly the speed of light. We study the dissociation and transitions between different charmonium states caused by the electric field as well as the hot medium in Sch\"odinger equation formalism.

The electric field with large magnitude generates significant effects in the charmonium production in the early stage of the collisions. The charmonium states are dissociated more strongly than the case without electric field.
 Due to the selection rule for electric dipole transitions, the electric field converts some $J/\psi$ to $\chi_c$ states and vice versa.

In the future we plan to extend our study to a realistic non-uniform electric field. We will also adopt the more realistic initial states from pQCD calculations and a realistic temperature evolution profile. The effect of strong magnetic field created in relativistic heavy-ion collisions will be considered in our future work as well.
In a similar way, this approach can be also used to study the evolution of bottomonia in QGP.

{\bf Acknowledgments}: We thank Xiaojian Du for his codes.
This work is supported by the new faculty startup funding by the Institute of Modern Physics, Chinese Academy of Sciences. This work is also supported by NSFC Grant No. 11705125.\\

\bibliographystyle{ieeetr}
\bibliography{ref}

\end{document}